\documentclass[fleqn]{article}
\usepackage{fleqn}
\usepackage{graphicx}
\usepackage{epsfig}
\usepackage{amsmath}
\usepackage{amsthm}
\usepackage{amssymb}
\usepackage{enumerate}
\usepackage{calc}
\usepackage{color}
\usepackage{upgreek}
\newcommand{\beq}{\begin{equation}}
\newcommand{\eeq}{\end{equation}}
\newcommand{\bea}{\begin{eqnarray}}
\newcommand{\eea}{\end{eqnarray}}
\theoremstyle{plain}
\newtheorem{lem}{Lemma}
\newtheorem{thm}{Theorem}
\theoremstyle{definition}
\newtheorem{defn}{Definition}
\theoremstyle{remark}

\newcommand{\Quad}{Q}

\usepackage{hyperref}
\hypersetup{
 pdfauthor={P\'eter L\'evay},
  pdftitle={A magic Veldkamp line for three qubits. Representations and geometry.},
   pdfproducer={LaTeX with hyperref},
    pdffitwindow=false,
     pdfstartview={FitH},
      colorlinks=true,
       linkcolor={blue},
        citecolor={blue},
	 filecolor={blue},
	  urlcolor={blue}
	  }
	  
\begin{document}
	  \Large
	  \begin{center}
	  {\bf The magic three-qubit Veldkamp line: A finite geometric underpinning for form theories of gravity and black hole entropy}
	  \end{center}
	  \large
	  \vspace*{-.1cm}
	  \begin{center}
	  P\'eter L\'evay$^{1,2}$ , Fr\'ed\'eric Holweck$^{3}$ and Metod Saniga$^{4}$ 
	   \end{center}
	   \vspace*{-.4cm} \normalsize
	   \begin{center}

$^{1}$Department of Theoretical Physics, Institute of Physics, Budapest University of\\
Technology and Economics 

$^{2}$MTA-BME Condensed Matter Research Group, H-1521 Budapest, Hungary

$^{3}$Laboratoire Interdisciplinaire Carnot de Bourgogne, ICB/UTBM, UMR 6303 CNRS, Universit\'e Bourgogne Franche-Comt\'e, 90010 Belfort Cedex, France

$^{4}$Astronomical Institute, Slovak Academy of Sciences, SK-05690 Tatransk\'a Lomnica, Slovak Republic

\vspace*{.0cm}

\vspace*{.2cm} (4 April 2017)

\end{center}

\vspace*{-.3cm} \noindent \hrulefill

\vspace*{.1cm} \noindent {\bf Abstract:}

We investigate the structure of the three-qubit magic Veldkamp line (MVL).
This mathematical notion has recently shown up as a tool for understanding the structures of the set of Mermin pentagrams, objects that are used
to rule out
certain classes of hidden variable theories.
Here we show that this object also provides a unifying
finite geometric underpinning for understanding the structure  of functionals used in form theories of gravity and black hole entropy.
We clarify the representation theoretic, finite geometric and physical meaning of the different parts of our MVL.
The upshot of our considerations is that the basic finite geometric objects enabling such a diversity of physical applications of the MVL are  the \emph{unique} generalized quadrangles with lines of size three, their one point extensions as well as their other extensions isomorphic to affine polar spaces of rank three and order two.
In a previous work we have already connected generalized quadrangles to the structure of cubic Jordan algebras 
related to
entropy fomulas of black holes and strings in \emph{five} dimensions.
In some respect the present paper can be regarded as a generalization of that analysis for also providing a finite geometric understanding of \emph{four}-dimensional black hole entropy formulas.
However, we find many more structures whose physical meaning is yet to be explored.
 As a familiar special case our work provides a finite geometric representation of the algebraic extension from cubic Jordan algebras to Freudenthal systems based on such algebras.

 \vspace*{.3cm} \noindent
 \hspace*{1.95cm} --

 \vspace*{-.2cm} \noindent \hrulefill

 \section{Introduction}

In quantum information instead of bits we use qubits.
Qubits are elments of a two dimensional complex vector space $\mathbb{C}^2$.
The basic observables for a single qubit are the Pauli operators $I,X,Y,Z$ where $I$ is the identity operator and the remaining ones are the usual operators represented by the Pauli spin matrices.
For a system consisting of $N$ qubits quantum states correspond to the rays of the $N$-fold tensor product space
$\mathbb{C}^2\otimes\cdots \otimes\mathbb{C}^2$ and the simplest type of observables being the $N$-fold tensor products of the single qubit Pauli operators. Since the algebra of these simple $N$-qubit observables is a non-commutative one, commuting subsets of observables enjoy a special status.
Special arrangements of observables containing such commuting subsets are widely used in quantum theory.

Perhaps the most famous arrangements of that type are the ones that show up in considerations
revisiting the famous proofs of the Kochen-Specker\cite{KS} and Bell theorems\cite{Bell}.
Using special configurations of two, three and four qubits Peres\cite{Peres}, Mermin\cite{Mermin} and Greenberger, Horne and Zeilinger\cite{Horne} have provided a new way of looking at these theorems. A remarkable feature appearing in
these works was that they were able to rule out
certain classes of hidden variable theories without the use of probabilities.
For the special configurations featuring commuting subsets of simple observables, the terms Mermin squares and pentagrams were coined.
Since the advent of quantum information theory similar structures have been under an intense
scrutiny\cite{Waegell,Waegell3qbit,Saniga1,Levay1,Planat,SLev}.

Another important topic where special commuting sets of Pauli operators are of basic importance is the theory of quantum error-correcting codes.
The construction of such codes is naturally facilitated within the so-called stabilizer formalism\cite{Nielsen,Gottesman,Sloane}.
Here it is recognized that the basic properties of error-correcting codes are related to the fact that two operators in the Pauli group are either commuting or anticommuting.
A quantum error control code
is a
subspace of the $N$-qubit state space. In the theory the code subspace is defined by a set of mutually commuting simple Pauli operators stabilizing it.
Correctable errors are implemented by a special set of operators anticommuting with the generators taken from the commuting subset.

Surprisingly, the third field where Pauli  observables of simple qubit systems turned out to be useful is black hole physics within string theory.
In the so called Black-hole/qubit correspondence\cite{BHQC} it has been observed that simple entangled qubit systems and certain extremal black hole solutions sometimes share identical patterns of symmetry.
In particular, certain macroscopic black hole entropy formulas on the string theoretic side turned out to be identical to certain multiqubit measures of entanglement\cite{DuffCayley}. In the string theoretic context the group of continuous transformations leaving invariant such formulas turned out to contain physically interesting discrete subgroups named the U-duality groups\cite{Town}.  
For example, in the special case of compactifying type IIA string theory on the six-dimensional torus one obtains a classical low energy theory which has on shell continuous $E_{7(7)}$ symmetry\cite{Julia}.
In the quantum theory this symmetry breaks down\cite{Town} to the discrete U-duality group $E_7(\mathbb{Z})$.
This group, in turn, contains the physically important subgroup $W(E_7)$, the Weyl group of the exceptional group $E_7$, implementing a generalization of the electric-magnetic duality group known from Maxwell-theory\cite{Obers}.
Now $W(E_7)/{\mathbb Z}_2$ is isomorphic to\cite{Bour} $Sp(6,\mathbb{Z}_2)$, which is the symplectic group encapsulating the commutation properties of the three-qubit Pauli observables.
This observation provided a new way of understanding the mathematical structure of the $E_7$-symmetric black hole entropy formula in terms of three-qubit quantum gates\cite{Levfin,Geemen}.

Recent work also attempted to relate configurations like Mermin squares to finite geometric structures.
In finite geometry the basic notion is that of \emph{incidence}. We have two disjoint sets of objects called \emph{points} and \emph{lines} and incidence is a relation between these sets. For simple incidence structures the lines are comprising certain subsets of the set of points, and incidence is just the set-theoretic membership relation.
Regarding the nontrivial Pauli observables as points and observing that any pair of observables is either commuting or anticommuting, one can define incidence either via commuting or anticommuting.
For $N$-qubit systems an approach of that kind was initiated in Ref.\cite{San1} with the incidence structure arising from commuting called
${\mathcal W}(2N-1,2)$, the symplectic polar space of rank $N$ and order two\cite{Buek}. 
In this spirit it has been realized that certain subconfigurations 
of ${\mathcal W}(2N-1,2)$,
called geometric hyperplanes\cite{Shult}, are also worth studying. For example, for the case of ${\mathcal W}(3,2)$ one particular class of its geometric hyperplanes features the $10$ possible Mermin squares one can construct from two-qubit Pauli operators.

Geometric hyperplanes turned out to have an interesting relevance to the structure of black hole entropy formulas as well.
One particular type of geometric hyperplane of an incidence structure related to  ${\mathcal W}(5,2)$ is featuring $27$ points and $45$ lines and having the incidence geometry of a generalized quadrangle\cite{Payne}  $GQ(2,4)$, with the automorphism group $W(E_6)$. In Ref.\cite{Levfin2} it has been shown
how $GQ(2,4))$ encodes information about the structure of the $E_{6(6)}$-symmetric black hole entropy formula.
It has been also observed\cite{Levfin2,BHQC} that certain truncations of this entropy formula correspond to truncations to further interesting subconfigurations.
For example, the $27$ points of $GQ(2,4)$ can be partitioned into three sets of Mermin {\it squares} with $9$ points each.
This partitioning corresponds to the reduction of the $27$-dimensional irreducible representation of $E_{6(6)}$ to a substructure arising from three copies of $3$ dimensional irreps of three $SL(3,{\mathbb R})$s.
The configuration related to this truncation has an interesting physical interpretation in terms of wrapped membrane configurations and is known in the literature as the bipartite entanglement of three qutrits\cite{Duff,BHQC}.

Sometimes it is useful to form a new incidence structure with points being geometric hyperplanes. In this picture certain geometric hyperplanes, regarded as points, form lines called Veldkamp lines. These lines and points are in turn organized into the so-called Veldkamp space\cite{Veld,Shult}.
Applying this notion to the simplest nontrivial case the structure of the Veldkamp space of ${\mathcal W}(3,2)$ has been thoroughly investigated, 
the physical meaning of the geometric hyperplanes clarified, and pictorially illustrated\cite{San2}.
For an arbitrary number of qubits the diagrammatic approach of Ref.\cite{San2} is not feasible. However, a  later study\cite{VranLev} has shown how the structure of the Veldkamp space of ${\mathcal W}(2N-1,2)$ can be revealed in a purely algebraic fashion.

In a recent paper\cite{Zsolt} it has been shown that the space of possible Mermin pentagrams of cardinality 12\,096 (see \cite{Saniga1}) can be organized into $1008$ families, each of them containing  $12$ pentagrams. 
Surprisingly, the $1008$ \emph{families} can be mapped bijectively to the $1008$ \emph{members} of a subclass of Veldkamp lines of the Veldkamp space for three-qubits\cite{Zsolt}.
For the families comprising $12$ pentagrams the term  \emph{double-sixes of pentagrams} has been coined.
Due to the transitive action of the symplectic group $Sp(6,\mathbb{Z}_2)$ on this class of Veldkamp lines\cite{VranLev}, it is enough to study merely one particular family, called the canonical one. 
It turned out that the structure of the canonical double-six is encapsulated in the weight diagram for the $20$-dimensional irreducible representation of the group $SU(6)$.

For three qubits ($N=3$) this class of Veldkamp lines associated with the space of Mermin pentagrams is of a very special kind. 
For reasons to be clarified later we will call this line \emph{the magical Veldkamp line}.
The canonical member from this magical class of Veldkamp lines is featuring three geometric hyperplanes. Two of them are quadrics of physical importance. 
One of them is containing $35$ points. Its incidence structure is that of the so called Klein quadric over $\mathbb{Z}_2$. In physical terms the points of this quadric form the set of nontrivial \emph{symmetric} Pauli observables (i.e. the ones containing an even number of $Y$ operators, the trivial one $III$ excluded).
The other one is containing $27$ points. Its incidence structure is that of a generalized quadrangle $GQ(2,4)$.
In physical terms the points of this quadric form the set of nontrivial operators that are: either symmetric and commuting ($15$ ones), or antisymmetric and anticommuting ($12$ ones) with the special operator $YYY$.
In entanglement theory these 27 Pauli observables are precisely the nontrivial ones that are left invariant with respect to the so-called Wootters spin flip operation\cite{Wootters}.
The third geometric hyperplane comprising our Veldkamp line is arising from the $31$ nontrivial observables that are \emph{commuting} with our fixed special observable $YYY$.

For three qubits one has $63$ nontrivial Pauli observables. All of our geometric hyperplanes featuring the magical Veldkamp line are intersecting in the $15$-element \emph{core-set}  of \emph{symmetric} operators, that are at the same time \emph{commuting} with the fixed one $YYY$. It can be shown that this set displays the incidence structure of a generalized quadrangle $GQ(2,2)$. In physical terms this incidence structure is precisely the one of the $15$ nontrivial two-qubit Pauli observables. The core set and the three complements with respect to the three geometric hyperplanes give rise to a partitioning of the $63$ nontrivial observables of the form:
$63=15+12+20+16$.

The results of \cite{Zsolt} and \cite{Levfin2} clearly demonstrate that apart from information concerning incidence, our magic Veldkamp line also carries information concerning representation theory of certain groups and their invariants.
Indeed, the $15+12=27$ point $GQ(2,4)$-part encapsulates information on the structure of the cubic invariant
of the $27$ dimensional irreducible representation  of the exceptional group $E_6$,
with the physical meaning being black hole entropy in five dimensions. On the other hand, the $20$-point double-six of pentagrams part
encapsulates information on the $20$-dimensional irreducible representation associated with the action of the group $A_5=SU(6)$ on three-forms in a six dimensional vector space.
Moreover, we will show that this part of our Veldkamp line also encodes information on the structure of Hitchin's quartic invariant
for three forms\cite{Hitchin}, and certain black hole entropy formulas in four dimensions\cite{FormTH}. Amusingly, this invariant also coincides with the entanglement measure used for three fermions with six single particle states\cite{LevVran}, a system of importance in the history of the $N$-representability problem\cite{Borland}.

Motivated by these interesting observations coming from different research fields, in this paper we would like to answer the following three questions. What is the representation theoretic meaning of the different parts of our magic Veldkamp line?
What kind of finite geometric structures does this Veldkamp line encode? And, finally, how are these geometric structures related to special invariants that show up as black hole entropy formulas and Hitchin functionals in four, five, six and seven dimensions?

The organization of this paper is as follows. For the convenience of high energy physicists not familiar with the slightly unusual language  of finite geometry, we devoted Section 2. to presenting the background material on incidence structures.
In this section the main objects of scrutiny appear: generalized quadrangles, extended generalized quadrangles and Veldkamp spaces.
Following the current trend  of high energy physicists also adopting the language of quantum information and quantum entanglement we gently introduce the reader to these abstract concepts via the language of Pauli groups of multiqubit systems.
In Section 3. we introduce our main finite geometric object of physical relevance: the magic Veldkamp line (MVL).
We have chosen the word \emph{magic} in reference to 
objects called magic configurations (like Mermin squares and pentagrams) that are used in the literature
to rule out
certain classes of hidden variable theories.
As we will see, these objects are intimately connected to the structure of our Veldkamp line, justifying our nomenclature. In the main body of the paper in different subsections of Section 3. we study different components of our MVL.
In each of these subsections (Sections 3.1.-3.7) our considerations involve studying the interplay between representation theoretic, finite geometric and invariant theoretic aspects of the corresponding part.
As we will demonstrate, each part can be associated with a natural invariant of physical meaning.
These invariants are the ones showing up in Hitchin functionals of form theories of gravity and certain entropy formulas of black hole solutions in string theory, hence contain the physical meaning.
Of course, the physical role of these invariants is well-known but their natural appearance in concert within a nice and unified finite geometric picture is new.   
In developing our ideas one can see that the finite geometric picture helps to reformulate some of the known results in an instructive  new way. At the same time this approach also establishes some new connections between functionals of form theories of gravity.
We are convinced that in the long run these results might help establishing further new results within the field of generalized exceptional geometry.

Throughout the paper we emphasized the role of grids (i.e. generalized quadrangles of type $GQ(2,1)$) labelled by 
Pauli observables, alias Mermin squares, as basic building blocks (geometric hyperplanes) comprising certain Veldkamp lines.
In concluding, in Section 4. we also hint at a nested structure of Veldkamp lines for three and four qubits with grids sitting in their cores.
In light of this basic role for these simple objects, it is natural to ask: What is the physical meaning of this building block?
Originally, these objects
were used to rule out
certain classes of hidden variable theories.
Since they are now appearing in a new role this question is of basic importance.
However, apart from presenting some speculations at the end of Section 4, in this paper we are not attempting to answer this interesting question.
Here we are content with the aim of demonstrating that these building blocks can be used for establishing a unified finite geometric underpinning for form theories of gravity.
The possible physical implications of the unified picture provided by our MVL we would like to explore in future work.

\section{Background}

The aim of this section is to present the basic definitions, and refer to the necessary results already presented elsewhere.
In the following we conform with the conventions of Refs.\cite{VranLev,Levay1}. 
The basic object we will be working with
is defined as follows:
\begin{defn}\label{defn:inci}
The triple $({\mathcal P},{\mathcal L},{\mathcal I})$ is called an incidence structure (or point-line incidence
geometry) if ${\mathcal P}$ and ${\mathcal L}$ are disjoint sets and ${\mathcal I}\subseteq {\mathcal P}\times {\mathcal L}$ is an incidence
relation. The elements of ${\mathcal P}$ and ${\mathcal L}$ are called points and lines, respectively.
We say that $p\in {\mathcal P}$ is incident with $l\in {\mathcal L}$ if $(p,l)\in {\mathcal I}$.
Two points incident with the same line are called \emph{collinear}.
\end{defn}\label{defn:geomhyp}
In the following we consider 
merely those incidence geometries that are called simple. In simple incidence structures the
 lines may be identified with the sets of points they are incident with, so we
 can think of these as a set ${\mathcal P}$ together with a subset ${\mathcal L}\subseteq 2^{\mathcal P}$ of the
 power set of ${\mathcal P}$. Then $({\mathcal P},{\mathcal L},\in)$ is an incidence structure 
 i.e. the points incident with
 a line will be called the elements of that line.
In a point-line geometry there are distinguished sets of points called
{\it geometric hyperplanes} \cite{Shult}:
\begin{defn}
Let $({\mathcal P},{\mathcal L},{\mathcal I})$ be an incidence structure. A subset ${\mathcal H}\subseteq {\mathcal P}$ of ${\mathcal P}$ is
called a geometric hyperplane if the following two conditions hold:
\begin{enumerate}[(H1)]
\item $(\forall l\in {\mathcal L}):(|{\mathcal H}\cap l|=1\textrm{ or }l\subseteq {\mathcal H})$,
\item $ {\mathcal H}\neq {\mathcal P}$.
\end{enumerate}
\end{defn}

Our aim is to associate a point-line incidence geometry to the $N$-qubit observables forming the Pauli
group $P_N$.
In order to do this we summarize the background concerning  $P_N$.

	Let us define the $2\times 2$ matrices
\beq
X=\begin{pmatrix}
0 & 1  \\
1 & 0
\end{pmatrix}
\qquad
Z=\begin{pmatrix}
1 & 0  \\
0 & -1
\end{pmatrix}.
\end{equation}
Observe that these matrices satisfy $X^2=Z^2=I$ where $I$ is the $2\times 2$
identity matrix. The product of the two will be denoted by $iY=ZX=-XZ$. 
The $N$-qubit Pauli group, $P_N$, is the subgroup of $GL(2^{N},\mathbb{C})$ consisting of
the $N$-fold tensor (Kronecker) products of the matrices $\{\pm I,\pm iI, \pm X, \pm iX, \pm Y, \pm iY, \pm Z, \pm iZ\}$. Usually the shorthand notation $AB\ldots C$ will be used for the tensor product
$A\otimes B\otimes\ldots\otimes C$ of one-qubit Pauli group elements $A,B,\ldots,C$.  The center of this group
is the same as its commutator subgroup, it is the subgroup of the fourth roots of unity, i.e. 
\beq
\mathbb{G}_4\equiv\{\pm 1,\pm i\}\subset{\mathbb C}^{\times}.
\label{mu}
\eeq
\noindent

It is useful to restrict to the $N=3$ case, our main concern here.
The $N$-qubit case can be obtained by rewriting the expressions below in a trivial manner.
An arbitrary element of $p\in P_3$ can be written in the form
\beq
p=s
Z^{\mu_1}X^{\nu_1}\otimes Z^{\mu_2}X^{\nu_2}\otimes Z^{\mu_3}X^{\nu_3},\qquad s\in\mathbb{G}_4,\qquad (\mu_1,\nu_1,\mu_2,\nu_2,\mu_3,\nu_3)\in{\mathbb Z}_2^6.
\label{prod1}
\eeq
\noindent
Hence if $p$ is parametrized as
\beq
p\leftrightarrow(s;\mu_1,\dots,\nu_3)
\eeq
then
the product of two elements
$p,p^{\prime}\in P_3$ corresponds to\beq
pp^{\prime}\leftrightarrow(ss^{\prime}(-1)^{\sum_{j=1}^3\mu_j^{\prime}\nu_j};\mu_1+\mu_1^{\prime},\dots,
\nu_3+\nu_3^{\prime}). \label{prod} \eeq \noindent Hence two elements
commute, if and only if, \beq
\sum_{j=1}^{3}(\mu_j\nu_j^{\prime}+\mu_j^{\prime}\nu_j)=0. \label{komm}
\eeq The commutator subgroup of $P_3$ coincides with its
center $Z(P_3)$, which is $\mathbb{G}_4$ of Eq.(\ref{mu}).
Hence the central quotient
$V_3=P_3/Z(P_3)$ is an Abelian group which, by
virtue of (\ref{prod1}), is also a six-dimensional vector space over
${\mathbb Z}_2$, i.e. $V_3\equiv {\mathbb Z}_2^6$. Moreover, on $V_3$
the left-hand-side of (\ref{komm}) defines a symplectic form\beq
\langle\cdot,\cdot\rangle:V_3\times V_3\to {\mathbb Z}_2,\qquad
(p,p^{\prime})\mapsto \langle p,p^{\prime}\rangle\equiv
\sum_{j=1}^3 (\mu_j\nu_j^{\prime}+\nu_j\mu_j^{\prime}).\label{ezittaszimpl} \eeq \noindent The
elements of the vector space $(V_3,\langle\cdot,\cdot\rangle)$ are
equivalence classes corresponding to quadruplets of the form
$\{\pm\mathcal{O}_1\otimes\mathcal{O}_2\otimes \mathcal{O}_3,
\pm i\mathcal{O}_1\otimes\mathcal{O}_2\otimes \mathcal{O}_3
\}$ where ${\mathcal O}_j\in\{I,X,Y,Z\},\quad j=1,2,3$. 
We choose $\mathcal{O}_1\otimes\mathcal{O}_2\otimes \mathcal{O}_3$ 
(or $\mathcal{O}_1\mathcal{O}_2\mathcal{O}_3$ in short)
as the canonical representative of the corresponding equivalence class. 
This representative $\mathcal{O}_1\mathcal{O}_2\mathcal{O}_3$ is Hermitian, hence will be called
a {\it three-qubit observable}. 

In our geometric considerations
the role of the (\ref{ezittaszimpl}) symplectic form is of utmost importance.
It is taking its values in ${\mathbb Z}_2$ according to whether the corresponding representative Pauli operators 
are commuting $(0)$ or not commuting $(1)$. In our geometric considerations Pauli operators commuting or not will correspond to the points in the relevant geometry being collinear or not.

 According to (\ref{prod1}), for a single qubit the equivalence classes are represented as \beq I\mapsto (00),\qquad
 X\mapsto (01),\qquad Y\mapsto (11),\qquad Z\mapsto (10).
 \label{corr1} \eeq Adopting  
 the ordering convention \beq
 \mathcal{O}_1\mathcal{O}_2\mathcal{O}_3\leftrightarrow
 p\equiv(\mu_1,\mu_2,\mu_3,\nu_1,\nu_2,\nu_3)\in V_3 \label{konvencio} \eeq \noindent
  the canonical basis vectors in $V_3$ are
 associated to equivalence classes as follows \beq
 ZII\leftrightarrow e_1=(1,0,0,0,0,0),\qquad\dots \qquad
 IIX\leftrightarrow e_6=(0,0,0,0,0,1). \label{bazis} \eeq \noindent
 With respect to this basis the matrix of the symplectic form
 is\beq J_{ab}\equiv \langle e_{a},e_{b}\rangle
 =\begin{pmatrix}0&0&0&1&0&0\\0&0&0&0&1&0
 \\0&0&0&0&0&1\\
         1&0&0&0&0&0
	     \\  0&1&0&0&0&0
	         \\  0&0&1&0&0&0\end{pmatrix}, \qquad a,b=1,2,\dots 6.
		         \label{simplmatr}
			         \eeq
				         \noindent

Since $V_3$ has even dimension and the symplectic form is
nondegenerate, the invariance group of the symplectic form is the
symplectic group $Sp(6,{\mathbb Z}_2)$. This group is acting on the row
vectors of $V_3$ via $6\times 6$ matrices $S\in Sp(6,{\mathbb{Z}_2})\equiv Sp(6,2)$ from the
right, leaving the matrix $J$ of the symplectic form invariant \beq
v\mapsto vS,\qquad SJS^t=J. \label{transz} \eeq \noindent It is
known that $\vert Sp(6,2)\vert =1451520=2^9\cdot 3^4\cdot 5\cdot
7$ and this group is generated by transvections\cite{Geemen}
$T_p\in Sp(6,2), p\in V_3$ of the form \beq T_p:V_3\to V_3,\qquad
q\mapsto T_pq=q+\langle q,p\rangle p \label{transvections} \eeq
\noindent and they are indeed symplectic, i.e. \beq \langle
T_pq,T_pq^{\prime}\rangle=\langle q,q^{\prime}\rangle. \label{szimpltulajd} \eeq
\noindent There is a surjective homomorphism\cite{Bour} from $W(E_7)$, i.e.
the Weyl group of the exceptional group $E_7$, to $Sp(6,2)$ with
kernel ${\mathbb Z}_2$.

The projective space $PG(2N-1,2)$ consists of the nonzero subspaces
of the $2N$-dimensional vector space $V_N$ over $\mathbb{Z}_2$.
The points of the projective space are one-dimensional subspaces of the vector
space, and more generally, $k$-dimensional subspaces of the vector space are
$(k-1)$-dimensional subspaces of the corresponding projective space.
A subspace of $(V_N,\langle\cdot,\cdot\rangle)$  (and also the subspace in the corresponding
projective space) is called isotropic if there is a vector in it which is orthogonal (with respect to the symplectic form)
to the whole subspace, and totally isotropic if the subspace is orthogonal to
itself.
The space of totally isotropic subspaces of $(PG(2N-1,2),\langle\cdot,\cdot\rangle)$ is called the {\it symplectic polar space} of rank $N$, and order two, denoted by ${\mathcal W}(2N-1,2)$.
The maximal totally isotropic subspaces are called Lagrangian subspaces.

For an element $x\in V_3$ represented as in (\ref{konvencio}), let us define the quadratic form \beq
Q_0(p)\equiv \sum_{j=1}^3 \mu_j\nu_j. \label{kankvadrat} \eeq \noindent
It is easy to check that for vectors representing symmetric
observables $Q_0(p)=0$ (the ones containing an {\it even} number of $Y$s) and for antisymmetric ones $Q_0(p)=1$
(the ones containing an {\it odd} number of $Y$s). Moreover, we have the relation 
\beq \langle p,p^{\prime}\rangle =Q_0(p+p^{\prime})+Q_0(p)+Q_0(p^{\prime}).
\label{kvdratosszef} \eeq \noindent The (\ref{kankvadrat}) quadratic form will be
regarded as the one labelled by the $0$-element of $V_3$
with representative observable $III$. There are however,
$63$ other quadratic forms $Q_p$ compatible with the symplectic
form $\langle\cdot\vert\cdot\rangle$ labelled by the nontrivial
elements $q$ of $V_3$ also satisfying \beq \langle p,p^{\prime}\rangle =Q_q(p+p^{\prime})+Q_q(p)+Q_q(p^{\prime}).
\label{kvdratosszef2} \eeq \noindent They are defined as \beq
Q_q(p)\equiv Q_0(p)+\langle q,p\rangle^2
\label{ujkvadforms}\eeq\noindent and, since we are over the two-element
field, the square can be omitted. 

For more information on these
quadratic forms we orient the reader to \cite{VranLev,Geemen}. Here
we merely elaborate on the important fact that there are two classes of
such quadratic forms.
They are the ones that are labelled by symmetric
observables ($Q_0(q)=0$), and antisymmetric ones ($Q_0(q)=1$). The
locus of points in $PG(5,2)$ satisfying $Q_q(p)=0$ for $Q_0(q)=0$
is called a {\it hyperbolic} quadric and the locus $Q_q(p)=0$ for
which $Q_0(q)=1$ is called an {\it elliptic} one. The space of the former type of quadrics
will be denoted by $Q^+(5,2)$ and the latter type by $Q^-(5,2)$. 
Looking at Eq.(\ref{ujkvadforms}) one can see that in terms of three-qubit observables (modulo elements of $\mathbb{G}_4$)
one can characterize the quadrics $Q(5,2)$ as follows.
The three-qubit observables $p\in Q(5,2)$ characterized by $Q_q(p)=0$ are the ones that are either symmetric and commuting with $q$ or antisymmetric and anticommuting with $q$.
It can be shown\cite{VranLev,Geemen} that we have $36$ quadrics of type $Q^+(5,2)$ and $28$ ones of type $Q^-(5,2)$, with the former containing $35$ and the latter containing $27$ points of $PG(5,2)$.
A quadric of $Q^+(5,2)$ type in $PG(5,2)$ is called the Klein-quadric.
Note that the points lying on the Klein quadric $Q_0\in Q^+(5,2)$ given by the equation $Q_0(p)=0$ can be represented by symmetric observables, i.e. ones that contain an \emph{even} number of $Y$s.  

On the other hand, a quadric of $Q^-(5,2)$ type can be shown to display the structure of a generalized quadrangle\cite{Payne} $GQ(2,4)$, an object we already mentioned in the introduction and define below.

\begin{defn}\label{defn:genquadr}
A \emph{Generalized Quadrangle} $GQ(s,t)$ of order $(s,t)$ is an incidence structure of points and lines (blocks) where every point is on $t+1$ lines ($t>0$), and every line contains $s+1$ points ($s>0$) such that if $p$ is a point and $L$ is a line, $p$ not on $L$, then there is a unique point $q$ on $L$ such that $p$ and $q$ are collinear.
\end{defn}

It is easy to prove that in a $GQ(s,t)$ there are $(s+1)(st+1)$ points and $(t+1)(st+1)$ lines\cite{Payne}.
In  what  follows,  we  shall  be  uniquely  concerned  with  generalized  quadrangles  having
lines of size
three, $GQ(2,t)$ and $t\geq 1$. One
readily sees\cite{Payne} that these quadrangles are of three distinct kinds, namely $GQ(2,1)$, $GQ(2,2)$ and $GQ(2,4)$.
A $GQ(s,1)$ is called a \emph{grid}.
In this paper 
$GQ(2,1)$ grids, with their points labelled by Pauli observables, will play an important role. Their points correspond to $9$ observables commuting along their $6$ lines. Clearly, every observable is on two lines and every line contains three observables. Since the observables are commuting along the lines, one can take their product unambiguously.
We are interested in lines labelled by observables producing plus or minus the identity when multiplied.
Such lines will be called \emph{positive} or \emph{negative} lines.
A \emph{Mermin square} is a $GQ(2,1)$ labelled by Pauli observables having an \emph{odd} number of negative lines.
It can be shown\cite{FredMetod} that any grid labelled by multiqubit Pauli observables has an odd number of negative lines.
Hence, any $GQ(2,1)$ labelled by multiqubit Pauli observables is a Mermin square.

\begin{figure}[pth!]
\centerline{\includegraphics[width=4truecm,clip=]{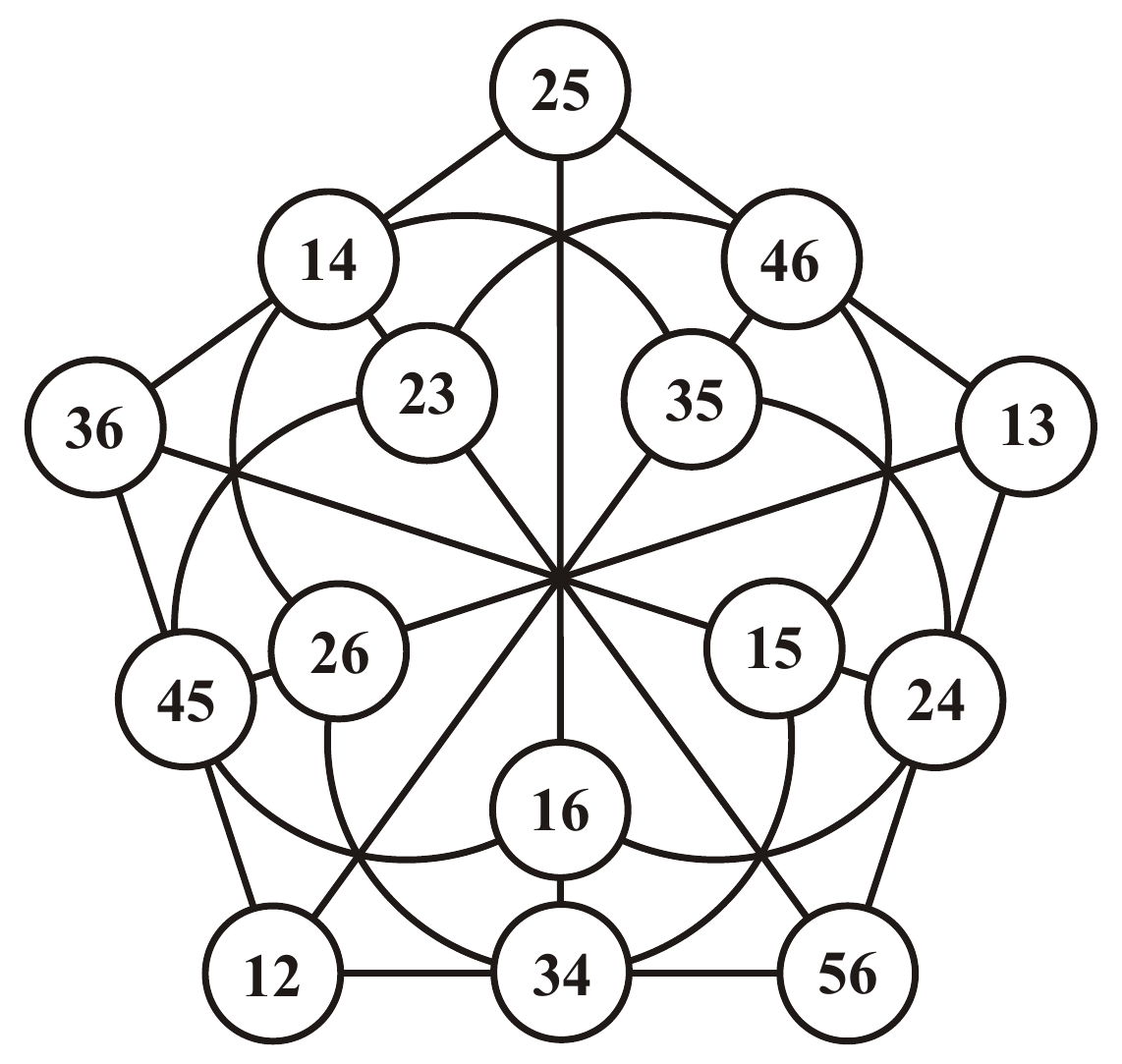}}
\caption{The doily with the duad labelling.}
\label{DoilyDuad}
\end{figure}

A generalized quadrangle of type $GQ(2,2)$ is also called the \emph{doily}\cite{Payne,Polster}. It has $15$ points and $15$ lines. Its simplest representation can be obtained by the so-called \emph{duad construction} as follows. 
Take the $15$ two-element subsets (duads) of the set $S=\{1,2,3,4,5,6\}$ and regard triples of such duads collinear whenever their pairwise intersection is the empty set: e.g. $\{\{12\},\{34\},\{56\}\}$ is such a line.  
A visualisation of this construction is given in Figure \ref{DoilyDuad}.

\begin{figure}[pth!]
\centerline{\includegraphics[width=4truecm,clip=]{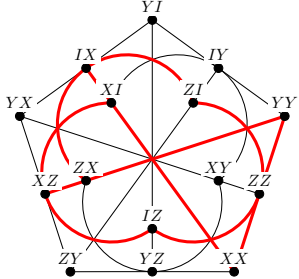}}
\caption{The \emph{doily} labelled by nontrivial two-qubit Pauli observables.
Inside the \emph{doily} a geometric hyperplane (see Definition \ref{defn:geomhyp}), a \emph{grid} is shown.}
\label{doilygrid}
\end{figure}

\begin{figure}
\begin{center}
 \includegraphics[width=3cm]{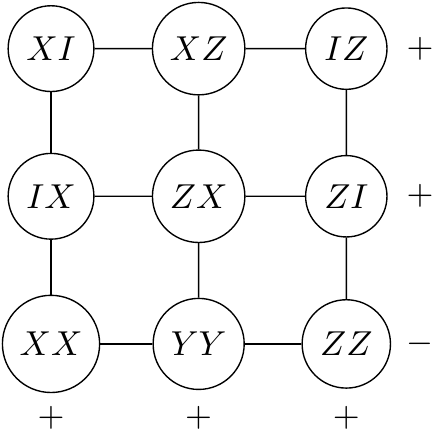}
 \includegraphics[width=3cm]{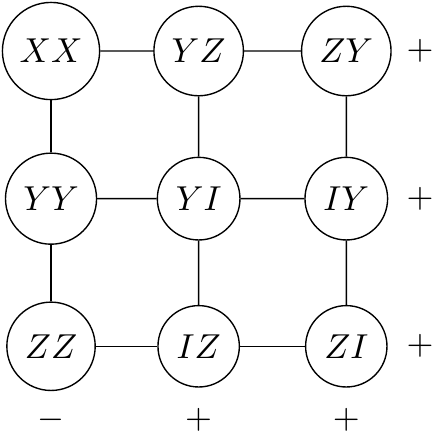}
 \includegraphics[width=3cm]{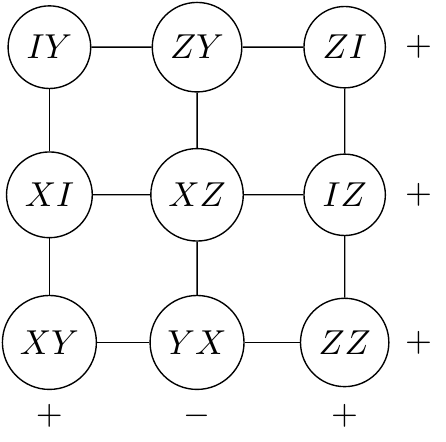}
 \includegraphics[width=3cm]{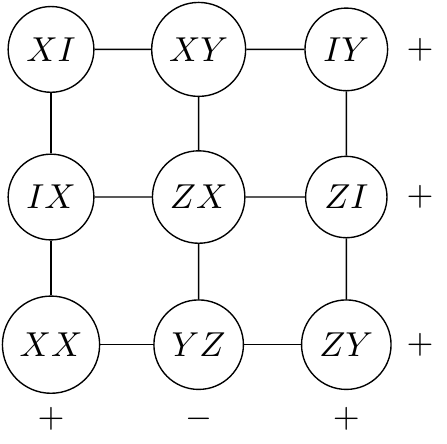}
 \includegraphics[width=3cm]{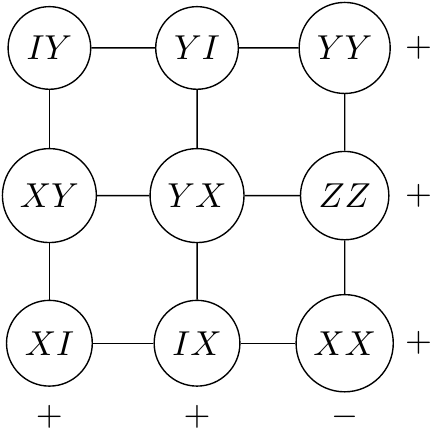}
 \includegraphics[width=3cm]{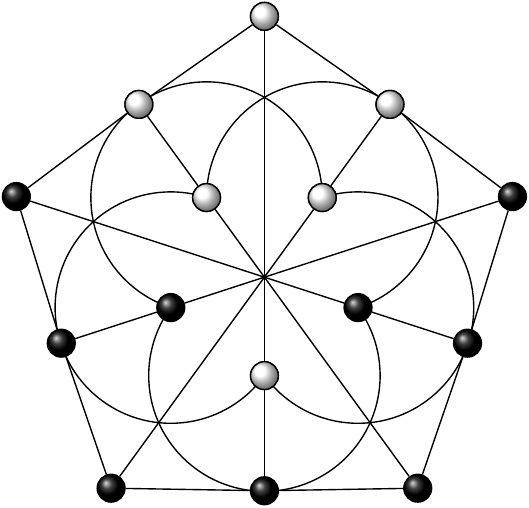}
 \includegraphics[width=3cm]{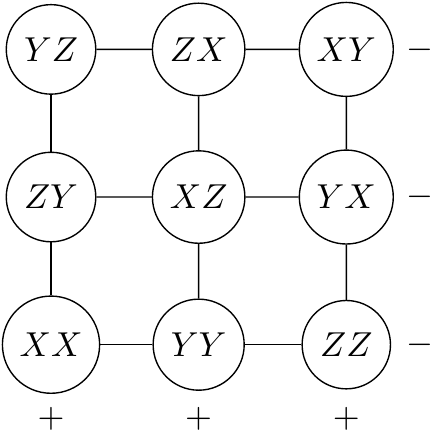}
 \includegraphics[width=3cm]{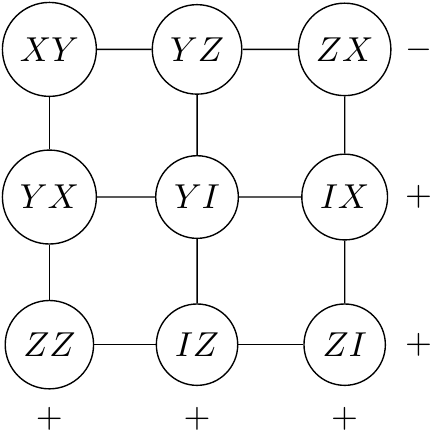}
 \includegraphics[width=3cm]{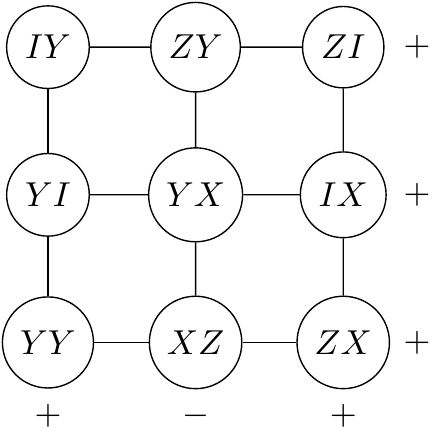}
 \includegraphics[width=3cm]{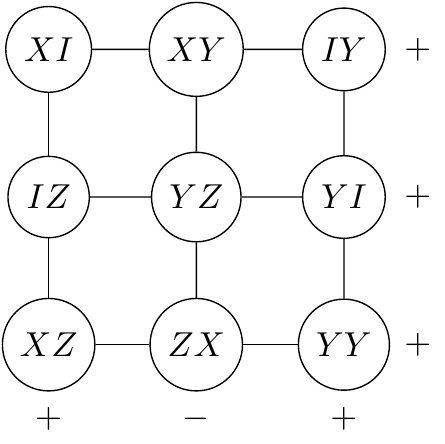}
 \includegraphics[width=3cm]{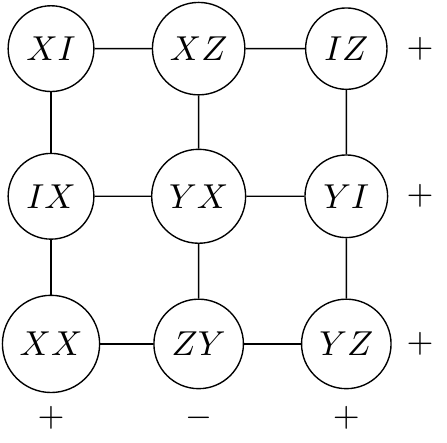}
 \includegraphics[width=3cm]{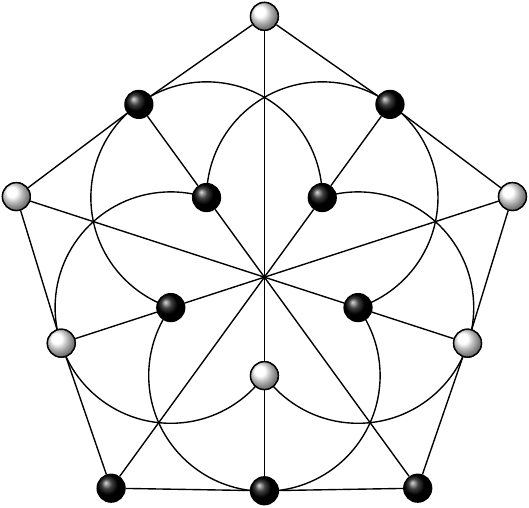}
\caption{The full set of Mermin squares living inside the doily.
The $10$ copies of relevant grids can be identified after successive rotations by $72$ degrees of the embedded patterns of grids seen in the second and fourth rows. The first Mermin square, up to an automorphism, is the grid of Figure \ref{doilygrid}. It is embedded in the doily after a counter clock-wise rotation by $72$ degrees of the second pattern seen in the lower right corner.}
\label{mermins}
\end{center}
\end{figure}

An alternative realization of the doily, depicted in Figure \ref{doilygrid}, is obtained by noticing that we have precisely $15$ nontrivial (identity removed) two-qubit Pauli observables\cite{San1}, and also $15$ pairwise commuting triples of them.
It can be shown that there are precisely $10$ grids, i.e. $GQ(2,1)$s, living as geometric hyperplanes inside the doily\cite{San1,VranLev}. A particular example of a grid inside the doily is shown in Figure \ref{doilygrid}.
All of these grids give rise to Mermin squares as shown in Figure \ref{mermins}.

The final item in the line of generalized quadrangles with $s=2$ is $GQ(2,4)$, i.e. our elliptic quadric $Q^-(5,2)$. In order to label this object by Pauli observables three-qubits are needed.
 A pictorial representation of $GQ(2,4)$ having $27$ points and $45$ lines labelled by three-qubit observables can be found in Ref.\cite{Levfin2}.
$GQ(2,4)$ contains $36$ copies of doilies as geometric hyperplanes.
It also contains
grids, though they are
not
geometric hyperplanes of $GQ(2,4)$. It can be shown that there are $40$ triples
of pairwise disjoint grids\cite{Levfin2} inside $GQ(2,4)$.
Grids giving rise to Mermin squares labelled by three-qubit Pauli observables are arising in groups of $10$ living inside doilies with three-qubit labels.
A trivial example of that kind can be obtained by adjoining as a third observable the identity to all the two-qubit labels of Figure \ref{mermins}.

For our purposes it will be important to know that the notion of generalized quadrangles can be extended\cite{Cameron,Fra}.
Let us consider an incidence structure $\mathcal{S}$ consisting of \emph{points} and \emph{blocks} (lines). 
For any point $p$, 
let us then define $\mathcal{S}_p$ as the structure of all the points different from $p$ that are on a block on $p$, and all the blocks on $p$.
$\mathcal{S}_p$ is called the \emph{residue} of $p$. Then we have the following definition.

\begin{defn}\label{defn:extgenquadr}
An \emph{Extended Generalized Quadrangle} $EGQ(s,t)$ of order $(s,t)$ is a finite connected incidence structure $\mathcal{S}$, such that for any point $p$ its residue $\mathcal{S}_p$ is a generalized quadrangle of order $(s,t)$.
\end{defn}

We have seen that incidence structures labelled by three commuting observables giving rise to the identity up to sign are of special status.
This motivates the introduction of the following point-line incidence structure.

\begin{defn}\label{defn:polar}
Let $N\in\mathbb{N}+1$ be a positive integer, and $V_{N}$
be the symplectic $\mathbb{Z}_{2}$-linear space.
The incidence structure $\mathcal{G}_{N}$ of the $N$-qubit Pauli
group is $({\mathcal P},{\mathcal L},\in)$ where ${\mathcal P}=V_{N}\setminus\{0\}$,
\begin{equation}
{\mathcal L}=\{\{p,q,p+q\}|p,q\in {\mathcal P},p\neq q,\langle p,q\rangle=0\}
\end{equation}
and $\in$ is the set theoretic membership relation.
\end{defn}

Clearly the points and lines of $\mathcal{G}_{N}$ are the ones of the symplectic polar space ${\mathcal W}(2N-1,2)$.
Of course our main concern here is the $N=3$ case. 
In this case $\mathcal{G}_{3}$ has $63$ points and $315$ lines.

Our next task is to recall the properties of the geometric hyperplanes of $\mathcal{G}_{N}$.
The following lemma was proved in Ref.\cite{VranLev}.

\begin{lem}\label{lem:hypplanes}
Let $N\in\mathbb{N}+1$ be a positive integer, $\mathcal{G}_{N}=({\mathcal P},{\mathcal L},\in)$ and
$q\in V_{N}$ be any vector. Then the sets
\begin{equation}
C_{q}=\{p\in {\mathcal P}|\langle p,q\rangle=0\}
\end{equation}
and
\begin{equation}
H_{q}=\{p\in {\mathcal P}|\Quad_{q}(p)=0\}
\end{equation}
satisfy (H1).
\end{lem}

This lemma shows that apart from $C_0$, all of the sets above are geometric hyperplanes of the geometry $\mathcal{G}_{N}$.
The set $C_q$ is called the  \emph{perp-set}, or the \emph{quadratic cone} of $q\in V_N$.  Modulo an element of $\mathbb{G}_4$, $C_q$ represents the set of observables commuting with a fixed one $q$. 
Back to the implications of our lemma one can show that in fact more is true, {\it all geometric
hyperplanes} arise in this form\cite{VranLev}:
\begin{thm}
Let $N\in\mathbb{N}+1$, $\mathcal{G}_{n}=({\mathcal P},{\mathcal L},\in)$, and ${\mathcal H}\in {\mathcal P}$ a subset
satisfying (H1). Then either ${\mathcal H}=C_{p}$ or ${\mathcal H}=H_{p}$ for some $p\in V_{N}$.
\end{thm}

One can prove that for $N\geq 2$ no geometric hyperplane is contained in another one, more precisely\cite{VranLev}:
\begin{thm}\label{thm:Vpoints}
Let $N\in\mathbb{N}+2$, $\mathcal{G}_{n}=({\mathcal P},{\mathcal L},\in)$ and suppose that
$A,B\subset {\mathcal P}$ are two geometric hyperplanes. Then $A\subseteq B$ implies $A=B$.
\end{thm}
Another property of two different geometric hyperplanes is that the complement of their symmetric difference
gives rise to a third geometric hyperplane i.e.

\begin{lem}
For $A\neq B$ geometric hyperplanes in $\mathcal{G}_{N}=({\mathcal P},{\mathcal L},\in)$
with $N\ge 1$ the set
\begin{equation}
A\boxplus B:=\overline{A\triangle B}=(A\cap B)\cup(\overline{A}\cap\overline{B})
\label{figkell1}
\end{equation} is also a geometric hyperplane.
\end{lem}
One can also check that by using the notation $C\equiv A\boxplus B$
\beq
A\cap C
 =A\cap B,\qquad B\cap C= A\cap B, \qquad A\boxplus C = B.
\label{figkell2}
    \eeq
    \noindent
A corollary of this is that any two of the triple $(A,B,A\boxplus B)$ of hyperplanes determine the third.

Sometimes it is also possible to associate to a particular incidence geometry another one
called its Veldkamp space whose points are geometric hyperplanes of the original geometry\cite{Shult}:
\begin{defn}\label{defn:veldkamp}
Let $\Gamma=({\mathcal P},{\mathcal L},{\mathcal I})$ be a point-line geometry. We say that $\Gamma$ has
Veldkamp points and Veldkamp lines if it satisfies the conditions:
\begin{enumerate}[(V1)]
\item For any hyperplane $A$ it is not properly contained in any other hyperplane $B$.
\item For any three distinct hyperplanes $A$, $B$ and $C$, $A\cap B\subseteq C$ implies $A\cap B=A\cap C$.
\end{enumerate}

If $\Gamma$ has Veldkamp points and Veldkamp lines, then we can form the
Veldkamp space $V(\Gamma)=({\mathcal P}_{V},{\mathcal L}_{V},\supseteq)$ of $\Gamma$, where ${\mathcal P}_{V}$ is
the set of geometric hyperplanes of $\Gamma$, and ${\mathcal L}_{V}$ is the set of
intersections of pairs of distinct hyperplanes.
\end{defn}

Clearly, by Theorem \ref{thm:Vpoints},  $\mathcal{G}_{N}$ contains Veldkamp points for $N\geq 2$, hence in this case V1 is satisfied.
In order to see that V2 holds as well, we note\cite{VranLev}:
\begin{lem}\label{lem:Vlines}
Let $N\in\mathbb{N}+1$, $p,q\in V_{N}$ and $\mathcal{G}_{N}=({\mathcal P},{\mathcal L},\in)$. Then the
following formulas hold:
\begin{eqnarray}
C_{p}\boxplus C_{q} & = & C_{p+q},  \nonumber  \\
H_{p}\boxplus H_{q} & = & C_{p+q},  \\
C_{p}\boxplus H_{q} & = & H_{p+q}.  \nonumber
\end{eqnarray}
\end{lem}
From this it follows that for any three geometric hyperplanes $A,B,C$ we have $A\cap B=A\cap C=B\cap C$.
One can however show more\cite{VranLev}, namely that there is no other possibility i.e.  $A\cap B\subseteq C$ implies $C\in\{A,B,A\boxplus B\}$.
\begin{thm}
Let $N\in\mathbb{N}+3$, and suppose that $A,B,C$ are distinct geometric
hyperplanes of $\mathcal{G}_{N}=({\mathcal P},{\mathcal L},\in)$ such that $A\cap B\subseteq C$.
Then $A\cap B=A\cap C$.
\end{thm}
Notice that the statement is not true for $N=2$.

From these results it follows that there are two different types of Veldkamp lines incident with three $C$-hyperplanes and three types of lines
which are incident with one $C$-hyperplane and two $H$-hyperplanes. 
Indeed, the two types are arising from the possibilities for $C_p$ and $C_q$ having $\langle p,q\rangle=0$ or $\langle p,q\rangle=1$. For the three types featuring also two $H$-type hyperplanes we mean 
\begin{eqnarray}
\{\{H_{p},H_{q}\}|p,q\in V_{N},p\neq q,\Quad_{0}(p)=\Quad_{0}(q)=0\},  \\
\{\{H_{p},H_{q}\}|p,q\in V_{N},p\neq q,\Quad_{0}(p)=\Quad_{0}(q)=1\},  \\
\{\{H_{p},H_{q}\}|p,q\in V_{N},\Quad_{0}(p)\neq\Quad_{0}(q)\}.
\end{eqnarray}

\section{The magic Veldkamp line}

From the previous section we know that for the incidence geometry $\mathcal{G}_{N}$ we have five different classes of Veldkamp lines. 
Three classes contain lines featuring two quadrics and a perp-set as geometric hyperplanes.
These lines are defined by the triple of the form $(H_p,H_q,C_{p+q})$.

Let us now consider $N=3$ and the choice $\{\{H_{p},H_{q}\}|p,q\in V_3,\Quad_{0}(p)\neq\Quad_{0}(q)\}$.
Hence, one of our quadrics should be an elliptic and the other a hyperbolic one.
For $N=3$ we have 36 possibilities for choosing the hyperbolic and $28$ ones for choosing the elliptic one, hence altogether this class contains $28\times 36=1008$ lines.
Let us now consider the special case
of $p=III$ and $q=YYY$. In the following we will call the
corresponding Veldkamp line
 $\{H_{III},H_{YYY},C_{YYY}\}$ \emph{the canonical magic Veldkamp line}.
By transitivity in our class of Veldkamp lines from the canonical one we can reach any of the $1008$ lines via applying a set of suitable symplectic transvections of the (\ref{transvections}) form.
For the construction of the explicit form of such transvections see Refs\cite{VranLev,Zsolt}.

\begin{figure}[pth!]
\centerline{\includegraphics[width=6truecm,clip=]{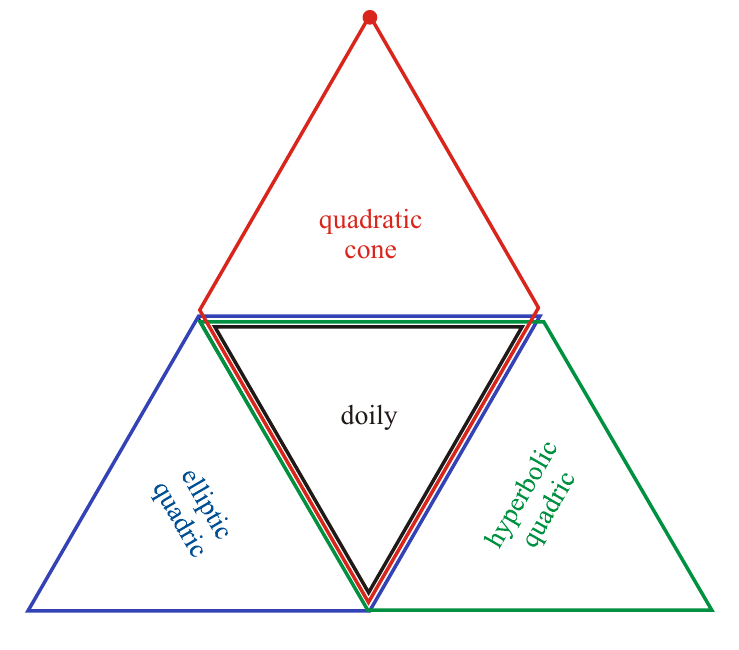}}
\caption{The structure of the magic Veldkamp line. The coloured parallelograms are geometric hyperplanes with characteristic cardinalities (number of points) as follows: red $31$ (perp-set), blue $27$ (elliptic quadric), green $35$ (hyperbolic quadric). 
Their common intersection is the core set of $15$ points which forms a doily. The three hyperplanes are satisfying
the properties of Eqs.(\ref{figkell1})-(\ref{figkell2}). The red dot on the top of the triangle corresponds to the special point defining the perp-set. For the canonical magic Veldkamp line this point is labelled by the observable $YYY$.}
\label{gods-eye-composition}
\end{figure}

\begin{figure}[pth!]
\centerline{\includegraphics[width=6truecm,clip=]{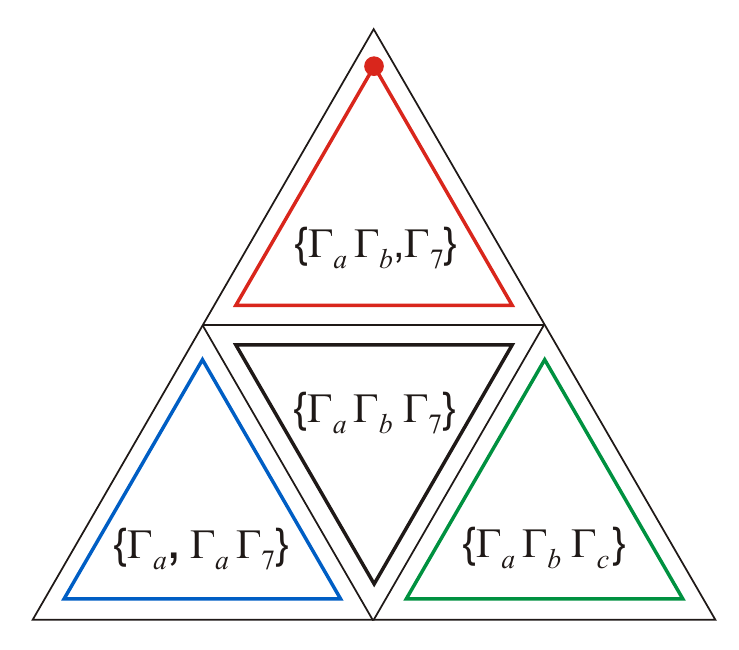}}
\caption{Decomposing the magic Veldkamp line into triangles of characteristic substructure via Clifford labelling.
In this decomposition the basis vector $\Gamma_7$ corresponding to the red dot enjoys a special status.
It belongs to the perp-set part of the Veldkamp line.}
\label{gods-eye-clifford}
\end{figure}

According to \cite{Zsolt}, to our Veldkamp line one can associate subsets of Pauli observables of cardinalities: $15$ (core set, a generalized quadrangle $GQ(2,2)$-doily), $27$ (elliptic quadric, a generalized quadrangle $GQ(2,4)$),
$35$ (hyperbolic quadric, i.e. the Klein quadric), $31$ (perp-set, a quadratic cone).
In addition to these basic cardinalities one also has the characteristic numbers: $12$ (Schl\"afli's double-six\cite{Polster,Levfin2}),
$20$ (the double-six of Mermin pentagrams\cite{Zsolt}), and $16$ (the complement of the core in the perp-set).
These sets are displayed in Figure \ref{gods-eye-composition} and \ref{gods-eye-clifford}.

In \cite{Zsolt} the
complement of the doily in the hyperbolic-quadric-part of this Veldkamp line, i.e. the green triangle of Figure \ref{gods-eye-composition}, has been studied.
It was shown
that this cardinality $20$ part
forms a very special configuration of $12$ Mermin pentagrams. For this structure the term \emph{double-six of Mermin pentagrams} has been coined.
In \cite{Zsolt}
the representation theoretic meaning of this part has been clarified.
Our aim in this paper is to achieve a unified representation theoretic understanding for \emph{all} parts
of this Veldkamp line and connect these findings to the structure of black hole entropy formulas and Hitchin invariants.
As we will see the Veldkamp line of Figure \ref{gods-eye-composition} acts as an agent for arriving at a \emph{unified framework} for a finite geometric understanding of Hitchin functionals
giving rise to form theories of gravity\cite{FormTH}.

As we stressed, the hint for using the notion of a Veldkamp line for arriving at this unified framework came from a totally unrelated field: a recent study of the space of Mermin pentagrams. 
The basic idea of \cite{Zsolt} was to establish a bijective correspondence between the $20$ Pauli observables of the double-six of pentagrams part and the weights of the $20$-dimensional irrep of $A_5$ in  such a way, that the notion of commuting observables translates to weights having a particular angle between them.
Then the notion of four observables comprising a line translates into the notion that the sum of four incident weights being zero.
As a result of that procedure, a labelling of the Dynkin diagram and the highest weight vector with three-qubit Pauli observables of $A_5$ was found. Then, due to the correpondence between the Weyl reflections and the symplectic transvections, the weight diagram labelled with observables can also be found. 
Hence, as the main actors for the role of understanding the geometry of the space of Mermin pentagrams the approach of
\cite{Zsolt} employed finite geometry and representation theory.
In this paper we add new actors to the mix. They are certain invariants that are inherently connected to the finite geometric and representation theoretic details.

In order to arrive at a similar level of understanding for \emph{all} parts of our Veldkamp line as in \cite{Zsolt} we proceed as follows. 
First we  employ a labelling scheme which is displaying the geometric content more transparently than the one in terms of observables.
A convenient labelling of that kind is provided by using the structure of a seven-dimensional Clifford algebra.
As a particular realization we consider the following set of generators 
\beq
(\Gamma_1,\Gamma_2,\Gamma_3,\Gamma_4,\Gamma_5,\Gamma_6,\Gamma_7)=(ZYI,YIX,XYI,IXY,YIZ,IZY,YYY) 
\label{choice}
\eeq
satisfying
\beq \{\Gamma_I,\Gamma_J\}=2\delta_{IJ},\qquad I,J=1,2,\dots 7, \eeq
\noindent
and
\beq
i\Gamma_1\Gamma_2\Gamma_3\Gamma_4\Gamma_5\Gamma_6\Gamma_7=III.
\label{gamszorz}
\eeq
\noindent
Let us then consider the following three sets of operators
\beq
\Gamma_I,\qquad \Gamma_I\Gamma_J,\qquad \Gamma_I\Gamma_J\Gamma_K,\qquad 1\leq I<J<K\leq 7. \eeq
It is easy to check that the first two sets contain $7+21=28$ antisymmetric operators and the third set contains $35$ symmetric ones. Consider now the relations above modulo elements of $\mathbb{G}_4$. Using a labelling based on this Clifford algebra one can derive an explicit list of Pauli operators featuring our magic Veldkamp line. Indeed the relevant subsets, corresponding to the triangles of Figure \ref{gods-eye-clifford}, of cardinalities: $20,12,16,15$  can be labelled as
\beq
\{\Gamma_a\Gamma_b\Gamma_c\},\qquad \{\Gamma_a,\Gamma_a\Gamma_7\},\qquad \{\Gamma_a\Gamma_b,\Gamma_7\},\qquad \{\Gamma_a\Gamma_b\Gamma_7\}\qquad 1\leq a<b<c\leq 6. 
\label{setsofveldkamp}
\eeq
\noindent

One can then check that the particular choice of (\ref{choice}) automatically reproduces the set of Pauli observables of \cite{Zsolt} that make up the ones of the double-six of pentagrams in the form $\{\Gamma_i\Gamma_j\Gamma_k\}$. This procedure results in a labelling of the $20$ operators of the canonical set in terms of the $3$ element subsets of the set $\{1,2,3,4,5,6\}=\{1,2,3,\overline{1},\overline{2},\overline{3}\}$. Indeed, we have
\beq
\Gamma^{(123)}\leftrightarrow IIX,\qquad
\begin{pmatrix}\Gamma^{(\overline{1}23)}&\Gamma^{(\overline{1}31)}&\Gamma^{(\overline{1}12)}\\\Gamma^{(\overline{2}23)}&\Gamma^{(\overline{2}31)}&\Gamma^{(\overline{2}12)}\\\Gamma^{(\overline{3}23)}&\Gamma^{(\overline{3}31)}&\Gamma^{(\overline{3}12)}\end{pmatrix}
\leftrightarrow
\begin{pmatrix}ZZZ&YXY&XZZ\\XYY&IIZ&ZYY\\ZXZ&YZY&XXZ\end{pmatrix},
\label{pocok1}
\eeq
\beq
\Gamma^{(\overline{123})}\leftrightarrow YYZ,\qquad
\begin{pmatrix}\Gamma^{(1\overline{23})}&\Gamma^{(1\overline{31})}&\Gamma^{(1\overline{12})}\\\Gamma^{(2\overline{23})}&\Gamma^{(2\overline{31})}&\Gamma^{(2\overline{12})}\\\Gamma^{(3\overline{23})}&\Gamma^{(3\overline{31})}&\Gamma^{(3\overline{12})}\end{pmatrix}
\leftrightarrow
\begin{pmatrix}XXX&ZII&XZX\\IZI&YYX&IXI\\ZXX&XII&ZZX\end{pmatrix},
\label{pocok2}
\eeq
where we employed the notation $\Gamma^{(\mu\nu\rho)}\equiv\Gamma_{\mu}\Gamma_{\nu}\Gamma_{\rho}$.
Note that the transvection $T_{YYY}=T_{\Gamma_7}$ acts as the involution of taking the complement in the set $\{1,2,3,\overline{1},\overline{2},\overline{3}\}$, since for example,
\beq
\Gamma_7\Gamma^{(\overline{1}23)}=\Gamma_7\Gamma_{2}\Gamma_3\Gamma_4\simeq \Gamma_1\Gamma_5\Gamma_6=\Gamma^{(1\overline{23})}\simeq XXX,
\eeq
where $\simeq$ means equality modulo an element of $\mathbb{G}_4$. One can also immediately verify that the transvection $T_{\Gamma_7}$ exchanges the two components of the set $\{\Gamma_{\mu},\Gamma_{\mu 7}\}$ (Schl\"afli's double-six\cite{Polster,Levfin2}) and the $15$ operators $\{\Gamma_{\mu}\Gamma_{\nu}\Gamma_7\}$ and
 $\{\Gamma_{\mu}\Gamma_{\nu}\}$
provide two different labellings for the doily ($GQ(2,2)$), the first giving a labelling for the core set\cite{Zsolt}, the second one provides the duad labelling corresponding to Figure \ref{DoilyDuad}.

Now, according to (\ref{setsofveldkamp}), the special structure of our canonical Veldkamp line seems to be related to the special realization of our Clifford algebra. Indeed, all of the operators of (\ref{choice}) are \emph{antisymmetric} ones. However, since all what is important for us is merely commutation properties, we could have used \emph{any} such realization of the algebra. Hence our labelling convention suggests that one should be able to recast all the relevant information concerning our Veldkamp line entirely in terms of one, two, and three element subsets of the set $\{1,2,\dots,7\}$. This is indeed the case.

Let us elaborate on that point\cite{Richter,Ron,Ron2}.
Let $S\equiv\{1,2,3,4,5,6,7\}$. Then we are interested in incidence structures defined on certain sets of elements of $\mathcal{P}(S)\equiv 2^S$ with cardinality $1,2$ and $3$.
$\mathcal{P}(S)$ can be given the structure of a vector space over $\mathbb{Z}_2$. Addition is defined by taking the symmetric difference of two elements
$\mathcal{A},\mathcal{B}\in \mathcal{P}(S)$, i.e.
\beq
\mathcal{A}+\mathcal{B}=(\mathcal{A}\cup \mathcal{B})-(\mathcal{A}\cap \mathcal{B})
\eeq
\noindent
and $1\cdot\mathcal{A}=\mathcal{A}$ and $0\cdot\mathcal{A}=\{0\}$.
Let us denote by $\vert\mathcal{A}\vert$ the cardinality of $\mathcal{A}$ modulo $2$.
Then one can define a symplectic form $\langle\cdot\vert\cdot\rangle:\mathcal{P}(S)\times\mathcal{P}(S)\to\mathbb{Z}_2$ by
\beq
\langle\mathcal{A}\vert\mathcal{B}\rangle=\vert\mathcal{A}\vert\cdot\vert\mathcal{B}\vert +\vert\mathcal{A}\cap\mathcal{B}\vert.
\label{alldjacentsympl}
\eeq
\noindent
One can again define the symplectic transvections, as  involutive $\mathbb{Z}_2$ linear maps given by the expression
\beq
T_{\mathcal{B}}\mathcal{A}=\mathcal{A}+\langle\mathcal{A}\vert\mathcal{B}\rangle\mathcal{B}.
\label{transset}
\eeq
\noindent

Now it is easy to connect this formalism to our description of Pauli observables in terms of a seven-dimensional Clifford algebra.
Define the map $f: \mathcal{P}(S)\to Cliff(7)$ by
\beq
f(\mathcal{A})=\Gamma^{(\mathcal{A})}
\label{bijmap}
\eeq
\noindent
where, for example, for $\mathcal{A}=\{134\}$ we have $f(\mathcal{A})=\Gamma^{(134)}=\Gamma_1\Gamma_3\Gamma_4$ etc.
Now it is easy to prove that\cite{Richter}
\beq
\Gamma^{(\mathcal{A})}\Gamma^{(\mathcal{B})}=(-1)^{\langle\mathcal{A}\vert\mathcal{B}\rangle}\Gamma^{(\mathcal{B})}\Gamma^{(\mathcal{A})}.
\label{commanticomm}
\eeq
\noindent
Hence, for example, for sets $\mathcal{A}$ and $\mathcal{B}$ of cardinality $3$ if $\vert\mathcal{A}\cap\mathcal{B}\vert=1$ the observables $\Gamma^{(\mathcal{A})}$ and  $\Gamma^{(\mathcal{B})}$ commute, otherwise they  anticommute.
Now one can check that all of the relevant information on the commutation properties of observables, and also information on the action of the symplectic group can be nicely expressed in terms of data concerning $1,2$ and $3$-element subsets of $S$.

\subsection{The Doily and Hitchin's symplectic functional}

In our magic Veldkamp line we have two subsets of $15$ observables.
One belongs to the core set of our Veldkamp line (black triangle of Figure \ref{gods-eye-clifford}), and the other is belonging to the complement of the core in the perp-set of the special observable $\Gamma_7=YYY$ (red triangle in Figure \ref{gods-eye-clifford}).
Let us concentrate on the latter subset. According to Figure \ref{gods-eye-clifford} this set is described by an $\mathcal{A}\in \mathcal{P}(S)$ of the form $\mathcal{A}=\{a,b\vert 1\leq a<b\leq 6\}$. The corresponding observables are $i\Gamma^{(\mathcal{A})}\leftrightarrow i\Gamma_a\Gamma_b$. They are represented by antisymmetric, Hermitian matrices.
It is easy to establish a bijective correspondence between these $15$ observables and the $15$ weights of the $15$-dimensional irrep of $A_5$.

As is well-known\cite{Slansky} these weights are living in a five-dimensional hyperplane, with normal vector $n\equiv(1,1,1,1,1,1)^T$
of $\mathbb{R}^6$.
Knowing that the Dynkin labels\cite{Slansky} of the highest weight vector of this representation are $(01000)$, an analysis based on the Cartan matrix of $A_5$ similar to the detailed one that can be found in \cite{Zsolt}, yields the following set of $15$ weights
\beq
\Lambda^{(ab)}=e_a+e_b-\frac{1}{3}n,\qquad n=(1,1,1,1,1,1)^T,\qquad  1\leq a<b\leq 6.
\label{w15}
\eeq
\noindent
One can immediately check that the weights $\Lambda^{(ab)}$ are orthogonal to $n$ and satisfy the scalar product relations
\beq
(\Lambda^{(\mathcal A)},\Lambda^{(\mathcal B)})=\begin{cases}-\frac{2}{3}&,\quad\vert{\mathcal A}\cap{\mathcal B}\vert =0,\\
+\frac{1}{3}&,\quad\vert{\mathcal A}\cap{\mathcal B}\vert =1,\\
+\frac{4}{3}&,\quad\vert{\mathcal A}\cap{\mathcal B}\vert =2\equiv 0\quad {\rm mod}\quad 2.
\end{cases}
\label{skalarszorz}
\eeq
\noindent
Hence, by virtue of (\ref{commanticomm}), if $\vert{\mathcal A}\cap{\mathcal B}\vert =0$ the corresponding observables are commuting otherwise anticommuting ones.
Notice that for two different commuting observables the corresponding weights have an angle of $120$ degrees.
Three different mutually commuting observables represented by three weights that satisfy the sum rule
\beq
\Lambda^{({\mathcal A}_1)}+
\Lambda^{({\mathcal A}_2)}+\Lambda^{({\mathcal A}_3)}={\bf 0},\qquad \vert{\mathcal A}_{\alpha}\cap {\mathcal A}_{\beta}\vert =0,\quad \alpha,\beta=1,2,3,
\eeq
are of special status.
	Indeed, it is easy to show that the $15$ weights regarded as points and the $15$ triples of points satisfying our sum rule, regarded as lines, give rise to an incidence structure of a doily, $GQ(2,2)$.
For example, the weights $\Lambda^{(12)},\Lambda^{(34)},\Lambda^{(56)}$ satisfy the sum rule and give rise to the line $(12,34,56)$ of $GQ(2,2)$.
This $GQ(2,2)$ structure can also be realized as a distribution of $15$ points on the surface of a four-sphere with the radius $2/\sqrt{3}$
lying in the five dimensional hyperplane with a normal vector $(1,1,1,1,1,1)$.
The lines are then formed by any three equidistant points connected by a geodesic on the surface of that four-sphere (the three points are corresponding to three vectors with an angle of $120$ degrees lying on a great circle). 
Alternatively, triples of points representing lines correspond to three mutually commuting observables with their poducts giving rise to $\Gamma_7$ modulo $\mathbb{G}_4$.

There is, however, an alternative way of producing the incidence structure of the doily.
This way is arising from regarding the weights of Eq.(\ref{w15}) as ones labelled by four-element subsets $\mathcal{A}$ of $S$.
Hence, for example the highest weight $\Lambda^{(12)}$ can alternatively be labelled as $\Lambda^{(3456)}$.
Moreover, since $\Gamma^{(3456)}\simeq \Gamma^{(127)}$ this labelling by four-element subsets can be converted to three-element ones. This means that we can dually label the points 
of a $GQ(2,2)$ by triples of the form $\{ab7\}$. Since according to Figure \ref{gods-eye-clifford} this set covers precisely the core set of our Veldkamp line, we conclude that the finite geometric structure of the core is just another copy of the doily.
An example of a line of this doily is $(127,347,567)$. By virtue of Eq.(\ref{gamszorz}) to any such triples of triads there corresponds a triple of mutually commuting \emph{symmetric} Pauli observables such that their product equals the identity $III$ (modulo $\mathbb{G}_4$).

\begin{figure}[pth!]
\centerline{\includegraphics[width=5truecm,clip=]{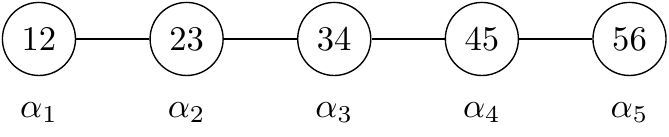}}
\caption{The $A_5$ Dynkin diagram labelled by duads.}
\label{A5diagram}
\end{figure}

\begin{figure}[pth!]
\centerline{\includegraphics[width=4truecm,clip=]{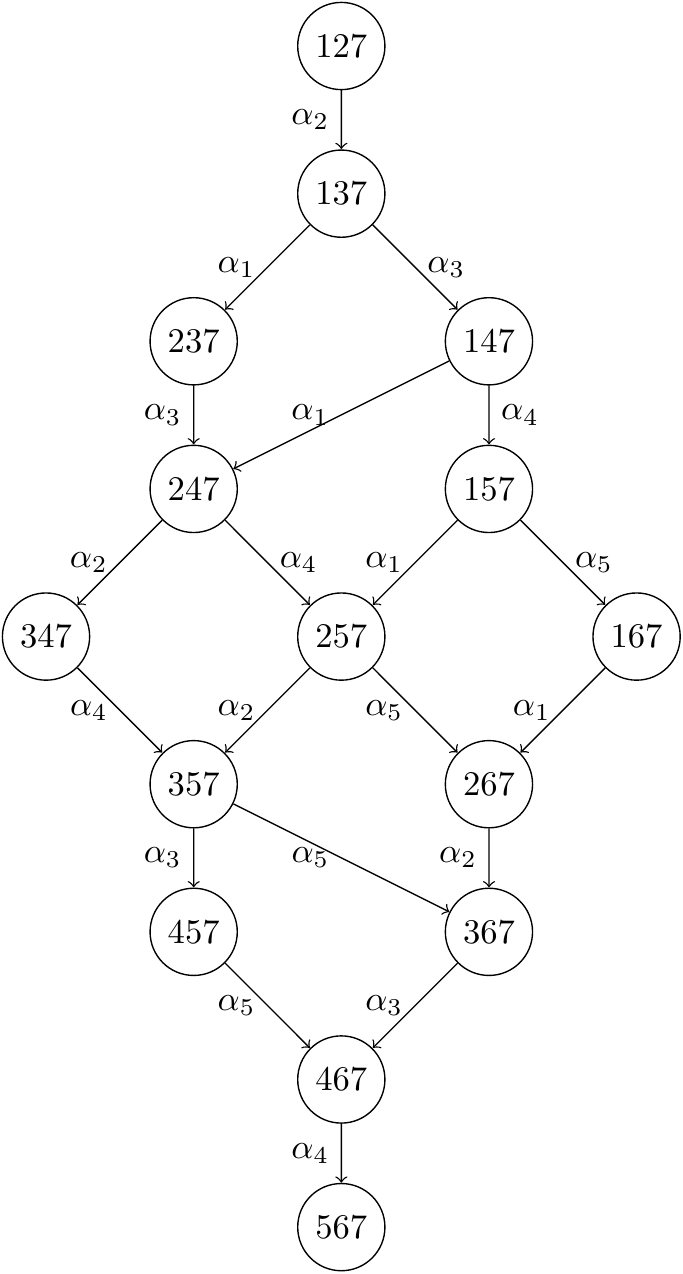}}
\caption{The weight diagram for the $15$ of $A_5$.
The weights labelled by three-element subsets as $\{ab7\}$ correspond to the core set
of Figure \ref{gods-eye-clifford} of our Veldkamp line labelled by $\Gamma_a\Gamma_b\Gamma_7$.}
\label{15A5}
\end{figure}

If we label the nodes of the $A_5$-Dynkin diagram by the simple roots $\alpha_n$, $n=1,2,3,4,5$, as can be seen in Figure \ref{A5diagram} and apply the (\ref{transset}) transvections
$T_{\mathcal{A}}$ with $\mathcal{A}$ taken from the five subsets $\{12,23,34,45,56\}$ to the weights, then starting from the highest weight $\mathcal{B}=\{127\}$ the weight diagram can be generated.
The result can be seen in Figure \ref{15A5}.
Notice that according to the (\ref{choice}) dictionary and the bijective mapping of Eq.(\ref{bijmap}) this labelling via elements of $\mathcal{P}(S)$ of the Dynkin and the weight diagrams automatically gives rise to a labelling in terms of Pauli observables. Similar labelling schemes can be found in \cite{Geemen,Zsolt}.
This result establishes a correspondence between a representation theoretic and a finite geometric structure (namely the doily).

Let us now show that the incidence structure of the $15$-element core set of our Veldkamp line encodes the structure of a cubic $SL(6)$ invariant related to the $15$ of $A_5$. As is well-known this invariant is the Pfaffian of an antisymmetric $6\times 6$ matrix $\omega_{ij}$
\beq
{\rm Pf}(\omega)=\frac{1}{3!2^3}\varepsilon^{abcdef}\omega_{ab}\omega_{cd}\omega_{ef}=\omega_{12}\omega_{34}\omega_{56}-
\omega_{13}\omega_{24}\omega_{56}+\dots.
\label{Pfaff}
\eeq
\noindent
Clearly, according to Figure \ref{DoilyDuad}, the $15$ monomials of this cubic polynomial
can be mapped bijectively to the $15$ lines of the doily residing in the core of our Veldkamp line.
What about the signs of the monomials?

In order to tackle the problem of signs we relate this invariant to Pauli observables of the form $\mathcal{O}_{ab}=
\mathcal{O}_{ab}^{\dagger}=
i\Gamma_a\Gamma_b\Gamma_7$ with $1\leq a<b\leq 6$
and define the $8\times 8$ matrix $\Omega$
\beq
\Omega=i\sum_{a<b}\omega_{ab}\Gamma_a\Gamma_b\Gamma_7=\omega_{12}(i\Gamma_1\Gamma_2\Gamma_7)+\omega_{13}(i\Gamma_1\Gamma_3\Gamma_7)+\dots +\omega_{56}(i\Gamma_5\Gamma_6\Gamma_7).
\label{OmegaMermin}
\eeq
\noindent
Then the Pfaffian can also be written in the form
\beq
{\rm Pf}(\omega)=\frac{-1}{3!2^3}{\rm Tr}(\Omega^3).
\label{Pfaff2}
\eeq
\noindent
Indeed, in this new version of the Pfaffian the $15$ basis vectors in the expansion of $\Omega$ are three-qubit observables.
Since, according to Eq.(\ref{gamszorz}), the product of commuting triples of observables corresponding to lines in Figure \ref{DoilyDuad} results in $\pm 1$ times the $8\times 8$ identity matrix, these terms give rise to the $15$ signed monomials of Eq.(\ref{Pfaff}). The remaining triples give trace zero terms hence the result of Eq.(\ref{Pfaff2}) follows.

A dual of this invariant is obtained by
considering the dual $6\times 6$ matrix
\beq
	\tilde{\omega}^{ef}=\frac{1}{8}{\varepsilon}^{efabcd}\omega_{ab}\omega_{cd}\equiv \frac{1}{24}\varepsilon^{efabcd}Q_{abcd}.
\label{Q}
\eeq
\noindent
Note that, for example,
\beq
\tilde{\omega}^{56}=Q_{1234}=\omega_{12}\omega_{34}-\omega_{13}\omega_{24}+\omega_{14}\omega_{23}.
\label{perpdoily}
\eeq
\noindent
Then 
\beq
{\rm Pf}(\tilde{\omega})=[{\rm Pf}(\omega)]^2.
\eeq
\noindent
Introducing a new $8\times 8$ matrix
\beq
\tilde{\Omega}=
Q_{3456}\Gamma_3\Gamma_4\Gamma_5\Gamma_6-Q_{2456}\Gamma_2\Gamma_4\Gamma_5\Gamma_6+\dots =
i\tilde{\omega}^{12}\Gamma_1\Gamma_2\Gamma_7+i\tilde{\omega}^{13}\Gamma_1\Gamma_3\Gamma_7+\dots
\label{Omegatilde}
\eeq
\noindent
one can write
\beq
{\rm Pf}(\tilde{\omega})=\frac{-1}{3!2^3}{\rm Tr}(\tilde{\Omega}^3).
\label{Pfaff3}
\eeq
If ${\rm Pf}(\omega)>0$, then
\beq
{\rm Pf}(\omega)=\sqrt{{\rm Pf}(\tilde{\omega})}\equiv\sqrt{{\mathcal{F}}(Q)}.
\label{Hitchinfunct1}
\eeq
\noindent
The quantity $\sqrt{{\mathcal{F}}(Q)}$ is the invariant which is used to define a functional\cite{FormTH} on a closed orientable six-manifold $M$ equipped with a (nondegenerate) four-form $Q=\frac{1}{2}\omega\wedge\omega$
\beq
V_{SH}[Q]=\int_{M}\sqrt{{\mathcal{F}}(Q)}d^6x.
\label{symplhitch}
\eeq
\noindent
The critical point of this Hitchin functional\cite{Hitchin2,FormTH} defines a symplectic structure for the six-manifold.

Let us elaborate on the structure of ${\mathcal{F}}(Q)$ underlying this Hitchin functional.
Clearly, its structure is encoded into the one of the matrix of Eq.(\ref{Omegatilde}).
It can be regarded as an expansion in terms of three-qubit Pauli observables. Now the labels $\{abcd\}$, with $a<b<c<d$ can be regarded as dual ones to the familiar labels $\{mn7\}$ of our core doily.
Indeed, a line of the doily like $127-347-567$ can be labelled dually as $3456-1256-1234$.
In the cubic expression of Eq.(\ref{Pfaff3}) this line gives rise to a term proportional to $\Gamma_{3456}\Gamma_{1256}\Gamma_{1234}=-III$, i.e. the negative of the $8\times 8$ identity matrix.
Alternatively, one can regard this identity as the one between three commuting Pauli observables with product being the negative of the identity.
Now all the lines giving contribution to ${\mathcal{F}}(Q)$ will be featuring triples of commuting observables giving rise to either $-III$ or $+III$. These lines will be called \emph{positive} or \emph{negative lines}.
Now from Figure \ref{mermins} we know that Mermin squares are geometric hyperplanes of the doily with $9$ points and $6$ lines, and a particular distribution of signs for the $15$ lines of the doily governed by 
${\mathcal{F}}(Q)$ 
implies a distribution for the six lines of the $10$ possible Mermin squares.
It is easy to check that all of them contain an odd number of negative lines, hence can furnish a proof for ruling out noncontextual hidden variable theorems.

As an illustration
let us use again the notation $\{1,2,3,4,5,6\}=\{1,2,3,\overline{1},\overline{2},\overline{3}\}$ and keep only $9$ terms from the expression of $\tilde{\Omega}$ of (\ref{Omegatilde}) defining the matrix
\beq
\mathcal{M}=i\sum_{\alpha=1}^3\sum_{\overline{\beta}=\overline{1}}^{\overline{3}}\tilde{\omega}^{\alpha\overline{\beta}}\Gamma_{\alpha\overline{\beta}7}=
i\tilde{\omega}^{14}\Gamma_{147}+\dots=Q_{2356}\Gamma_{2356}+\dots.
\label{MerminGrid}
\eeq
\noindent
Then it is easy to check that the $9$ observables showing up in the expansion of $\mathcal{M}$ form a $3\times 3$ grid labelled by Pauli observables such that the ones along its six lines are commuting.
A short calculation shows that we have three negative and three positive lines.
Hence, the object we obtained is an example of a Mermin square\cite{Mermin}.

Let us now calculate the restriction of (\ref{Pfaff3}) to $\mathcal{M}$.
The result is
\beq
\frac{1}{3!2^3}{\rm Tr}(\mathcal{M}^3)={\rm Det}(\tilde{\omega}^{\alpha\overline{\beta}}).
\label{szukul}
\eeq
\noindent
Since it is the determinant of a $3\times 3$ matrix, it has three monomials with a plus and three ones with a minus 
sign,
hence reproducing the distribution of signs of a Mermin square via a substructure of the Pfaffian.

Summing up: to an incidence structure (doily) forming the core of our magic Veldkamp line one can associate an invariant which encodes information on the structure of its lines and also on the distribution of signs for these lines.
Moreover, Eqs. (\ref{Hitchinfunct1}) and (\ref{szukul}) also show that substructures like Mermin squares live naturally inside the expression for Hitchin's invariant ${\mathcal{F}}(Q)$.

We note in closing that from (\ref{perpdoily}) we see that the $15$ independent components of $Q_{abcd}$ are built up from those components of $\omega_{ab}$ that are labelling the perp-sets (geometric hyperplanes again) of the doily. For example, in the duad labelling the perp-set of $(56)$ consists of
the points labelled by the ones: $(12),(34),(13),(24),(14),(23)$. In terms of observables these correspond to the ones commuting with the fixed one $-i\Gamma_5\Gamma_6=YZX$.
This occurrence of the perp-sets inside the core doily can be understood yet another way.
We already know from Figure \ref{15A5} that the weights of the $15$ of $A_5$ are labelled as $\{ab7\}$.
We can decompose this irrep with respect to the subgroup $A_1\times A_3$.
More precisely let us consider the real form  $SU(6)$ of $A_5$. Then
under the subgroup $SU(2)\times SU(4)\times U(1)$ the $15$ of $SU(6)$ decomposes as\cite{Slansky}
\beq
15=(1,1)(4)+(1,6)(-2)+(2,4)(1).
\label{decompperp}
\eeq
Let us make a split of the set $\{ab7\}$ as follows: $\{\alpha\beta7\}$ where $1\leq \alpha<\beta\leq 4$, $\{567\}$ and
$\{\alpha 57\}$, $\{\alpha 67\}$. 
Let us delete the node of the Dynkin diagram labelled by $\alpha_4$. Then we are left with the two Dynkin diagrams of an $A_3$ and an $A_1$.
Now $\{\alpha\beta7\}$ corresponds to the $6$ of $SU(4)$.
Indeed, starting from the highest weight $\{127\}$ the six weights of this representation are obtained using the roots $\alpha_1,\alpha_2,\alpha_3$, comprising the $SU(4)$ part of the Dynkin diagram.
The $\{567\}$ part forms a singlet both under $SU(2)$ and $SU(4)$.
In the language of Pauli observables, the singlet part corresponds to the observable $-i\Gamma_5\Gamma_6\Gamma_7=IXZ$, and the weights of the six-dimensional irrep correspond to observables commuting with $IXZ$. These $7$ observables form a geometric hyperplane which is the perp-set of the doily. 
The complements of this perp-set in the doily decompose into the two sets of four observables namely $\{\alpha 57\}$ and $\{\alpha 67\}$. Each of them forms a $4$-dimensional irrep under $SU(4)$. They are exchanged by the transvection $T_{56}$, hence they form an $SU(2)$-dublet.

For the sake of completeness, we should also mention that one more type of geometric hyperplane of $GQ(2,2)$, an \emph{ovoid}
(that is a set of points of $GQ(2,2)$ such that each line of $GQ(2,2)$ is incident with exactly one point of the set) is represented by
$\{ab7\}$, where $b$  is fixed. For example, for $b=6$ we have the set $\{167,267,367,467,567\}$. The five observables corresponding to these triples $i\Gamma_a\Gamma_b\Gamma_7$ are mutually \emph{anticommuting}, i.e. form a five dimensional Clifford algebra.
In terms of the duad version $\{16,26,36,46,56\}$ of this five-tuple, Figure \ref{DoilyDuad} clearly shows the ovoid property of the corresponding five points.

\subsection{An Extended Generalized Quadrangle $\bf{EGQ(2,1)}$ and Hitchin's functional}

Let us now revisit the results of \cite{Zsolt} from a different perspective.
Let $S=\{1,2,3,4,5,6\}$.
We consider the green triangle part of Figure \ref{gods-eye-clifford}. This part is labelled by subsets $\mathcal{A}\in\mathcal{S}\subset\mathcal{P}(S)$ of the form $\mathcal{A}=\{abc\}$
where $1\leq a<b<c\leq 6$. As was shown in \cite{Zsolt} starting from the highest weight $(00100)$ of the $20$-dimensional irrep of $A_5$, the $20$ weights can be constructed. They are residing in the hyperplane through the origin of $\mathbb{R}^6$ with normal $n=(1,1,1,1,1,1)^T$. They take the following form
\beq
\Lambda^{(abc)}=e_a+e_b+e_c-\frac{1}{2}n.
\label{w20}
\eeq
\noindent
According to Eqs.(\ref{pocok1})-(\ref{pocok2}) and (\ref{commanticomm}), if the intersection sizes of weight vector labels are odd (even) the corresponding operators are commuting (not commuting).
This information translates to an incidence structure between weight vectors. Namely: having scalar product $-\frac{1}{2},\frac{3}{2}$ corresponds to incident vectors (commuting operators), and
$\frac{1}{2},-\frac{3}{2}$  to not incident vectors (not commuting operators).
This is summarized as
\beq
(\Lambda^{(\mathcal A)},\Lambda^{(\mathcal B)})=\begin{cases}-\frac{3}{2}&,\quad\vert{\mathcal A}\cap{\mathcal B}\vert =0\\
-\frac{1}{2}&,\quad\vert{\mathcal A}\cap{\mathcal B}\vert =1\\
+\frac{1}{2}&,\quad\vert{\mathcal A}\cap{\mathcal B}\vert =2\equiv 0 \quad{\rm mod} \quad 2\\
+\frac{3}{2}&,\quad\vert{\mathcal A}\cap{\mathcal B}\vert =3\equiv 1\quad {\rm mod} \quad 2.
\end{cases}
\eeq
\noindent
Since norm-squared for weight vectors equals $\frac{3}{2}$, two different weights $\Lambda^{(\mathcal A)}$ and
$\Lambda^{(\mathcal B)}$
are incident when the angle between them satisfies
$\cos\theta_{{\mathcal A}{\mathcal B}}=-1/3$.
Weights with labels satisfying $\vert\mathcal{A}\cap\mathcal{B}\vert =0$ will be called \emph{antipodal}. Indeed, for such pairs
(e.g. $123$ and $456$) we have $\cos\theta_{{\mathcal A}{\mathcal B}}=-1$.

Let us now consider four \emph{different} weights, called \emph{quadruplets}.  Subsets $\mathcal{A}_s$, $s=1,2,3,4$, with the corresponding quadruplets satisfying
\beq
\Lambda^{({\mathcal A}_1)}+
\Lambda^{({\mathcal A}_2)}+\Lambda^{({\mathcal A}_3)}+\Lambda^{({\mathcal A}_4)}=0,\qquad \vert{\mathcal A}_s\cap {\mathcal A}_t\vert =1,\quad s,t=1,2,3,4,
\label{blockrule}
\eeq
will be called \emph{blocks}.
An example of a block is
\beq
({\mathcal A}_1,{\mathcal A}_2,{\mathcal A}_3,{\mathcal A}_4)=
(123,156,246,345).
\label{subidubi}
\eeq
\noindent
Hence, apart from the constraint $\vert{\mathcal A}_s\cap {\mathcal A}_t\vert =1$, a block is characterized by a \emph{double occurrence} of all elements of $S=\{1,2,\dots 6\}$.
We note that in terms of the $20$ observables $\mathcal{O}_{abc}=i\Gamma_a\Gamma_b\Gamma_c=\mathcal{O}^{\dagger}$  the blocks bijectively correspond to the lines of a double-six of Mermin pentagrams of \cite{Zsolt}.
In this language the (\ref{blockrule}) rule defining the blocks translates to the fact that the product of four commuting observables is (up to a crucial sign) the identity.

Let us now choose any of the triples, e.g. $123$. This triple shows up in $6$ blocks. These blocks can be described by adjoining to $123$ the following $3\times 3$ arrangement of triples,
\beq
123\mapsto
\mathcal{S}_{123}\equiv\begin{pmatrix}156&146&145\\
256&246&245\\
356&346&345
\label{kanvan}
\end{pmatrix}.
\eeq
Regard temporarily these triples as numbers and the arrangement as a $3\times 3$ matrix.
Then multiplying the triple $123$ with the determinant of the matrix $\mathcal{S}_{123}$ we get $6$ terms. The $6$ terms showing up (signs are not important) in this quartic polynomial are the $6$ blocks featuring $123$. 
In particular, the block of Eq.(\ref{subidubi}) is arising from the diagonal of $\mathcal{S}_{123}$.
One can furnish $\mathcal{S}_{123}$ with the structure of: all the \emph{points} $\mathcal{A}\neq 123$ that are on a block on $123$, and all the \emph{blocks} on $123$.
We will call $\mathcal{S}_{123}$ equipped with this structure the \emph{residue}\cite{Fra} of the point $123$.

Since we have $20$ points, we have $20$ residues $\mathcal{S}_{\mathcal{A}}$. One can generate all residues from the one of Eq.(\ref{kanvan}), dubbed the \emph{canonical one}, as follows.
First, notice that to our canonical residue one can associate its \emph{antipode}
\beq
456\mapsto
\mathcal{S}_{456}\equiv\begin{pmatrix}234&134&124\\
235&135&125\\
236&136&126
\label{kanvananti}
\end{pmatrix}.
\eeq
Now the symmetric group $S_6$ clearly acts on $\mathcal{S}$ via permutations.
The canonical residue and its antipode are left invariant by the group $S_3\times S_3$ acting via separate permutations of the numbers $1,2,3$ and $4,5,6$. Hence, the $9$ transpositions of the form $\{14\},\{15\},\{16\},\dots,\{36\}$ generate $9$ new residues from the canonical one. Combining this with the antipodal map $18$ new residues are obtained. Taken together with the canonical one and its antipode all of the $20$ residues can be obtained.
For example, after applying the transposition $\{14\}$ the new residues are
\beq
\mathcal{S}_{234}\equiv\begin{pmatrix}456&146&145\\
256&126&125\\
356&136&135
\end{pmatrix},\qquad
\mathcal{S}_{156}\equiv\begin{pmatrix}123&134&124\\
235&345&245\\
236&346&246
\end{pmatrix}.
\label{kantrans}
\eeq
One can then check the following. The number of blocks is $\vert\mathcal{B}\vert =30$.
Moreover, two distinct blocks meet in $0,1$ or $2$ points. On the other hand, two distinct points are either on no common block or on $2$ common blocks.
An illustration of this structure can be found in Figure 4 of Ref.\cite{Zsolt} depicting the double-six structure of Mermin pentagrams. We have $12$ pentagrams in this configuration with each pentagram having $5$ lines. However, certain pairs of pentagrams are having lines in common. As a result we will have merely $30$ lines in this configuration. After identifying the lines of the double-sixes with our blocks one can check that the incidence structures are isomorphic.

\begin{figure}[pth!]

\centerline{\includegraphics[width=9truecm,clip=]{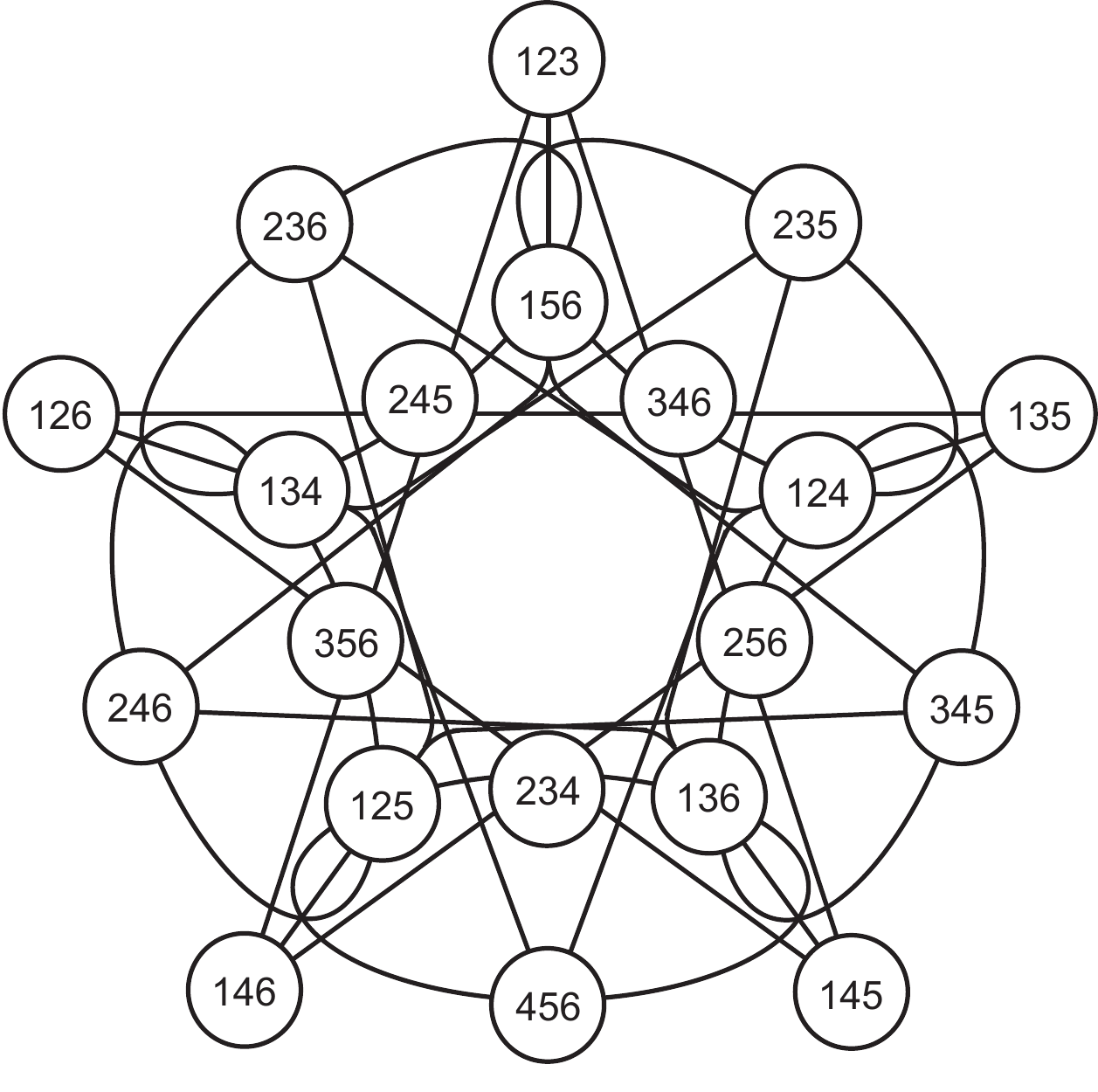}}
\centerline{\includegraphics[width=9truecm,clip=]{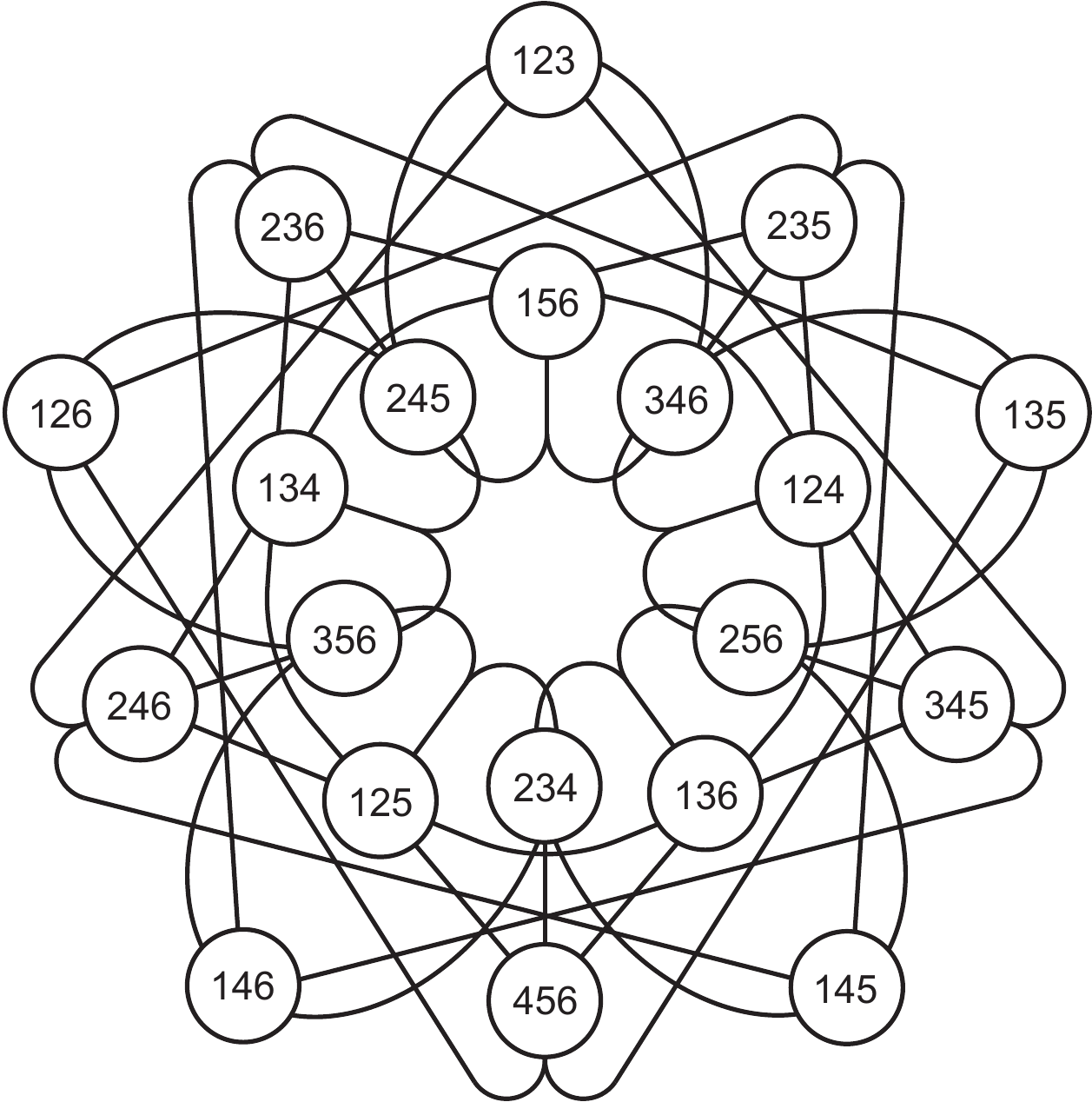}}
\caption{The twin Steiner-Pl\"ucker configurations illustrating the $30$ blocks of $EGQ(2,1)$
related to the $20$ of $A_5$. Our $EGQ(2,1)$, which is the affine polar space of order two and type $D_2^+$ \cite{Fra}, can
be viewed as the union of twin Steiner-Pl\"ucker configurations. The two configurations are
identical as point-sets, their points being represented by unordered triples of elements from
the set $S = \{1, 2, 3, 4, 5, 6\}$. Lines (blocks) of the configurations are represented by four
points that pairwise share one element.
The name "Steiner-Pl\"ucker" configuration comes from the fact\cite{Boben} that it is a $(20_3,15_4)$-configuration that consists of $20$ Steiner points and $15$ Pl\"ucker lines of the famous \emph{Hexagrammum Mysticum} of Pascal (see e. g., \cite{Conway,Ladd}).
}
\label{STPL}
\end{figure}

Let us now turn back to our construction of this incidence structure based on residues.
Recalling Definition \ref{defn:genquadr} one can observe that each residue is having the structure of a generalized quadrangle of type $GQ(2,1)$, i.e. a grid.
On the other hand, 
according to Definition \ref{defn:extgenquadr} a connected structure with two types, namely \emph{points} and \emph{blocks}, such that each residue $\mathcal{S}_p$ of a point $p$, is a generalized quadrangle $GQ(s,t)$ is called an 
\emph{Extended Generalized Quadrangle}: $EGQ(s,t)$.
Hence, in  our case we have found two interesting applications of this concept. Namely, we have verified that the block structure defined on the set of weights of the $20$ of $A_5$ by Eq.(\ref{blockrule}),  and the double-six structure of pentagrams of Ref.\cite{Zsolt} with blocks defined via commuting quadruplets of observables give rise to two realizations of an $EGQ(2,1)$.
A nice way of illustrating the structure of our $EGQ(2,1)$ can be obtained by observing that this configuration can be built from two copies of the so-called Steiner-Pl\"ucker configuration, see Figure \ref{STPL}.

Now as the most important application of this concept let us show that our $EGQ(2,1)$ encapsulates the geometry of Hitchin's functional\cite{Hitchin} in terms of the information encoded into the canonical residue $\mathcal{S}_{123}$ of Eq.(\ref{kanvan}).

Let us first give two altenative forms of Hitchin's functional.
The original one\cite{Hitchin}, widely used by string theorists\cite{FormTH,PW}, is
defined via introducing $K$, a $6\times 6$ matrix giving rise to an almost complex structure on $\mathcal{M}$, a closed, real, orientable six-manifold equipped with a (nondegenerate, negative) three-form $P$ \beq
{(K_P)^a}_b=\frac{1}{2!3!}\varepsilon^{ai_2i_3i_4i_5i_6}{P}_{bi_2i_3}{P}_{i_4i_5i_6},\qquad 1\leq a,b,i_2,\dots i_6\leq 6.
\label{KP} \eeq \noindent In terms of this quantity Hitchin's invariant can be
expressed as \beq
\mathcal{D}(P)=\frac{1}{6}{\rm Tr}(K_P^2).
\label{Hitchinform} \eeq \noindent
It is known that for real forms there are two nondegenerate classes of such forms, forms with $\mathcal{D}<0$ and 
$\mathcal{D}>0$.
Now Hitchin's functional is defined as \beq V_H[P]=\int_{\mathcal{M}}\sqrt{\vert{\cal
D}(P)\vert}d^6x.\label{Hitchinfunk} \eeq \noindent 
In the special case when $\mathcal{D}(P)<0$ varying this functional in a fixed cohomology class the Euler-Lagrange equations imply that the almost complex structure $K/\sqrt{-\mathcal{D}(P)}$, with $P$ being the one defining the critical point, is integrable\cite{Hitchin}. Hence, the critical points of this functional define complex structures on $\mathcal{M}$.
The quantum theory based on this functional was studied by Pestun and Witten\cite{PW}. It is related to the quantum theory of topological strings\cite{TopString}.

An alternative form of this functional is given by writing Hitchin's invariant in the following form\cite{LevSar}.
Recall our labelling convention $(1,2,3,4,5,6)=(1,2,3,\overline{1},\overline{2},\overline{3})$. Define
 \beq
 \eta\equiv P_{123},\qquad \xi\equiv P_{\overline{123}},
 \label{etaxi}
 \eeq
 \beq
 {\mathbf X}=\begin{pmatrix}X_{11}&X_{12}&X_{13}\\X_{21}&X_{22}&X_{23}\\X_{31}&X_{32}&X_{33}\end{pmatrix}
 \equiv\begin{pmatrix}P_{1\overline{23}}&P_{1\overline{31}}&P_{1\overline{12}}\\
 P_{2\overline{23}}&P_{2\overline{31}}&P_{2\overline{12}}\\P_{3\overline{23}}&P_{3\overline{31}}&P_{3\overline{12}}\end{pmatrix},
 \label{Xmatr}
 \eeq
 \beq
 {\mathbf Y}=\begin{pmatrix}Y^{11}&Y^{12}&Y^{13}\\Y^{21}&Y^{22}&Y^{23}\\Y^{31}&Y^{32}&Y^{33}\end{pmatrix}\equiv
 \begin{pmatrix}P_{\overline{1}23}&P_{\overline{1}31}&P_{\overline{1}12}\\P_{\overline{2}23}&P_{\overline{2}31}&P_{\overline{2}12}\\
 P_{\overline{3}23}&P_{\overline{3}31}&P_{\overline{3}12}\end{pmatrix}.
 \label{Ymatr}
 \eeq
With this notation Hitchin's invariant is
\beq
{\cal D}(P)=[\eta\xi -{\rm
 Tr}({\mathbf X}{\mathbf Y})]^2-4{\rm Tr}({\mathbf X}^{\sharp}{\mathbf Y}^{\sharp})+4\eta{\rm
  Det}({\mathbf X})+4\xi{\rm Det}({\mathbf Y}) \label{Cayleygen}
   \eeq
    \noindent where ${\mathbf X}^{\sharp}$ and ${\mathbf Y}^{\sharp}$ correspond to the
     regular adjoint matrices for ${\mathbf X}$ and ${\mathbf Y}$, hence, for example
      ${\mathbf X}{\mathbf X}^{\sharp}={\mathbf X}^{\sharp}{\mathbf X}={\rm Det }({\mathbf X}){\mathbf I}$
       with ${\mathbf I}$ the $3\times 3$ identity matrix.

Let us refer to this $1+9+9+1$ split via introducing the arrangement $P=(\eta,{\mathbf X},{\mathbf Y},\xi)$. Then one can define a \emph{dual} arrangement $\tilde{P}=(\tilde{\eta},\tilde{\mathbf X},\tilde{\mathbf Y},\tilde{\xi})$ as follows\cite{Krutelevich}

\beq
\frac{\tilde{\eta}}{2}=\eta\kappa+{\rm Det}{\mathbf Y},\quad
\frac{\tilde{\mathbf X}}{2}=\xi{\mathbf Y}^{\sharp}-2{\mathbf Y}\times {\mathbf X}^{\sharp}-\kappa {\mathbf X},\quad
\frac{\tilde{\mathbf Y}}{2}=-\eta{\mathbf X}^{\sharp}+2{\mathbf X}\times {\mathbf Y}^{\sharp}-\kappa {\mathbf Y},\quad
\frac{\tilde{\xi}}{2}=-\xi\kappa-{\rm Det}{\mathbf X},
\label{duálisP}
\eeq
\noindent
where
\beq 2\kappa=\eta\xi-{\rm Tr}({\mathbf X}{\mathbf Y}),\qquad 2({\mathbf X}\times {\mathbf Y})=({\mathbf X}+{\mathbf Y})^{\sharp}-{\mathbf X}^{\sharp}-{\mathbf Y}^{\sharp}.
\eeq
\noindent
Then by defining the symplectic form\cite{Krutelevich}
\beq
\{P_1,P_2\}\equiv \eta_1\xi_2-\eta_2\xi_1+{\rm Tr}({\mathbf X}_1{\mathbf Y}_2)-{\rm Tr}({\mathbf X}_2{\mathbf Y}_1)
\eeq
\noindent
one can alternatively write
\beq
\mathcal{D}(P)=\frac{1}{2}\{\tilde{P},P\}.
\label{DualD}
\eeq

The (\ref{Cayleygen}) and (\ref{DualD}) ways of writing Hitchin's invariant are very instructive. The reason for it is twofold. First, they reveal their intimate connection with the fermionic entanglement measure introduced in Ref.\cite{LevVran,LevSar}.
It turns out that the entanglement classes of a three-fermion state with six modes represented by $P$ are 
characterized by the quantities $\mathcal{D}(P)$, $K_P$, and $\tilde{P}$.
This observation connects issues concerning the Hitchin functional to the Black Hole/Qubit Correspondence\cite{BHQC}.

Second, one can immediately realize that ${\mathbf X}$ is just the matrix associated to the one of Eq.(\ref{kanvan}) i.e. up to a sign in the second column it is the residue\footnote{For issues of incidence the signs are not important. However, here for understanding the structure of Hitchin's invariant they turn out to be important. Clearly, sign flips are arising when "normal ordering" of labels like $P_{1\overline{31}}=P_{164}=-P_{146}$ is effected.} of $123$.
Now the term $\eta{\rm
  Det}({\mathbf X})$ is featuring precisely the $6$ terms defining the blocks on $123$.
Similarly, the matrix associated with the residue of $456$ is ${\mathbf Y}$ and the corresponding $6$ terms of $\xi{\rm Det}({\mathbf Y})$ encode the six antipodal blocks on
$456$.
Based on these observations one expects that the structure of Hitchin's invariant is encapsulated into the geometry of a \emph{single residue} of $EGQ(2,1)$, i.e. the canonical one of
Eq.(\ref{kanvan}).

In order to prove this recall that, according to Eqs.(\ref{kanvan})-(\ref{kanvananti}) and (\ref{kantrans}),
from the canonical residue one can obtain all of the $20$ residues by two types of operations.
One of them is the antipodal map relating e.g. Eq.(\ref{kanvan}) with (\ref{kanvananti}), and the other is an application  of $9$ transpositions of the form: $(\alpha\overline{\beta})$ where $1\leq \alpha,\beta\leq 3$.
As discussed in Eq.(\ref{transset}), these operations are neatly described by transvections: $T_7,T_{\alpha\overline{\beta}}$.
In order to understand how these transvections act on the $20$ $P_{abc}$, with $1\leq a<b<c\leq 6$, we have to lift the action of the transvections to three-qubit observables\cite{Geemen}. This lift associates to $T_7,T_{\alpha\overline{\beta}}$ the adjoint action of $8\times 8$ unitary matrices
\beq
\mathcal{U}(T_7)=\frac{1}{\sqrt{2}}(I_8+i\Gamma_7),\qquad
 \mathcal{U}(T_{\alpha\overline{\beta}})=\frac{1}{\sqrt{2}}(I_8+\Gamma_{\alpha}\Gamma_{\overline{\beta}}),
\label{unitaries}
\eeq
\noindent
on observables as follows
\beq
\mathcal{O}\mapsto \mathcal{U}^{\dagger}(T_7)\mathcal{O}\mathcal{U}(T_7),\qquad
\mathcal{O}\mapsto \mathcal{U}^{\dagger}(T_{\alpha\overline{\beta}})\mathcal{O}\mathcal{U}(T_{\alpha\overline{\beta}}).
\eeq
\noindent
Explicitly, for the $20$ observables of the form 
\beq
\mathcal{O}_{abc}=\mathcal{O}^{\dagger}_{abc}=i\Gamma_a\Gamma_b\Gamma_c,\qquad 1\leq a<b<c\leq 6,
\label{3observ}
\eeq
\noindent
we have
\beq
\mathcal{O}_{abc}\mapsto \mathcal{U}^{\dagger}(T_{\alpha\overline{\beta}})\mathcal{O}_{abc}\mathcal{U}(T_{\alpha\overline{\beta}})=\begin{cases}\mathcal{O}_{abc}&,\quad\vert\{\alpha\overline{\beta}\}\cap\{abc\}\vert\equiv 0\quad{\rm mod}\quad 2\\
-\Gamma_{\alpha}\Gamma_{\overline{\beta}}\mathcal{O}_{abc}&,\quad\vert\{\alpha\overline{\beta}\}\cap\{abc\}\vert\equiv 1\quad{\rm mod}\quad 2
.
\end{cases}
\eeq
\noindent
For example, choosing $T_{1\overline{1}}=T_{14}$ we have
\beq
\mathcal{O}_{346}\mapsto \mathcal{U}^{\dagger}(T_{1\overline{1}})\mathcal{O}_{346}\mathcal{U}(T_{1\overline{1}})=
-i\Gamma_1\Gamma_4(\Gamma_3\Gamma_4\Gamma_6)=i\Gamma_1\Gamma_3\Gamma_6=\mathcal{O}_{136}.
\label{pelda1}
\eeq
\noindent
Let us now define the following Hermitian $8\times 8$ matrix $\Pi$ associated to the three-form $P$ featuring $\mathcal{D}(P)$ of Eq.(\ref{Cayleygen})
\beq
\Pi=\sum_{1\leq a<b<c\leq 6}P_{abc}\mathcal{O}_{abc}.
\label{Pas3observ}
\eeq
\noindent
Then the action on the observable $\Pi\mapsto \Pi^{\prime}=\mathcal{U}^{\dagger}\Pi\mathcal{U}$ defines an action on the coefficients $P_{abc}$ as follows
\beq \Pi^{\prime}=
\sum_{1\leq a<b<c\leq 6}P_{abc}
\mathcal{U}^{\dagger}(T_{\alpha\overline{\beta}})
\mathcal{O}_{abc}
\mathcal{U}(T_{\alpha\overline{\beta}})=\sum_{1\leq a<b<c\leq 6}P^{\prime}_{abc}\mathcal{O}_{abc},\quad
P^{\prime}_{abc}\equiv[\mathcal{T}_{\alpha\overline{\beta}}(P)]_{abc}.
\label{Phatas}
\eeq
\noindent
As an example of the rules given by Eqs.(\ref{pelda1}) and (\ref{Phatas}), we give the explicit form of the action of the transvection $T_{1\overline{3}}=T_{16}$ on the $P_{abc}$s
\begin{eqnarray}
P_{123}\mapsto -P_{236}\mapsto -P_{123},\qquad
P_{456}\mapsto P_{145}\mapsto -P_{456},\qquad
P_{134}\mapsto -P_{346}\mapsto -P_{134},\\\nonumber
P_{135}\mapsto -P_{356}\mapsto -P_{135},\qquad
P_{124}\mapsto -P_{246}\mapsto -P_{124},\qquad
P_{125}\mapsto -P_{256}\mapsto -P_{125},
\end{eqnarray}
\noindent
and the remaining components are left invariant.
One can also check that the transformation rule of Eq.(\ref{Phatas}) for the map $\mathcal{U}(T_7)$ gives rise to the following transformation
\beq
\mathcal{U}(T_7):(\eta,{\mathbf X},{\mathbf Y},\xi)\mapsto (-\xi,-{\mathbf Y}^T,{\mathbf X}^T,\eta)
\label{antipodalmapampl}
\eeq
\noindent
which is the lift of the \emph{antipodal map}.
Eq.(\ref{Cayleygen}) clearly shows that under the (\ref{antipodalmapampl}) antipodal map  
$\mathcal{D}(P)$ is invariant.

Let us now define the following new quartic polynomial associated to our canonical residue and its antipode
\beq
\mathcal{G}(P)=(\eta\xi)^2-\eta\xi{\rm Tr}({\mathbf X}{\mathbf Y})+\eta{\rm Det}({\mathbf X})+\xi{\rm Det}({\mathbf Y})=\frac{1}{2}(\xi\tilde{\eta}-\eta\tilde{\xi}).
\label{atom}
\eeq
\noindent
One can immediately see that $\mathcal{G}(P)$ is invariant under the (\ref{antipodalmapampl}) antipodal map, and all of the transformations 
$P_{abc}\mapsto [\mathcal{T}_{\alpha\beta}(P)]_{abc}$, $P_{abc}\mapsto [\mathcal{T}_{\overline{\alpha}\overline{\beta}}(P)]_{abc}$ where $1\leq \alpha<\beta\leq 3$.
Indeed, the latter ones are effecting an exchange of either the rows or the columns of the matrices ${\bf X}$ and ${\bf Y}$ together with a
\emph{compensating sign change}. Under the latter two types of transformations the quantities $\eta$ and $\xi$, and the ${\rm Det}{\bf X}$, ${\rm Det}{\bf Y}$, ${\rm Tr}({\bf XY})$ factors are left invariant. 
Now the transformations $\{\mathcal{T}_{12},\mathcal{T}_{23}\}$ and 
$\{\mathcal{T}_{\overline{12}},\mathcal{T}_{\overline{23}}\}$
can be regarded as the generators of two copies of the group $S_3\times S_3$. Combining these transformations with a transposition of the corresponding matrices ${\bf X}$ and ${\bf Y}$ one obtains a representation of the automorphism group of our residue $GQ(2,1)$ which is the wreath product $S_3\wr S_2$. Since $\mathcal{G}(P)$ is left invariant under the automorphism group of $GQ(2,1)$ and the antipodal map, one suspects that this polynomial can be regarded as a seed for generating the polynomial $\mathcal{D}(P)$ invariant under the automorphism group $W(A_5)$ of the full extended geometry, i.e. $EGQ(2,1)$.
Indeed, since $W(A_5)=S_6$ one has $\vert S_6\vert/\vert S_3\wr S_2\vert =6!/2\cdot 3!\cdot 3!=10$ then one should be able to generate $\mathcal{D}(P)$ by acting on $\mathcal{G}(P)$ with suitable representatives of the coset $W(A_5)/S_3\wr S_2$.
These representatives are  precisely the nine unitaries of (\ref{unitaries}).
As a result of these considerations we obtain the following nice result
\beq
\mathcal{D}(P)=\mathcal{G}(P)+\sum_{\alpha,\beta=1}^3\mathcal{G}(\mathcal{T}_{\alpha\overline{\beta}}(P)).
\label{baromiklassz}
\eeq
\noindent
Or, in a more abstract notation
\beq
\mathcal{D}(P)=\sum_{\mathcal{A}\in G/H}\mathcal{G}(\mathcal{T}_{\mathcal{A}}P),\qquad G=W(A_5),\qquad H=S_3\wr S_2,
\label{abstract}
\eeq
\noindent
where, by an abuse of notation,  we referred to $\mathcal{A}\in\{\{0\},\{\alpha,\overline{\beta}\}\}\equiv G/H$.
Here $\mathcal{T}_{\{0\}}$ is the identity operator which represents the $H$-part of the coset. 

This compactified form of Hitchin's invariant clearly shows that it is geometrically underpinned by the smallest $EGQ(2,1)$ that is a one point extension of $GQ(2,1)$, related to Mermin squares.
The new (\ref{baromiklassz}) appearance of Hitchins invariant displays $10$ copies of the simple polynomial $\mathcal{G}(P)$.
Each copy is associated with a residue taken together with its antipodal version.
The antipodal map acts like a covering transformation via taking two copies: the canonical residue and its antipode (for a mathematical discussion on this point, see, e.g. Example 9.7 of Ref.\cite{Cameron}).
At first sight, in our treatise the pair $\eta,{\mathbf X}$ and its antipode $\xi,{\mathbf Y}$ seem to play a special role.
However, since independent of the residue chosen each of the summands in Eq.(\ref{abstract}) is having the same substructure, our new formula (\ref{abstract}) treats all of the $10$ doublets of residues democratically.
This is to be contrasted with the (\ref{Cayleygen}) version of $\mathcal{D}(P)$, where the distinguished role of the $1+9+9+1$ split to a quadruplet $(\eta,{\mathbf X},{\mathbf Y},\xi)$ is manifest. 

The explicit form of $\mathcal{D}(P)$ shows that it has $85$ monomials.
$30$ monomials are directly associated to the blocks of $EGQ(2,1)$. They are signed monomials ($16$ positive and $14$ negative ones) labelled by \emph{different quadruplets} of the form given by Eq.(\ref{subidubi}) and giving rise to terms like $P_{123}P_{156}P_{246}P_{345}$.
These blocks are illustrated by the lines of the twin Steiner-Pl\"ucker configurations of Figure \ref{STPL}.
In the language of Eq.(\ref{Cayleygen}), these monomials are coming from the $12$ terms of $\eta{\rm Det}{\bf X}$ and $\xi{\rm Det}{\bf Y}$ and, partly, from $18$ terms
contained in $-4{\rm Tr}({\bf X}^{\sharp}{\bf Y}^{\sharp})$.
However, in the new (\ref{baromiklassz}) formula each of these blocks appears on the same footing: they are ordinary ones showing up in $4$ different residues.
The remaining structure can be understood from the fact that the residues of $EGQ(2,1)$ are also organized to $10$ antipodal pairs. There are $10$ monomials coming from antipodal pairs with \emph{double} (e.g. $(P_{123}P_{456})^2$) and $45$ monomials from
\emph{single} occurrence (e.g. $(P_{123}P_{456})(P_{156}P_{234})$).

Let us elaborate on the physical meaning of the finite geometric structures found in connection with $\mathcal{D}(P)$.
As it is well-known from the literature, the value of Hitchin's functional at the critical point is related to black hole entropy\cite{FormTH,PW,LevSar}.
The simplest way to see this is to compactify type II string theory on a six-dimensional torus. Depending on whether we use IIA or IIB string theory, one can consider wrapped $D$-brane configurations of an even or odd type.
These configurations give rise to charges of electric and magnetic type in the effective four-dimensional supergravity theory.
In this theory one can consider static, extremal black hole solutions of Reissner-Nordstr\"om type and calculate the semiclassical Bekenstein-Hawking entropy.
For example, in type IIB theory one can consider wrapped $D3$-branes\cite{Moore}. The wrapping configurations then can be reinterpreted either as three qubits\cite{Borsten3q}, or more generally, as three-fermion states\cite{LevSar} related to our three-form $P$, or our observable $\Pi$ of Eq.(\ref{Pas3observ}).  
In this picture the $(\eta,{\bf X})$ , $(\xi, {\bf Y})$ split for the amplitudes $P_{jkl}$ is related to the physical split of charges to \emph{electric} and \emph{magnetic} type.
Our antipodal map of (\ref{antipodalmapampl}) then implements electric-magnetic \emph{duality} and $\mathcal{D}(P)$ is related to the semiclassical extremal black hole entropy as\cite{Moore,BHQC} ($\hbar=c=G_N=k_B=1$)
\beq
S=\pi\sqrt{\vert\mathcal{D}(P)\vert}.
\label{entropy}
\eeq
\noindent
According to whether $\mathcal{D}(P)$ is \emph{negative} or \emph{positive} there are charge configurations of BPS or non-BPS types\cite{BHQC}. Applying T-duality one can relate the $D3$-brane configurations of the type IIB theory to the combined $D0,D2,D4,D6$
-brane configurations of the type IIA one\cite{PW2}.
In this type IIA reinterpretation, after a convenient (STU) truncation\cite{STUunveiled}, the $(\eta,{\bf X},{\bf Y},\xi)$, featuring the canonical residue and its antipode yields $(D0,D4,D2,D6)$ brane charges. Keeping only the $D0,D6$ pairs we obtain just a single \emph{positive} term in the expression for $\mathcal{G}(P)$, namely: $(\eta\xi)^2=(P_{123}P_{456})^2$, hence this charge configuration is a non-BPS one\cite{Gimon,STUunveiled}.
Note the the $D0$ and $D6$ charges are related to each other by electric-magnetic duality, givin a special application
of our antipodal map of Eq.(\ref{unitaries}).
The $\eta{\rm Det}{\mathbf X}$ and $\xi{\rm Det}{\mathbf Y}$ terms of $\mathcal{G}(P)$ implement the well-known $D0D4$ and $D2D6$ systems\cite{Gimon,STUunveiled} which can be both BPS and non-BPS. 
The (\ref{entropy}) entropy formula is invariant under an infinite discrete group of U-duality transformations\cite{Town}.
In our case a special finite subgroup of these transformations is implemented by the Weyl reflections of our weight diagram. Their meaning has been identified as generalized electric-magnetic duality transformations\cite{Obers}. Now, our new formula of Eq.(\ref{baromiklassz}) shows that $\mathcal{D}(P)$ can be regarded as the image of the  special polynomial (\ref{atom}) under a subset of these Weyl reflections.

We also remark that there is a well-known connection\cite{Gaiotto,Pioline,Freudual} between the semiclassical entropy of 4D BPS black holes in type IIA theory compactified on a Calabi-Yau space $M$ and the entropy of spinning 5D BPS black holes in M-theory compactified on $M\times TN_{\eta}$, where
$TN_{\eta}$ is a Euclidean 4-dimensional Taub-NUT space with the NUT charge $\eta$.
If the 4D charges are represented by the arrangement $P=(\eta,{\mathbf X},{\mathbf Y},\xi)$, then there is a simple relationship between these quantities and the 5D black hole charge and spin (angular momenta) $\mathcal{J}_{\eta}$. It turns out that the latter quantity is related to $\tilde{\eta}$ by the simple formula
\beq
{J}_{\eta}=-\frac{\tilde{\eta}}{2}.
\label{angularmom}
\eeq
\noindent
There is also a dual connection between 4D black holes and 5D black \emph{strings}. In this case there is a relationship
between  the arrangement $P=(\eta,{\mathbf X},{\mathbf Y},\xi)$ and the 5D \emph{magnetic charges}. Moreover, in this case the corresponding angular momentum is related to the 4D quantities as
\beq
{J}_{\xi}=-\frac{\tilde{\xi}}{2}.
\eeq
\noindent
Amusingly in this 5D-lift the NUT-charges ($\eta,\xi$) and the corresponding angular momenta ($\mathcal{J}_{\eta},\mathcal{J}_{\xi}$), regarded as dual pairs, are related to our polynomial $\mathcal{G}(P)$ of finite geometric meaning as
\beq
\mathcal{G}(P)=\eta {J}_{\xi}-\xi {J}_{\eta}.
\eeq
\noindent 
Combining this formula with the new (\ref{abstract}) expression of Hitchins invariant
connects information concerning the canonical residue, Mermin squares and physical parameters characterizing certain black hole solutions in a striking way. The physical consequences of this interesting result should be explored further.

\subsection{An Extended Generalized Quadrangle {\bf EGQ(2,2)} and the Generalized Hitchin Functional}

Let us now consider the Schl\"afli double-six part taken together with the $EGQ(2,1)$ (twin Steiner-Pl\"ucker configuration) part known from the previous section.
The former is described by $12$ operators of the form $\Gamma_a,\Gamma_a\Gamma_7$ (the blue triangle of Figure \ref{gods-eye-clifford}) and the latter by $20$ ones of the form $\Gamma_a\Gamma_b\Gamma_c$ with $1\leq a<b<c\leq 6$ (green triangle of Figure \ref{gods-eye-clifford}).

It is easy to show that these two sets, taken together, describe the weights of the $32$ dimensional spinor representation of $Spin(12)$ with negative chirality.
Indeed, by virtue of $\Gamma_a\Gamma_7\simeq \Gamma_b\Gamma_c\Gamma_d\Gamma_e\Gamma_f$, with $1\leq b<c<d<e<f\leq 6$ and $a\neq \{b,c,d,e,f\}$, in the fermionic Fock space description of this representation\cite{SarLevFerm,LevHolw} this irreducible spinor representation is spanned by forms of an odd degree. We have a one-form with six ($v_a$), a three-form with twenty ($P_{abc}$) and a five-form converted to a vector ($w^a\simeq \varepsilon^{abcdef}w_{bcdef}$) with six components.

\begin{figure}[pth!]
\centerline{\includegraphics[width=5truecm,clip=]{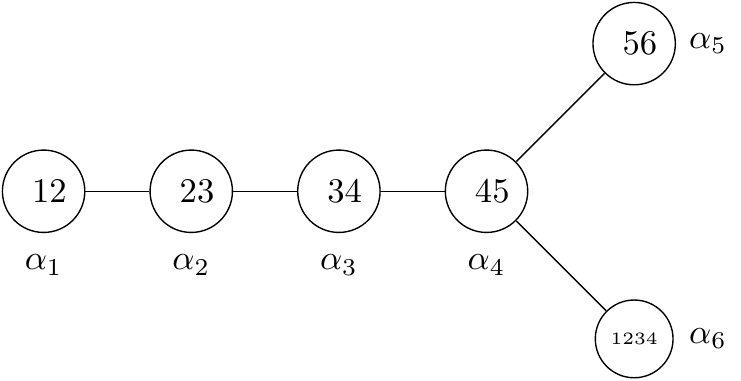}}
\caption{The $D_6$ Dynkin diagram with our labelling convention shown.}
\label{Dynk32}
\end{figure}

\begin{figure}[pth!]
\centerline{\includegraphics[width=11truecm,clip=]{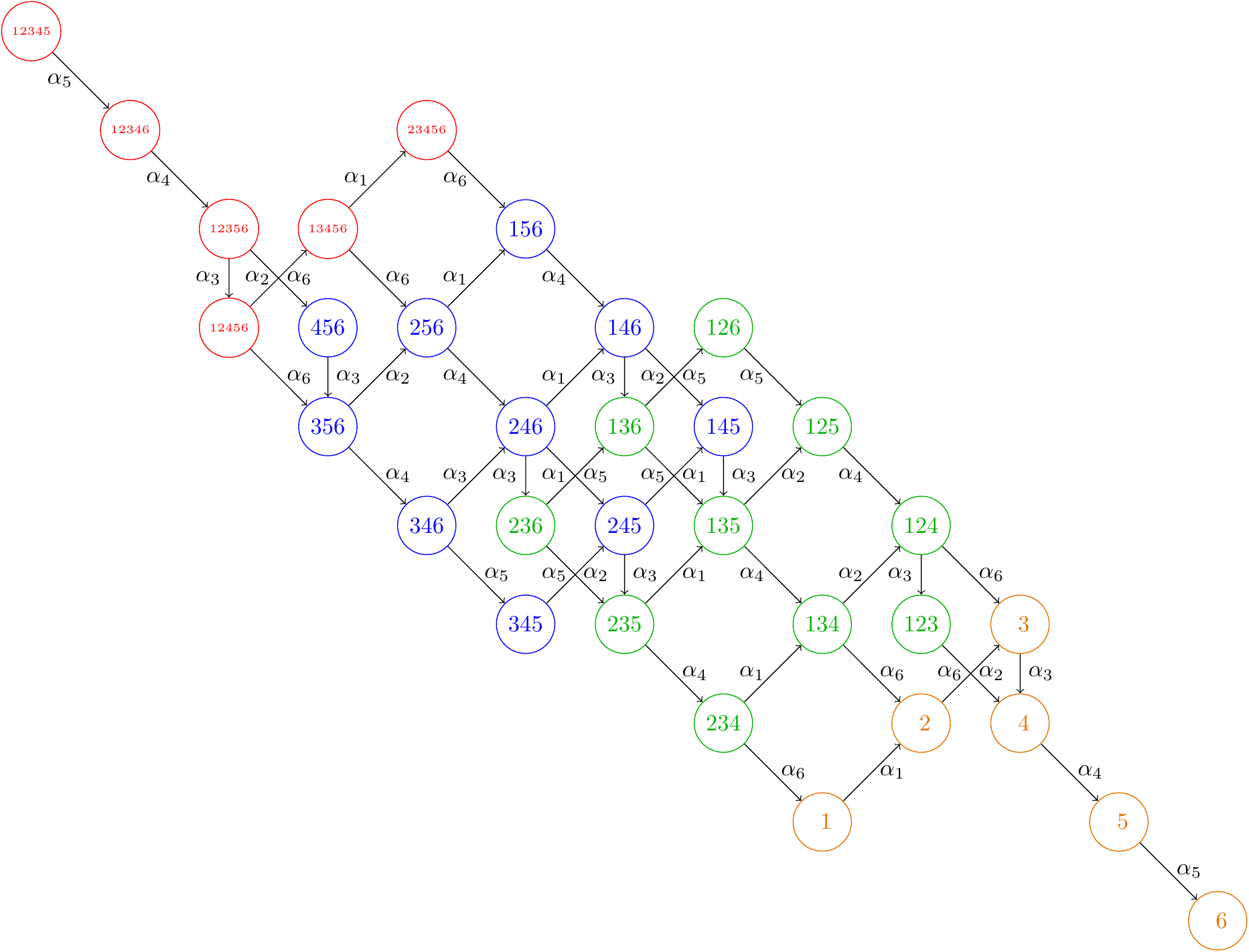}}
\caption{The weight diagram for the $32$ of $D_6$ labelled by $1$, $3$ and $5$-element subsets of $S$.}
\label{W32}
\end{figure}

\begin{figure}[pth!]
\centerline{\includegraphics[width=5truecm,clip=]{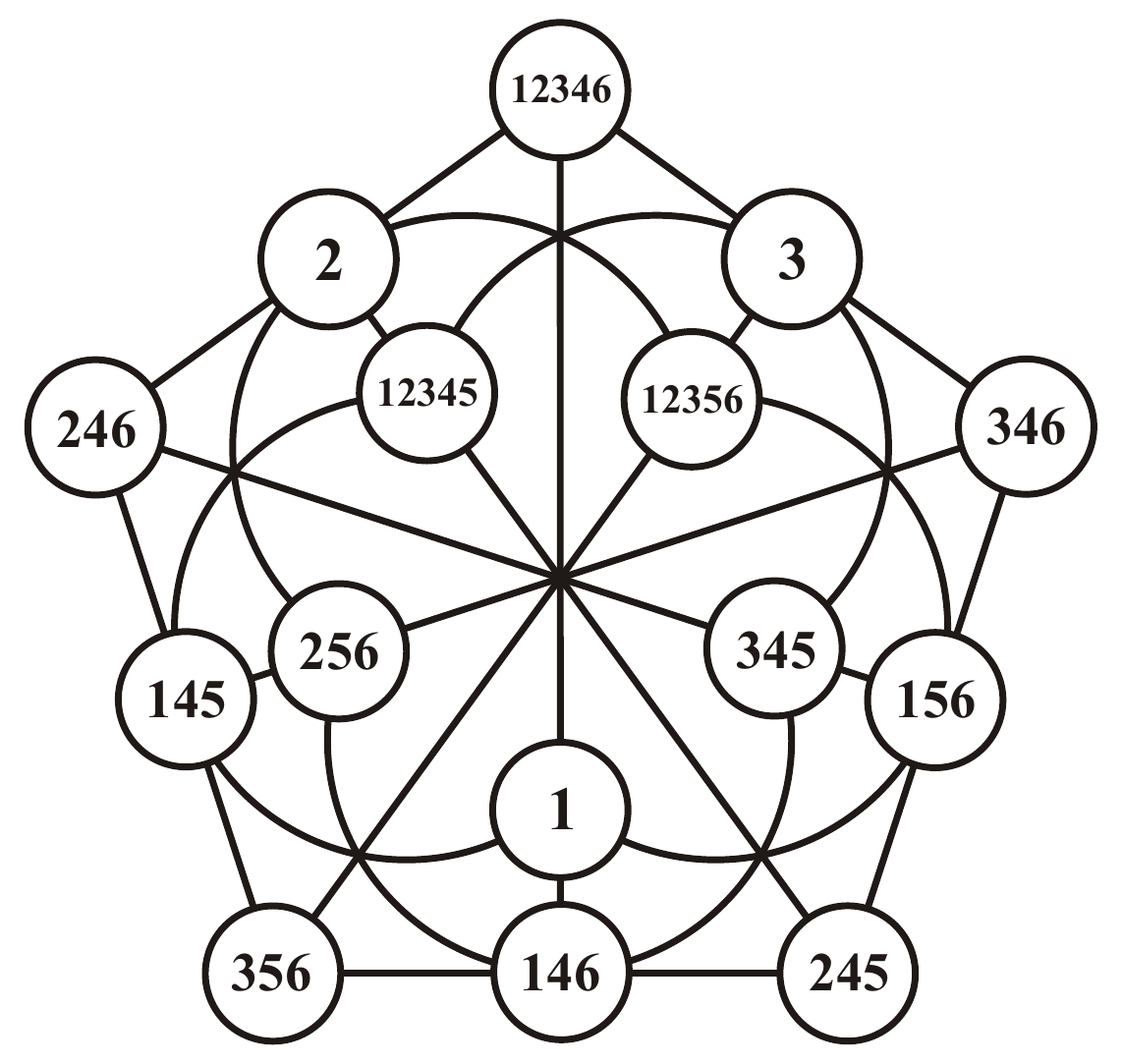}}
\caption{The doily corresponding to the residue $\mathcal{T}_{123}$.}
\label{PointedDoily}
\end{figure}

In order to construct the $32$ weights we label the $D_6$-Dynkin diagram as shown in Figure \ref{Dynk32}. Five nodes and their labels from the Dynkin diagram of $D_6$ coincide with the $A_5$-diagram and the extra node is labelled as: $\alpha_6\leftrightarrow 1234$. 
Then the Dynkin labels of the representation are $(000010)$. Using the explicit form of the Cartan matrix and its inverse and the explicit form\cite{Slansky}
\beq
\alpha_1=e_1-e_2,\quad
\alpha_2=e_2-e_3,\quad\alpha_3=e_3-e_4,\quad\alpha_1=e_4-e_5,\quad\alpha_5=e_5-e_6,\quad\alpha_6=e_5+e_6,
\nonumber
\eeq
\noindent
one obtains the weights
\beq
\Lambda^{(a)}=e_a-\frac{1}{2}n,\qquad
\Lambda^{(bcdef)}=\frac{1}{2}n-e_a\qquad
\Lambda^{(bcd)}=\frac{1}{2}n-e_b-e_c-e_d,
\nonumber
	\eeq
\noindent
where $n=(1,1,1,1,1,1)^T$,
$1\leq b<c<d<e<f\leq 6$ and $a\neq \{b,c,d,e,f\}$.
The weight diagram for the $32$ of $D_6$ takes the form as shown in Figure \ref{W32}.
We can split our $32$-element set of labels of these weights into two $16$-element ones as follows
\begin{eqnarray} \{1,2,3,12345,12356,12346,123,156,146,145,256,246,245,356,346,345\},\\
 \{4,5,6,12456,23456,13456,456,234,134,124,235,135,125,236,136,126\}.
\label{halmazka}
\end{eqnarray}
\noindent 
These combinations regarded as elements of $\mathcal{P}(S)$ will be denoted by $\mathcal{T}$.
Weights belonging to the two different $16$-element sets, with their corresponding labels satisfying $\vert\mathcal{A}\cap\mathcal{B}\vert =0$ and $\mathcal{A}\cup\mathcal{B}=S$ will be called \emph{antipodal}. Again, for such pairs we have $\cos\theta_{{\mathcal A}{\mathcal B}}=-1$.
As in the previous section  we consider four \emph{different} weights, called \emph{quadruplets}.  Quadruplets of subsets $\mathcal{A}_s$, $s=1,2,3,4$, taken from $\mathcal{T}$  will be called \emph{blocks} if they satisfy(\ref{blockrule}).
For an example of a block again Eq.(\ref{subidubi}) can be used. However, now we have blocks of a new type. For example,
apart from the $6$ blocks through $123$ we are familiar with from Eq.(\ref{kanvan}), one has $9$ extra blocks of the form 
\begin{eqnarray}
(123,145,1,12345),\qquad (123,146,1,12346),\qquad (123,156,1,12356),\\\nonumber
(123,256,2,12356),\qquad (123,245,2,12345),\qquad (123,246,2,12346),\\\nonumber
(123,356,3,12356),\qquad (123,345,3,12345),\qquad (123,346,3,12346)\nonumber.
\end{eqnarray}
\noindent
Taking the $15$ points collinear with $123$ and giving them the block structure via the $15$ blocks discussed above one obtains the \emph{residue} of $123$, namely $\mathcal{T}_{123}$. 
For any $\mathcal{A}\in\mathcal{T}$ one can define a residue $\mathcal{T}_{\mathcal{A}}$. 
Clearly, each residue can be given the incidence structure of a doily, i.e. a $GQ(2,2)$.
As an example, we show this incidence structure for $\mathcal{T}_{123}$ in Figure \ref{PointedDoily}. 

Each of these residues is containing $15$ blocks. One can show that altogether one has $(32\times 15)/4=120$ blocks.   
One can then check, that $\mathcal{T}$ containing $32$ points and equipped with the block structure as described above, gives rise to the structure of an extended generalized quadrangle of type $EGQ(2,2)$.
One can verify that the point graph of this structure is distance regular and of diameter $3$. It is known that an $EGQ(2,2)$ with these properties is unique. It is one of the seven affine polar spaces referred to in the literature as type-$A_2$\cite{Cameron,Fra}. 
Recalling our results from the previous section we can record: the weights of the $20$ of $A_5$ and the ones of the $32$
of $D_6$ with the block structure defined by (\ref{blockrule}) give rise to extended generalized quadrangles $EGQ(2,t)$
with $t=1,2$. Both of them are of diameter $3$ and distance regular. 
The grids regarded as residues of the $EGQ(2,1)$ are contained inside the doilies regarded as residues of the $EGQ(2,2)$s.
This connection between the point sets of the corresponding geometries is related to the embedding of the weights of $20$ of $A_5$ inside the weights of the $32$
of $D_6$. Indeed, according to Figure \ref{W32}, cutting the weight diagram along $\alpha_6$ one obtains the weight diagram of the $20$ of $A_5$.  

Let us now connect the $EGQ(2,2)$ structure we have found to the structure of the Generalized Hitchin Functional (GHF).
The GHF for a six-dimensional, closed orientable manifold $M$ is defined by replacing the three-form $P$
in the usual formulation of the Hitchin functional by a polyform of odd or even degree\cite{Hitchin3}. 
To an odd degree form
\beq
\varphi=u_adx^a+
\frac{1}{3!}P_{abc}dx^a\wedge dx^b\wedge dx^c+
\frac{1}{5!}v^a\varepsilon_{abcdef}dx^b\wedge dx^c\wedge dx^d\wedge dx^e\wedge dx^f
\eeq
\noindent
 one can associate a three-qubit operator $\Phi$ of the form
\beq
\Phi=u_a\Gamma_a+\frac{1}{3!}P_{abc}\Gamma_a\Gamma_b\Gamma_c+\frac{1}{5!}v^a\varepsilon_{abcdef}\Gamma_b\Gamma_c\Gamma_d\Gamma_e\Gamma_f.
\label{Fifi}
\eeq
\noindent
Here we dualized the five form part to a vector with components $v^a$.

Then our split of the $32$-element set of observables, labelled as in (\ref{halmazka}), gives rise to a split of the set of real-valued functions 
$(u_i,P_{ijk},v^j)$ on $M$ into two sets $(\eta, x)$ and $(\xi,y)$ of cardinalities $16$ each as follows\cite{LevHolw}
\beq
x^{ab}=\begin{pmatrix}0&-u_3&u_2&-P_{156}&
P_{146}&-P_{145}
\\u_3&0&-u_1&-P_{256}&P_{246}&
-P_{245}\\
-u_2&u_1&0&-P_{356}&P_{346}&
-P_{345}\\
P_{156}&P_{256}&
P_{356}&0&-v^{6}&v^{5}\\
-P_{146}&-P_{246}&
-P_{346}&v^{6}&0&-v^{4}\\
P_{145}&P_{245}&P_{345}&-v^{5}&v^{4}&0
\label{ixtilde}
\end{pmatrix},\qquad\eta=P_{123},
\eeq
\noindent
\beq
y_{ab}=\begin{pmatrix}0&-v^3&v^2&-P_{234}&
-P_{235}&-P_{236}
\\v^3&0&-v^1&P_{134}&
P_{135}&P_{136}\\
-v^2&v^1&0&-P_{124}&-P_{125}&
-P_{126}\\
P_{234}&-P_{134}&
P_{124}&0&-u_{6}&u_{5}\\
P_{235}&-P_{135}&
P_{125}&u_{6}&0&-u_{4}\\
P_{236}&-P_{136}&P_{126}&-u_{5}&u_{4}&0
\label{iytilde}
\end{pmatrix},
\qquad\xi=P_{456}.
\eeq
\noindent
Hence we have two scalars $\xi,\eta$ and two $6\times 6$ antisymmetric matrices $x^{ab},y_{ab}$ yielding the new split: $32=1+15+15+1$.

Let us define the following $12\times 12$ matrix
\beq
{\mathcal{K}^I}_J=2\begin{pmatrix}\kappa{\delta^a}_b-{(xy)^a}_b& (\eta x-\tilde{y})^{ad}\\
(\xi y-\tilde{x})_{cb}&-\kappa{\delta^c}_d+{(xy)^c}_d\end{pmatrix},
\label{genalmostcomplex}
\eeq
\noindent
where
\beq
2\kappa=\eta\xi-\sum_{a<b}x^{ab}y_{ab},\qquad \tilde{x}_{ab}=\frac{1}{8}\varepsilon_{abcdef}x^{cd}x^{ef},\qquad
\tilde{y}^{ab}=\frac{1}{8}\varepsilon^{abcdef}y_{cd}y_{ef}.
\eeq
\noindent
With these quantities we define the generalized Hitchin invariant\cite{Hitchin3,LevSar} as
\beq
\mathcal{C}(\varphi)=\frac{1}{12}{\rm Tr}(\mathcal{K}^2)=4[\kappa^2-\sum_{a<b}\tilde{x}_{ab}\tilde{y}^{ab}+\eta{\rm Pf}(x)+\xi{\rm Pf}(y)],
\label{GenHInv}
\eeq
\noindent
where $-6{\rm Pf}(x)={\rm Tr}(\tilde{x}x)$, see also Eq.(\ref{Pfaff}).
Now the Generalized Hitchin Functional is given by the formula
\beq
V_{GH}[\varphi]=\int_{M}\sqrt{\vert\mathcal{C}(\varphi)\vert}d^6x.
\label{genhitchfunc}
\eeq
\noindent
The generalized Hitchin functional is designed to produce \emph{generalized complex structures} on $M$.
Such a mathematical object, in some sense, combines the complex and K\"ahler structures of $M$ in an inherent way.
These structures are of utmost importance in string theory for Calabi-Yau three-folds. Their combination to a generalized complex structure gives rise to the important notion of generalized Calabi-Yau manifolds\cite{Hitchin3}.
For a polyform $\varphi$ with $\mathcal{C}(\varphi)<0$ the quantity $\mathcal{K}/\sqrt{-\mathcal{C}(\varphi)}$ defined using Eq.(\ref{genalmostcomplex}) gives rise to a generalized almost complex structure. Then it can be shown that critical points of (\ref{genhitchfunc})
give rise to \emph{integrable} generalized complex structures.

Note that for the special choice of $u_a=v^a=0$ we have
\beq
x=\begin{pmatrix}0&-{\bf X}\\{\bf X}^T&0\end{pmatrix},\qquad
y=\begin{pmatrix}0&-{\bf Y}^T\\{\bf Y}&0\end{pmatrix},
\eeq
\noindent 
where ${\bf X}$ and ${\bf Y}$ are given by Eq.(\ref{Xmatr})-(\ref{Ymatr}).
One can then check that in this special case the expression of $\mathcal{C}(\varphi)$ boils down to the (\ref{Cayleygen}) expression of $\mathcal{D}(P)$. In this way the generalized Hitchin functional boils down to the usual Hichin functional of Eq.(\ref{Hitchinfunk}).

Let us now define $\Sigma\subset\mathcal{P}(S)$ as
\beq
\Sigma=\{\{0\},\{mn\},\{1234\},\{1235\},\{1236\},\{1456\},\{2456\},\{3456\}\},
\eeq
\noindent
where  $m=1,2,3$, $n=4,5,6$ and $\{0\}$ is the empty set containing no elements.
Consider now the transvections $T_{\mathcal{A}}$, where $\mathcal{A}\in \Sigma$ and $T_{\{0\}}$ is the identity.
Notice that associating to the $16$ labels of $\Sigma$ observables
the set $\Sigma$ by itself can also be regarded as a "pointed doily", i.e. a residue.

Let us now define the quartic polynomial
\beq
\mathcal{E}(\varphi)=(\eta\xi)^2-\eta\xi\sum_{a<b}x^{ab}y_{ab}+\eta{\rm Pf}(x)+\xi{\rm Pf}(y).
\label{atomgenhitch}
\eeq
\noindent 
Then one can show that
\beq
\mathcal{C}(\varphi)=\sum_{\mathcal{A}\in\Sigma}\mathcal{E}(\mathcal{T}_{\mathcal{A}}\varphi)=\sum_{\mathcal{A}\in G/H}\mathcal{E}(\mathcal{T}_{\mathcal{A}}P)\qquad G=W(D_6)/\mathbb{Z}_2,\quad H=W(A_5),
\label{errorcorr}
\eeq
\noindent
where we have used that $W(D_6)/\mathbb{Z}_2\simeq 2^4\cdot Sp(6,2)$ and  $W(A_5)\simeq Sp(6,2)\simeq S_6$.
This result can be regarded as a generalization of the one encapsulated in Eq.(\ref{abstract}).
Here, similar to Eq.(\ref{Phatas}), the $\mathcal{U}(T_{\mathcal{A}})$ lifts of the transvections define an action $\mathcal{T}_{\mathcal{A}}$
on the $32$ functions $(u_a, P_{abc},v^a)$ with $1\leq a<b<c\leq 6$.
Then our new formula of Eq.(\ref{errorcorr}) clearly shows that $\mathcal{C}(\varphi)$ can be written as an average of a polynomial based on a single residue ($\mathcal{E}(\varphi)$)
over a residue ($\Sigma$) and, geometrically thus corresponds to a unique one-point extension of $GQ(2,2)$ \cite{Cameron}.
Our result demonstrates how the $EGQ(2,2)$ structure manifests itself in building up the generalized Hitchin invariant giving rise to the (\ref{genhitchfunc}) functional of physical importance.

Let us elaborate on the action of $\mathcal{T}_{\mathcal{A}}$, with $\mathcal{A}\in\Sigma$, on the polyform $\varphi$. First, in addition to $\mathcal{U}(T_{m\overline{n}})$ of Eq.(\ref{unitaries}), we define
\beq
\mathcal{U}(T_{abcd})=\frac{1}{\sqrt{2}}(I_8+\Gamma_a\Gamma_b\Gamma_c\Gamma_d),\qquad a<b<c<d,\quad \{abcd\}\in\Sigma,
\eeq
\noindent
then the action on (\ref{Fifi}) 
\beq
\Phi^{\prime}=\mathcal{U}^{\dagger}(T_{\mathcal{A}})\Phi\mathcal{U}(T_{\mathcal{A}}), \qquad \mathcal{A}\in \Sigma,
\eeq
\noindent
defines a set of transformations
\beq
\mathcal{T}_{\mathcal{A}}:(u_a,P_{abc},v^a)\mapsto (u_a^{\prime},P_{abc}^{\prime},{v^a}^{\prime})
\eeq
\noindent
or, alternatively, a set of $\mathcal{T}_{\mathcal{A}}:(\eta,x,y,\xi)\mapsto (\eta^{\prime},x^{\prime},y^{\prime},\xi^{\prime})$ where $\mathcal{A}\in\Sigma$.
This fixes the explicit form of the action on $\varphi$.

Just like in the previous section one can easily relate these considerations to structural issues concerning 4D semiclassical black hole entropy formulas.
The simplest way to uncover these connections is in the type IIA duality frame.
When compactifying type IIA supergravity on the six-torus $T^6$, one is left with a classical 4D theory with on-shell $E_{7(7)}$ duality symmetry\cite{Julia,CJ79}.
There are $U(1)$ charges associated with the Abelian gauge fields of this theory. They are transforming according to the $56$-dimensional representation of the $E_{7(7)}$ duality symmetry. 
There are also scalar fields (moduli) in the theory which are parametrizing the $70$-dimensional coset
$E_{7(7)}/SU(8)$. The $56$ charges can be represented in terms of the central charge matrix $\mathcal{Z}_{AB}$ of the $N=8$ supersymmetry algebra. This is an $8\times 8$ complex antisymmetric matrix. Partitioning this matrix into four $4\times 4$ blocks, the block-diagonal part gives rise to $12$ complex components which can be organized into the $24$ real NS-charges.
The remaining $16$ independent complex components are coming from one of the offdiagonal $4\times 4$ blocks.
They are comprising the $32$ real RR-charges.

Let us concentrate merely on this RR-sector.
When one writes the central charge matrix in an $SO(8)$ basis 
one has the form\cite{Vijay}
\beq
\frac{1}{\sqrt{2}}(x^{MN}+iy_{MN})=-\frac{1}{4}\mathcal{Z}_{AB}(\Gamma_{MN})^{AB},\qquad M,N,A,B=1,2,\dots 8,
\eeq
\noindent
In the case of the RR-truncation
the $8\times 8$ matrices $x^{MN}$ and $y_{MN}$ take the following form\cite{Pioline2}
\beq
x^{MN}=\begin{pmatrix}[D2]^{ab}&0&0\\
0&0&[D6]\\0&-[D6]&0\end{pmatrix},\qquad
y_{MN}=\begin{pmatrix}[D4]_{ab}&0&0\\
0&0&[D0]\\0&-[D0]&0\end{pmatrix}.
\label{Dbranecharges}
\eeq
\noindent
Here the quantities $[D0],[D2]^{ab},[D4]_{ab},[D6]$ are the $D$-brane charges. They are arising from wrapping configurations on cycles of $T^6$ of suitable dimensionality.
Now the unique quartic $E_{7(7)}$-invariant\cite{CJ79,KK96} is of the form
\beq
J(x,y)=-{\rm Tr}(xyxy)+\frac{1}{4}({\rm Tr}xy)^2-4[\rm{Pf}(x)+\rm{Pf}(y)],
\label{quarticCartan}
\eeq
\noindent
where
\beq
{\rm Pf}(x)=\frac{1}{2^44!}\varepsilon_{MNPQRSTU}x^{MN}x^{PQ}x^{RS}x^{TU}.
\label{buvos}
\eeq
\noindent
By virtue of (\ref{Dbranecharges}) a truncation of the quartic invariant to the RR-sector takes the form
\begin{eqnarray} J_{RR}=4[D6]{\rm Pf}([D2])+4[D0]{\rm Pf}([D4])-{\rm Tr}([D2][D4][D2][D4])-\\\nonumber
([D0][D6]-\frac{1}{2}{\rm Tr}[D2][D4])^2-2([D0][D6])^2.
\end{eqnarray} 
After the identifications 
\beq
\eta=-[D6],\qquad x^{ab}=[D2]^{ab},\qquad y_{ab}=[D4]_{ab},\qquad \xi=-[D0],
\label{kapcsolatHitch}
\eeq
\noindent
and
using the identity
\beq
4{\rm Tr}(\tilde{x}\tilde{y})=2{\rm Tr}(xyxy)-[{\rm Tr}(xy)]^2
\eeq
\noindent
the expression of $J_{RR}$ boils down to the \emph{negative} of the generalized Hitchin invariant of Eq.(\ref{GenHInv}),
i.e. $J_{RR}=-\mathcal{C}(\varphi)$.
In this special case the semiclassical black hole entropy formula takes the form
\beq
S=\pi\sqrt{\vert J_{RR}\vert}.
\eeq
\noindent
For more details on the connection between the critical points of the generalized Hitchin functional and black hole entropy we orient the reader to the paper of Pestun\cite{PW2}.
Interestingly, this entropy structure inherently connected to the generalized Hitchin functional has an alternative interpretation in terms of the $32$ amplitudes of $4$ real, unnormalized  three-qubit states built up from
six qubits\cite{BHQC,LevSar}.
This structure is coming from the "tripartite entanglement of seven qubits" interpretation of the (\ref{quarticCartan}) quartic invariant of Refs.\cite{DuffFano,LevFano} after truncating to the RR-sector.

\subsection{The Generalized Quadrangle GQ(2,4) and Cartan's cubic invariant}

Now we consider the the elliptic quadric part of our Veldkamp line, which corresponds to the blue parallelogram of
Figure \ref{gods-eye-composition}.
A detailed discussion of the finite geometric background, and its intimate link to the structure of the $5D$ semiclassical black hole entropy formula, of this case
can be found in Ref.\cite{Levfin2}. In this section we reformulate the results of that paper in a manner that  helps to elucidate the connections to the structure of our magic Veldkamp line.

In this case we have $27=6+6+15$ operators corresponding to the subsets $\{a\},\{a7\},\{ab7\}$.
The finite geometric interpretation of the $\{a\},\{a7\}$ part, depicted by the blue triangle of
Figure \ref{gods-eye-clifford}, corresponds to Schl\"afli's double-six configuration, and the black triangle represents our core configuration: the doily.
As it is known from Ref.\cite{Levfin2}, the operators $\Gamma^{({\mathcal{A}})}$ corresponding via (\ref{bijmap}) to the subsets $\mathcal{A}\in\{\{a\},\{a7\},\{ab7\}\}$ provide a noncommutative labelling for the generalized quadrangle $GQ(2,4)$. 
For an explicit labelling in terms of three-qubit operators see Figure 3 of \cite{Levfin2}.

\begin{figure}[pth!]
\centerline{\includegraphics[width=4truecm,clip=]{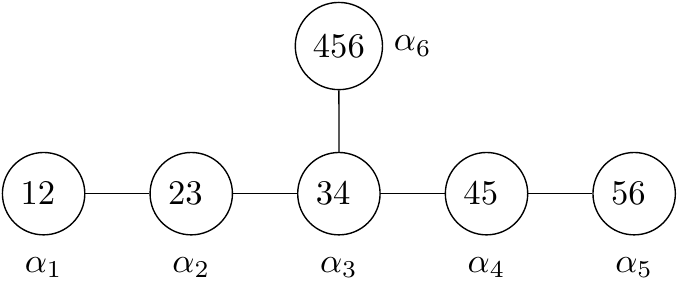}}
\caption{The $E_6$ Dynkin diagram labelled by subsets.}
\label{E6Dynk}
\end{figure}

One can elaborate on the representation theoretic meaning of the $GQ(2,4)$ structure as follows.
The $27$ points of $GQ(2,4)$ can be mapped to the $27$ weights of the fundamental irrep of $E_6$.
In order to see this one labels the nodes of the $E_6$-Dynkin diagram as shown in Figure \ref{E6Dynk}. 

\begin{figure}[pth!]
\centerline{\includegraphics[width=4truecm,clip=]{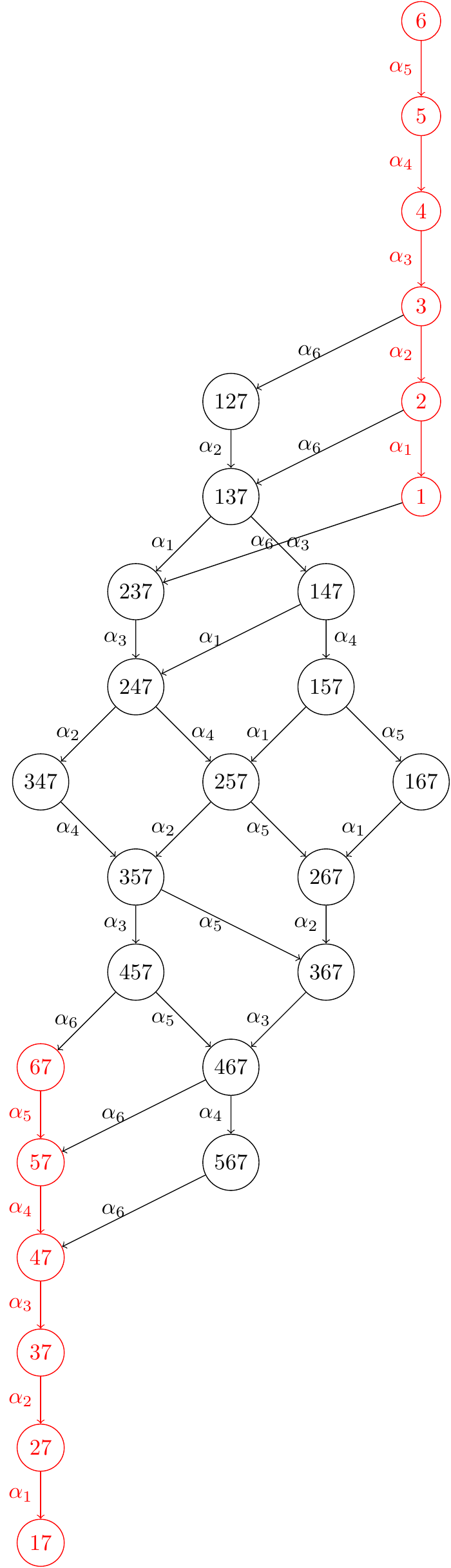}}
\caption{The weight diagram of the $27$ of $E_6$ labelled by the subsets $\{a\},\{a7\},$ and $\{ab7\}$.}
\label{E6Weight}
\end{figure}

\noindent
With this labelling convention the $E_6$ weight diagram takes the form as shown in Figure \ref{E6Weight}.

Note that the labelling of the weights is in accord with the usual labelling of exceptional vectors discussed in connection with $E_N$-lattices for $N=6$. 
In particular, the $27$ weights can be mapped to the seven component exceptional vectors $\{\Lambda^{(a)},\Lambda^{(a7)},\Lambda^{(ab7)}\}\in \mathbb{R}^{6,1}$. $\mathbb{R}^{6,1}$ is spanned by the canonical basis vectors $e_{\mu}$ with $\mu=0,1,\dots 6$ and it is equipped with a nondegenerate symmetric bilinear form with signature $(-1,1,1,1,1,1,1)$.
Explicitly, we have
\beq
\Lambda^{(a)}=e_a,\qquad \Lambda^{(a7)}=2e_0-e_1-\dots -e_6+e_a,\qquad \Lambda^{(ab7)}=a_0-e_a-e_b.
\eeq
\noindent
As it is well-known exceptional vectors are the ones that satisfy the constraints $k_N\cdot \Lambda=1$ and $\Lambda\cdot\Lambda =1$, where $k_N=-3e_0+\sum_{a=1}^Ne_a$. Our special choice conforms with the $N=6$ case.

Notice that due to the fact that the doily is embedded into $GQ(2,4)$ the weight diagram of the $27$ of $E_6$ contains the weight diagram of the $15$ of $A_5$ we are already familiar with from Figure \ref{15A5}.
This corresponds to the reduction
\beq
E_{6(6)}\supset SL(2)\times SL(6),\qquad {\bf 27}\to ({\bf 1},{\bf 15})+({\bf 2},{\bf 6}).
\eeq
\noindent

There is a famous $E_6$-invariant associated with the $GQ(2,4)$ structure. 
It is Cartan's cubic invariant\cite{Cartan}. As is well-known this invariant is connected to the geometry of smooth cubic surfaces in $\mathbb{C}P^3$. It is a classical result that the automorphism group of configurations of $27$ lines\cite{Manivel} on a cubic can be identified with $W(E_6)$, i.e. the Weyl group of $E_6$ of order $51840$. $W(E_6)$ is also the automorphism group of $GQ(2,4)$.
For a nice reference on the connection between cubic forms and generalized quadrangles we orient the reader to the paper of Faulkner\cite{Faulkner}.
In order to relate Cartan's invariant to our Veldkamp line we proceed as follows.

Let us define the observable
\begin{eqnarray}
\label{Psi}
\Psi=
u_a\Gamma_a+\frac{1}{2!4!}\omega_{ab}\varepsilon_{abcdef}\Gamma_c\Gamma_d\Gamma_e\Gamma_f+\frac{1}{5!}v^a\varepsilon_{abcdef}
\Gamma_b\Gamma_c\Gamma_d\Gamma_e\Gamma_f=\\\nonumber
(u_1\Gamma_1+\dots)+(\omega_{12}\Gamma_3\Gamma_4\Gamma_5\Gamma_6+\dots)+(
v^1\Gamma_2\Gamma_3\Gamma_4\Gamma_5\Gamma_6+\dots),
\end{eqnarray}
\noindent
where the real quantities $\omega_{ab},u_a,v^a$ are the ones already familiar from Eqs.(\ref{OmegaMermin}) and (\ref{Fifi}).
Here we also converted the $\{ab7\}$ and $\{a7\}$ index combinations
using the identities
\beq
\Gamma_{[a}\Gamma_{b]}\Gamma_7=\frac{i}{4!}\varepsilon_{abcdef}\Gamma_c\Gamma_d\Gamma_e\Gamma_f,\qquad
\Gamma_a\Gamma_7=\frac{i}{5!}\varepsilon_{abcdef}\Gamma_b\Gamma_c\Gamma_d\Gamma_e\Gamma_f.
\eeq
\noindent
Notice also that $\Psi$ is Hermitian.
Let us define $q\equiv(u_a,\omega_{ab},v^a)$, then Cartan's invariant is 
\beq
\mathcal{I}(q)=\frac{1}{48}{\rm Tr}(\Psi^3)={\rm Pf}(\omega)+v^T\omega u.
\label{Cartaninv}
\eeq
\noindent
The first term on the right-hand side contains $15$ cubic monomials corresponding to the $15$ lines of the doily, and the second term contains $15+15=30$ extra monomials. Hence, altogether we have $45$ cubic monomials in this invariant corresponding to the $45$ lines of $GQ(2,4)$.

In the physical interpretation the $27$ components  $q\equiv(u_a,\omega_{ab},v^a)$, corresponding to the points of our $GQ(2,4)$, describe electrical charges 
of black holes, or magnetic charges of black strings of the $N=2$, $D=5$ magic supergravities\cite{Gun1,Gun2}.
These configurations are related to the structures of cubic Jordan algebras represented by $3\times 3$ matrices over the division algebras (real and complex numbers, quaternions and octonions) or their split versions.
The (\ref{Cartaninv}) Cartan's invariant is then related to the cubic norm of a cubic Jordan algebra over the split octonions\cite{Levfin2}.
The corresponding magic supergravity theory has classically an $E_{6(6)}$ symmetry. In the quantum theory the black hole/string charges become integer-valued. Hence, in this case the classical symmetry group is  broken down to the U-duality group $E_{6(6)}({\mathbb Z})$. The Weyl group $W(E_6)$ can be regarded as a finite subgroup of the infinite U-duality group which is just the automorphism group of $GQ(2,4)$.

Cartan's invariant can also be given an interpretation in terms of the bipartite entanglement of three-qutrits\cite{Duff,Octonions}.
In this approach Cartan's invariant can be regarded as an entanglement measure encoding the charge configurations of the black hole solution in a triple of three qutrit states. Then semiclassical black hole entropy is related to this entanglement measure as
\beq
S=\pi\sqrt{\mathcal{I}(q)}.
\eeq
\noindent
Interestingly, in this qutrit approach the $27$ charges can be organized into three groups containing $9$ charges each.
The group theoretical reason for this rests on the decomposition\cite{Octonions}
\beq
E_{6(6)}\supset SL(3,\mathbb{R})_A\times SL(3,\mathbb{R})_B\times SL(3,\mathbb{R})_C,\qquad   {\bf 27}\to ({\bf 3}^{\prime},{\bf 3},{\bf 1})+({\bf 1},{\bf 3}^{\prime},{\bf 3}^{\prime})+({\bf 3}^{\prime},{\bf 1},{\bf 3}).
\eeq
\noindent
In our $GQ(2,4)$ picture this decomposition amounts to regarding $GQ(2,4)$ as a composite of three $GQ(2,1)$s, i.e. grids.
Since grids labelled by observables are just Mermin squares this gives rise to an alternative interpretation\cite{Levfin2} of 
describing the structure of Cartan's invariant as a special composite of three Mermin squares.
It is also known that there exists $40$ different ways for dissecting $GQ(2,4)$ into triples of Mermin squares, hence altogether there are $120$ possible Mermin squares lurking\cite{Levfin2} inside a particularly labelled $GQ(2,4)$.

A particular decomposition of $GQ(2,4)$ (with its points labelled by observables) to three Mermin squares can be given as follows.
Let us decompose our $6\times 6$ matrix $\omega$ and the two six component vectors $u,v$ into $3\times 3$ matrices and to a set of three-component vectors as follows
\beq
\omega=\begin{pmatrix}L_{\bf b}&-A^T\\A&L_{\bf c}\end{pmatrix},\qquad v^T=({\bf w}^T,{\bf r}^T),\qquad u^T=({\bf s}^T,{\bf z}^T),
\eeq
\noindent
where $L_{\bf b}({\bf w})={\bf b}\times{\bf w}$, i.e. the linear operator $L_{\bf b}$ implements the cross product on three-component vectors.
Using the three-component column vectors ${\bf b},{\bf c},{\bf w},{\bf z},{\bf r},{\bf s}$ one can form two  extra $3\times 3$ matrices
\beq
B=({\bf b},{\bf w},-{\bf s}),\qquad C^T=({\bf c},{\bf z},{\bf r}).
\eeq
\noindent
Then one can show that\cite{Faulkner}
\beq
\mathcal{I}(q)={\rm Det}A+{\rm Det}B+{\rm Det}C-{\rm Tr}(ABC).
\label{MerminesCartan}
\eeq
\noindent
From (\ref{MerminGrid}) it is clear that to the matrix $-A^T$, having index structure $\alpha\overline{\beta}$ with $\alpha,\beta =1,2,3$,  one can associate $9$ observables that can be arranged to a Mermin square. 
Moreover, due to Eq.(\ref{szukul}) the ${\rm Det}(A)$ part of $\mathcal{I}$ takes care of the sign distribution of this Mermin square. It is easy now to identify the remaining two Mermin squares. Indeed, referring to the corresponding observables in subset
notation these squares are of the form
\beq
\begin{pmatrix}2356&1346&1245\\1345&1256&2346\\1246&2345&1356\end{pmatrix},\quad
\begin{pmatrix}1&3456&13456\\12456&2&1456\\2456&23456&3\end{pmatrix},\quad
\begin{pmatrix}4&1236&12346\\12345&5&1234\\1235&12356&6\end{pmatrix}.
\label{cancomposite}
\eeq
\noindent
Along the lines and columns of these matrices we have commuting observables. The three lines plus three columns are comprising the six lines of a grid, i.e. a $GQ(2,1)$. The product of the observables is plus or minus the identity along the lines.  
Each square is having an odd number of negative lines.
The remaining Mermin square decompositions can be obtained in a straightforward manner via acting on (\ref{cancomposite}) with the transvections $T_{\alpha_j}$ generating $W(E_6)$. 

Let us also comment on the orign of the Hermiticity of $\Psi$ of Eq.(\ref{Psi}) within the framework of magic supergravities.
As is well-known there is a relation for such theories between the $D=4$ and $D=5$ dualities\cite{FerraraComposit}.
For $N=8$ supergravity the classic example of this is the relationship between the classical $E_{7(7)}$ symmetry\cite{Julia,CJ79} in four dimension and the $E_{6(6)}$ one in five dimensions. The former theory is featuring the (\ref{quarticCartan}) quartic\cite{CJ79,KK96} and the latter a cubic invariant for the semiclassical black hole entropy formula. The cubic invariant is just our Cartan invariant of Eq.(\ref{Cartaninv}).
For the $N=8$ theory we have the $8\times 8$ matrix of the central charge $Z_{AB}$ where $A,B=1,\dots 8$ of the supersymmetry algebra. It is a \emph{complex}
antisymmetric matrix.
This matrix is of the form
\beq
Z_{AB}=(x^{MN}+iy_{MN})(\Gamma^{MN})_{AB},
\label{centcharge}
\eeq
\noindent
	where summation is for $1\leq M<N\leq 8$ and the $x^{MN}, y_{MN}$ are antisymmetric $8\times 8$ matrices. The ${\Gamma}^{MN}$ are Hermitian antisymmetric matrices coming from the $28$ 
	combinations $\Gamma^{I8}\equiv\Gamma_I$ and $\Gamma^{IJ}\equiv i\Gamma_I\Gamma_J$ where $1\leq I<J\leq 7$, and they can be regarded as matrix-valued basis vectors for the expansion of $Z$.
We see that in $D=4$ we have $56$ expansion coefficients. These expansion coefficients are appearing in the quartic invariant of the $E_{7(7)}$ symmetric semiclassical black hole entropy formula. 
Now the usual way of obtaining the cubic invariant from the quartic one is via expanding $Z$ in an $USp(8)$-basis, which is appropriate since $USp(8)$ is the automorphism group of the $N=8$, $D=5$ supersymmetry algebra.
In order to do this one should chose the matrix of the symplectic form defining $USp(8)$. Let us choose $J=-i\Gamma_7$ as the matrix of this symplectic form. In our conventions this matrix is real, antisymmetric and has the form
$J=\epsilon\otimes\epsilon\otimes \epsilon$, where $\epsilon =i\sigma_2$.
Now the expansion of the $N=8$, $D=5$ central charge is obtained by imposing the constraints\cite{FerraraComposit}
\beq
{\rm Tr}(JZ)=0,\qquad \overline{Z}=JZJ^T.
\label{constraintssp8}
\eeq
\noindent
The first of these constraints is reducing the number of basis vectors in Eq.(\ref{centcharge}) from $28$ to $27$.
The second condition is a reality condition. It is easy to see that this condition demands that
\beq
y_{a8}=y_{a7}=x^{ab}=0.
\eeq
\noindent
Hence in the expansion of $Z$ only $27$ expansion coefficients are left. Renaming them as follows
\beq
x_{a8}\equiv -u_a,\qquad x_{a7}=v^a,\qquad y_{ab}\equiv\omega_{ab},
\eeq
\noindent
one can show that
\beq
\Psi=JZ.
\eeq
\noindent
In this language the reality condition means that $\Psi$ is Hermitian. Hence the origin of the Hermiticity of $\Psi$
can be traced back to the structure of the $N=8$ supersymmetry algebra.

\subsection{Klein's Quadric and the ${\mathbf G_2}$ Hitchin invariant}

Let us now consider the hyperbolic quadric (Klein quadric) part of our Veldkamp line. This part corresponds to the green parallelogram of Figure \ref{gods-eye-composition} and is labelled by the subsets $\{IJK\}$, where $1\leq I<J<K\leq 7$, and is split into two parts $\{abc\},\{ab7\}$ corresponding to the green and black triangles of Figure \ref{gods-eye-clifford}.
These triples can be used to label the weights of the $35$-dimensional irrep of $A_6$ as follows.
The simple roots of $A_6$ can be written as $\alpha_a=e_a-e_{a+1}, a=1,\dots 6$, where $e_I, I=1,2,\dots 7$ are the canonical basis vectors in $\mathbb{R}^7$. Using the Cartan matrix, its inverse and the fact that the Dynkin labels of this representation are encapsulated by the vector\cite{Slansky} $(001000)$, the $35$ weight vectors can be calculated.
They have the following form
\beq
\Lambda^{(IJK)}=e_I+e_J+e_K-\frac{3}{7}n,\qquad n=(1,1,1,1,1,1,1)^T,\qquad 1\leq I<J<K\leq 7.
\label{weightsA6}
\eeq
\noindent
The $A_6$ Dynkin diagram is just the $A_5$ Dynkin diagram of Figure \ref{A5diagram} with a consecutive extra node, labelled as $\alpha_6\leftrightarrow 67$, added.
The weight diagram of the $35$ of $A_6$ is obtained by glueing together the weight diagrams for the $15$ and $20$ of $A_5$ using figures \ref{15A5} and \ref{W32} as follows.
Consider that part of the weight diagram of Figure \ref{W32} which is labelled by triples $\{abc\}$.
Consider the seven triples: $(126,136,236,246,346,356,456)$. Connect these weights to the weights $(127,137,237,247,347,357,457)$ on the left-hand side of Figure \ref{15A5} by $\alpha_6$. Connect the remaining triples containing the letter $6$ to the corresponding weights of Figure \ref{15A5} by $\alpha_6$.
This construction corresponds to the fact that\cite{Slansky}
\beq
SU(7)\to SU(6)\times U(1),\qquad 35=15(-4)\oplus 20(3).
\eeq
\noindent

Let us now consider \emph{seven tuples} of weights with the property
\beq
\sum_{I=1}^7\Lambda^{(\mathcal{A}_I)}={\bf 0},\qquad \mathcal{A}_I\in\mathcal{P}_3(S).
\label{property}
\eeq
\noindent
It is easy to see that such seven tuples of weights are labelled by seven tuples of three element sets with a \emph{triple} occurrence for \emph{all} the numbers from $S=\{1,2,3,4,5,6,7\}$.
Two examples of such seven tuples are
\beq
\{(147),(257),(367),(123),(246),(356),(145)\},
\{(147),(257),(367),(123),(123),(456),(456)\}.
\nonumber
\eeq
\noindent
A special subset of such seven tuples is arising when, in addition to the property of Eq.(\ref{property}), the new one of
\beq
\vert\mathcal{A}_I\cap\mathcal{A}_J\vert =1,\qquad I\neq J,
\eeq
\noindent
is satisfied.
An example of such a seven-tuple is
\beq
\{(147),(257),(367),(123),(156),(246),(345)\}.
\label{Fanotriples}
\eeq
\noindent
Seven-tuples of the (\ref{Fanotriples}) form are called Steiner triples and they give rise to Fano planes. The seven points of the Fano plane are labelled by the triples, and its lines are labelled by the common intersection of such triples i.e. the elements of $S=\{1,2,3,4,5,6,7\}$.
One can prove that we have $30=7!/168$ such seven-tuples of triples, labelling different Fano planes. The number $168$ is the order of the Klein-group, i.e. $PSL(2,7)$, which is the automorphism group of the Fano plane.
Fano planes are planes in $PG(5,2)$, i.e. in the five-dimensional projective space over $\mathbb{Z}_2$.
One can apply the Klein correspondence\cite{Hirsch} and map these $30$ \emph{planes} of $PG(5,2)$ into $30$ heptads of mutually intersecting \emph{lines} of $PG(3,2)$.
It is well-known that there are two distinct sets of such heptads, each having
$15$ elements: a heptad of one set comprises seven lines passing through a point, whereas a heptad of other set consists of seven lines on a plane.
This $30=15+15$ split corresponds to a split of our $30$ Fano planes.
Fano planes belonging to the \emph{same class} intersect in a point, on the other hand Fano planes belonging to different classes either have zero intersection or they intersect in a line.   

In terms of three-qubit observables the meaning of these properties is as follows.
The $30$ Fano planes correspond to seven-tuples of mutually commuting observables represented by \emph{symmetric} $8\times 8$ matrices.
They represent $30$ from the $135$  maximal totally-isotropic subspaces (Lagrange subspaces defined after Eq.(\ref{szimpltulajd})), lying on the Klein quadric (i.e. the zero locus of the form defined in Eq.(\ref{kankvadrat})).
The $15+15$ split means that we have two different classes of such mutually commuting seven tuples of observables.
Seven tuples belonging to different classes are either disjoint or intersect in a \emph{triple} of mutually commuting observables. On the other hand, seven-tuples from the same class are intersecting in a single common observable.
The Klein group, $PSL(2,7)\simeq SL(3,2)$, as a subgroup of $Sp(6,2)$ acts transitively on this set of $30$ Fano planes.
It is well-known that the Klein group has a generator of order seven\cite{Levfin,Levay1} corresponding to the cyclic permutation $(1234567)$.
As a result one can easily provide a list of \emph{all} Fano planes lying on the Klein quadric\cite{Levay1}
\beq\{124,235,346,457,156,267,137\},\qquad
\{126,237,134,245,356,467,157\},
\label{specifanok}
\eeq
\beq
\{147,257,367,123,156,246,345\},\qquad
\{127,347,567,135,146,236,245\},
\label{set1}
\eeq
\noindent
\beq
\{157,247,367,456,235,134,126\},\qquad
\{137,257,467,124,156,236,345\}.
\label{set2}
\eeq
\noindent
The two sets of Eq.(\ref{specifanok}) are invariant under the cyclic shift $(1234567)$.
On the other hand, an application of this cyclic shift to the remaining four seven-tuples of Eqs.(\ref{set1})-(\ref{set2})
generates the remaining $24$ seven-tuples.
Notice that the \emph{first} and \emph{second} seven-tuples from Eqs.(\ref{specifanok})-(\ref{set2}) correspond to representatives of the \emph{first} and \emph{second} class, respectively.
For example, the first element of (\ref{specifanok}) and the second element of (\ref{set2}) intersect in the triple $(124,156,137)$ of commuting observables, hence they belong to different classes of Fano planes.

In the following we will be presenting a finite geometric understanding of Hitchin's $G_2$-functional introduced in\cite{Hitchin,Hitchin2}
and extensively used in string theory see for example\cite{FormTH}.
Let us consider a three-form on a real seven-dimensional orientable manifold $\mathcal{M}$,
\beq
\mathcal{P}=\frac{1}{3!}\mathcal{P}_{IJK}dx^I\wedge dx^J\wedge dx^K=\frac{1}{3!}P_{abc}dx^a\wedge dx^b\wedge dx^c+\frac{1}{2!}\omega_{ab}dx^a\wedge dx^b\wedge dx^7.
\label{faki}
\eeq
\noindent
We associate to this the observable
\beq
\Delta\equiv\Pi+\Omega,
\label{ujpi}
\eeq
\noindent
where we used the definitions (\ref{OmegaMermin}) and (\ref{Pas3observ}).
Notice that the Weyl group of $A_6$, i.e. $S_7$, acting on the weights lifts naturally to an action on our observable $\Delta$
according to the pattern as explained in Eq.(\ref{Phatas}). This action gives rise to the one on the coefficients $\mathcal{P}_{IJK}$ of our three-form. 

In addition to this discrete group action, there is the action of the continuous group $GL(7,\mathbb{R})$ at each point $x\in\mathcal{M}$
as follows
\beq \mathcal{P}_{IJK}\mapsto
{S_I}^{I^{\prime}}
{S_J}^{J^{\prime}}{S_K}^{K^{\prime}}\mathcal{P}_{I^{\prime}J^{\prime}K^{\prime}},\qquad
S\in GL(7,\mathbb{R}). \label{traf7} \eeq\noindent
The basic
covariant under (\ref{traf7}) is\cite{Hitchin,FormTH,SarLev2}
\beq
N_{IJ}=\frac{1}{24}\varepsilon^{A_1A_2A_3A_4A_5A_6A_7}\mathcal{P}_{IA_1A_2}\mathcal{P}_{JA_3A_4}\mathcal{P}_{A_5A_6A_7}
\label{basiccov7}
\eeq
\noindent
with transformation property
\beq
 N_{IJ}\mapsto ({\rm Det}S^{\prime})
 {S^{\prime}_I}^{I^{\prime}} {S^{\prime}_J}^{J^{\prime}}
 N_{I^{\prime}J^{\prime}}, \label{Nkovtr} \eeq
\noindent
where $S^{\prime}=(S^{-1})^T$.

Let us look at the structure of this covariant for three-forms with seven nonvanishing coefficients labelled by the triples giving rise to our $30$ Fano planes of Eqs.(\ref{specifanok})-(\ref{set2}). It is easy to see that for these heptads $N_{IJ}$ will be a diagonal matrix.  
Indeed for $I\neq J$, with $I$ and $J$ taken from different triples from any of our heptads, the complements of $I$ and $J$ with respect to their respective triples should have a common element. Hence $\varepsilon^{A_1A_2A_3A_4A_5A_6A_7}$ will have repeated indices giving zero.
Let us take any of the \emph{diagonal} elements, e.g. calculate $N_{11}$. Due to the Fano plane structure the number $1$ occurs in three triples. Let us employ any two of them. Then in order to have a nonvanishing term one has to employ the third one as well, since for a nonvanishing value of $\varepsilon^{A_1A_2A_3A_4A_5A_6A_7}$ all of the numbers from $1$ to $7$ have to show up.
Hence, the three terms in this cubic monomial will be featuring labels corresponding to a \emph{line} of our Fano plane.
For example, in the special case of Eq.(\ref{Fanotriples}) $N_{11}=\mathcal{P}_{123}\mathcal{P}_{147}\mathcal{P}_{156}$.
Hence, in this case, the line is $(123,147,156)$, its triples intersect in $1$, which is the label of the line and also the label of the diagonal element of $N_{IJ}$.
The net result is that the diagonal elements of $N_{IJ}$ will be featuring an ordered list of all seven lines of the corresponding Fano plane.
An easy way to build up an invariant from our covariant is just taking the determinant of $N_{IJ}$. However, this quantity will be only a relative invariant, i.e.
\beq
{\rm Det}N\mapsto ({\rm Det} S^{\prime})^9({\rm Det}N),
\label{compare}
\eeq
\noindent
so ${\rm Det}N$ is invariant only under the $SL(7,\mathbb{R})$ subgroup.

For the special case of $30$ heptads corresponding to Fano planes on the Klein quadric, ${\rm Det}N$, which is a monomial of order $21$, can be written as a cube of a monomial of order seven.
Clearly this monomial of order seven is just coming from the product of the seven nonvanishing coefficients of our special three-form. For example, for the Fano plane labels of Eq.(\ref{Fanotriples})
\beq
{\rm Det}N=(\mathcal{P}_{147}\mathcal{P}_{257}\mathcal{P}_{367}\mathcal{P}_{123}\mathcal{P}_{156}\mathcal{P}_{246}\mathcal{P}_{345})^3.
\label{Fanocov}
\eeq
\noindent
Hence, we conclude that there should be a relative invariant of order seven in the $35$ coefficients of $\mathcal{P}$.

In order to write down explicitely this invariant we introduce
additional covariants\cite{SarLev2}
\beq
{(M^I)^J}_K=\frac{1}{12}\varepsilon^{IJA_1A_2A_3A_4A_5}\Psi_{KA_1A_2}\Psi_{A_3A_4A_5},
\quad L^{IJ}={(M^I)^{A_1}}_{A_2}{(M^J)^{A_2}}_{A_1}. \label{Mkovar} \eeq
\noindent 
Under the (\ref{traf7}) transformations these
transform as follows 
\beq {(M^I)^J}_K\mapsto ({\rm Det}
S^{\prime}){S^I}_{I^{\prime}}
{S^J}_{J^{\prime}}{S^{\prime}_K}^{K^{\prime}}{(M^{I^{\prime}})^{J^{\prime}}}_{K^{\prime}}\quad
L^{IJ}\mapsto ({\rm Det}S^{\prime})^2
{S^I}_{I^{\prime}}{S^J}_{J^{\prime}}L^{I^{\prime}J^{\prime}},
\label{Lkovtr} \eeq \noindent where $S^{\prime}=(S^{-1})^T$. Notice
that the $7\times 7$ matrices $N_{IJ}$ and $L^{IJ}$ are symmetric.

From our covariants one can form a unique algebraically
independent relative invariant \beq
\mathcal{J}(\mathcal{P})=\frac{1}{2^4\cdot 3^2\cdot 7}{\rm Tr}(NL).
\label{jeinv}\eeq\noindent
Under (\ref{traf7}) $\mathcal{J}$ picks up the determinant factor, \beq
\mathcal{J}(\mathcal{P})\mapsto ({\rm Det} S^{\prime})^3\mathcal{J}(\mathcal{P})
\label{transje}, \eeq
\noindent
hence it is invariant under $SL(7,\mathbb{C})$.
Comparing with the transformation rule (\ref{compare}) one conjectures that ${\rm Det}N\simeq (\mathcal{J}(\mathcal{P}))^3$.
 Defining \beq
 \mathcal{B}_{IJ}=-\frac{1}{6}N_{IJ}\label{calbe} \eeq\noindent one can indeed prove that
 \beq (\mathcal{J}(\mathcal{P}))^3={\rm
 Det}\mathcal{B}. \label{kobinv} \eeq \noindent

A three-form with the property $\mathcal{J}(\mathcal{P})\neq 0$ is called nondegenerate\cite{Hitchin}.
In this case one can define the functional
\beq
V_{HG2}[\mathcal{P}]=\int_{\mathcal{M}}(\mathcal{J}(\mathcal{P}))^{1/3}d^7x.
\label{G2Hitch}
\eeq
\noindent
We will refer to this functional as the $G_2$-Hitchin functional.
The reason for this name is coming from the well-known fact (for a summary on related issues see \cite{ilka}) that the stabilizer in $GL(7,\mathbb{R})$ of three-forms associated with our 30 heptads gives rise to a particular real form of the exceptional group $G_2^{\mathbb{C}}$. For example, for the choice
\beq
\mathcal{P}_{\mp}=dx^{123}\mp dx^{156}\pm dx^{246}\mp dx^{345}\pm dx^{147}\pm dx^{257}\pm dx^{367},\qquad dx^{IJK}\equiv dx^I\wedge dx^J\wedge dx^K,
\eeq
\noindent
the stabilizers are  the compact real form $G_2$ ($\mathcal{P}_-$), which is the automorphism group of the octonions,
and the noncompact real form  $\tilde{G}_2$ ($\mathcal{P}_+$), which is the automorphism group of the split octonions.
For any fixed $x\in\mathcal{M}$ in the space of real three-forms the two orbits of $\mathcal{P}_{\mp}$ under the (\ref{traf7}) action is dense\cite{SatoKimura}.

For nondegenerate forms one can define a symmetric tensor field, i.e. a \emph{metric} on $\mathcal{M}$,
as
\beq
G_{\mathcal{P}IJ}\equiv(\mathcal{J}(\mathcal{P}))^{-1/3}\mathcal{B}_{IJ}.
\label{G2metrika}
\eeq
\noindent
Indeed, by virtue of Eqs.(\ref{Nkovtr}), (\ref{transje}) and (\ref{calbe}) $G_{IJ}$ transforms without determinant factors.
For the nondegenerate orbit of $\mathcal{P}_-$ the metric is a Riemannian one and the $G_2$-Hitchin functional can be written in the alternative form\cite{Hitchin,FormTH}
\beq
V_{HG2}[\mathcal{P}]=\int_{\mathcal{M}}\sqrt{G_{\mathcal{P}}}d^7x,\qquad G_{\mathcal{P}}={\rm Det}G_{\mathcal{P}IJ},
\eeq
\noindent
meaning that this functional is a volume form defined by $\mathcal{P}$.
The critical points of this functional in a fixed cohomology class give rise to metrics in $\mathcal{M}$ of $G_2$-holonomy\cite{Hitchin,Hitchin2}.
Such manifolds are of basic importance in obtaining compactifications of M-theory with realistic phenomenology\cite{BeckerBecker}.
In analogy with topological string theory related to Calabi-Yau manifolds, one can consider topological M-theory\cite{FormTH,Sinkovics} related to manifolds with $G_2$-holonomy. The classical effective description of topological M-theory is provided by $V_{HG2}[\mathcal{P}]$.

The expressions (\ref{Fanocov}), (\ref{calbe}) and (\ref{kobinv}) show that in identifying the finite geometric structure of $\mathcal{J}(\mathcal{P})$ the  $30$ Fano planes of our Klein quadric play a fundamental role. In other words we have a seventh-order invariant with $30$ of its monomials directly associated with Lagrangian subspaces (Fano planes) of a hyperbolic quadric in $PG(5,2)$.
The Klein quadric has $35$ points and $30$ Fano planes on it, with each Fano plane containing $7$ points. It can be regarded as a combination of a $GQ(2,2)$  and an $EQG(2,1)$ (black and green triangles of Figure \ref{gods-eye-clifford}). The former has $15$ points and $15$ lines, with each line containing $3$ points,
and the latter has $20$ points and $30$ blocks with each block containing $4$ points.
The former contains $10$ grids ($GQ(2,1)$s) as hyperplanes (see Figure \ref{mermins}), the latter contains $20$ grids as residues.
Moreover, the former object is associated with the \emph{cubic} invariant of (\ref{Pfaff}), and the latter with a \emph{quartic} one
of (\ref{Hitchinform}).
What we need is a finite geometric method for entangling the finite geometric structures
of $GQ(2,2)$ and $EGQ(2,1)$ in a way which also combines the cubic and quartic invariants to our (\ref{jeinv}) seventh order one.

At the time of writing this paper we are not aware of any finite geometric method of the above kind.
However, we are convinced that this method of entangling the structures of $GQ(2,2)$ and $EGQ(2,1)$ should be based on grids,
i.e. $GQ(2,1)$s. Since at the level of observables grids are associated with Mermin squares, this idea stresses the relevance of Mermin squares as universal building blocks for \emph{any} of our invariants discussed in this paper.
Let us share with the reader some solid piece of evidence in favour of this conjecture.

In the case of the canonical grid related to the decomposition of Eq.(\ref{MerminGrid}) we have the arrangements
\beq
i\begin{pmatrix}
147&157&167\\247&257&267\\347&357&367\end{pmatrix}=123\cdot\begin{pmatrix}156&-146&145\\256&-246&245\\356&-346&345\end{pmatrix}
=456\cdot\begin{pmatrix}234&-134&124\\235&-135&125\\236&-136&126\end{pmatrix}.
\label{szamok}
\eeq
\noindent
Here the triples mean products of gamma matrices hence, for example, $123\cdot 156 =2356=i147$.
Writing formally the determinant of the $3\times 3$ matrix on the left-hand side, the six monomials give rise to the lines of a grid of the $GQ(2,2)$-part labelled with observables, i.e. a Mermin square.
On the other hand, we see how the labels of the two antipodal residues of Eqs.(\ref{kanvan})-(\ref{kanvananti}) of the $EQG(2,1)$-part give rise to the \emph{same} Mermin square of the $GQ(2,2)$-part. This construction relates the $6$ lines of a grid in  $GQ(2,2)$ with $12$ blocks of the $EGQ(2,1)$.
Moreover this correspondence between lines of a grid (e.g. take the diagonal entries of the matrix on the left : $(147,257,367)$),
and the  blocks of a residue (e.g. take the diagonal entries of the middle and right matrices together with the residue labels
: $(123,156,246,345)$ and $(456,234,135,126)$) are based on Fano heptads.
Indeed the two heptads $(147,257,367,123,156,246,345)$ and $(147,257,367,456,234,135,126)$ are the ones belonging to the different classes of Fano planes intersecting in the common line $(147,257,367)$.
One can repeat this construction for any of the residues. Then we can relate the $10$ pairs of antipodal residues of $EGQ(2,1)$ to the $10$ grids living inside $GQ(2,2)$. Clearly, this method also establishes a correspondence between the $15$ lines of the doily to the $30$ blocks of $EGQ(2,1)$.

Let us now connect these observations to the structure of our seventh-order invariant of (\ref{jeinv}).
We would like to see how this invariant incorporates  the cubic and quartic invariants of Eqs.(\ref{Pfaff}) and (\ref{Cayleygen}) associated with the structures showing up in Eq.(\ref{szamok}).
We start with the $35$ components of Eq.(\ref{faki}). We decompose the $15$ components of $\omega_{ab}$ to $3\times 3$ matrices $\omega_{\alpha\beta},\omega_{\overline{\alpha}\overline{\beta}},\omega_{\alpha\overline{\beta}}$, with the first two of them being antisymmetric ones. 
The remaining $20$ components enjoy the decomposition of Eqs.(\ref{etaxi})-(\ref{Ymatr}) with two scalars $\eta,\xi$ and the relevant $3\times 3$ matrices having the index structure:
${{\bf X}_{\alpha}}^{\overline{\beta}}, {{\bf Y}_{\overline{\alpha}}}^{\beta}$. Notice that according to the left and middle matrices of (\ref{szamok}), the quantities $(\omega_{\alpha\overline{\beta}},\eta,{{\bf X}_{\alpha}}^{\overline{\beta}})$ correspond to a grid of $GQ(2,2)$ and the canonical residue associated to $\eta$ of $EGQ(2,1)$.
Our aim is to use these quantities to arrive at a new form of $(\ref{jeinv})$.

Let us define
\beq
{\mathcal{Q}_{\alpha}}^{\overline{\beta}}={(\eta {\bf X} -{\bf Y}^{\sharp})_{\alpha}}^{\overline{\beta}},\qquad
V_{\alpha\overline{\beta}}=\omega_{\alpha\overline{\beta}}-\frac{1}{\eta}\omega_{\alpha\gamma}{{\bf Y}^{\gamma}}_{\overline{\beta}},\qquad \omega^{\alpha}=\frac{1}{2}\varepsilon^{\alpha\beta\gamma}\omega_{\beta\gamma},
\label{fontidefi}
\eeq
\beq
U_{\overline{\alpha}\overline{\beta}}=\omega_{\overline{\alpha}\overline{\beta}}+\frac{2}{\eta}{{\bf Y}_{[\overline{\alpha}}}^{\gamma}V_{\gamma\vert\overline{\beta}]}-\frac{1}{\eta}\varepsilon_{\overline{\alpha}\overline{\beta}\overline{\gamma}}{{\bf X}^{\overline{\gamma}}}_{\delta}\omega^{\delta},\qquad G_{\alpha\beta}={\mathcal{Q}_{(\alpha}}^{\overline{\gamma}}V_{\overline{\gamma}\vert\beta)},
\label{fontosdefik}
\eeq
\noindent
where the notation $[\overline{\alpha}\gamma\vert\overline{\beta}]$ ,($({\alpha}\gamma\vert{\beta})$) means antisymmetrization (symmetrization) only in the index pair $\overline{\alpha},\overline{\beta}$, ($\alpha,\beta$). 
Let us also recal the (\ref{angularmom}) definition of $J_{\eta}$ in terms of the data of Eq.(\ref{duálisP}).
Employing the results of \cite{cluster} we obtain the compact expression 
\beq
\mathcal{J}(\mathcal{P})=-\frac{1}{\eta^2}{\rm Det}G-\frac{\eta}{8J_{\eta}}{\rm Det}\left(\mathcal{Q}U+\frac{2J_{\eta}}{\eta}V\right).
\label{cute}
\eeq
\noindent

Let us elaborate on this formula.
For $3\times 3$ matrices one can use the identity
\beq
{\rm Det}(A+B)={\rm Det}A+{\rm Tr}(AB^{\sharp})+{\rm Tr}(A^{\sharp}B)+{\rm Det}B
\label{detes}
\eeq
\noindent
to rewrite in the second term the determinant of the sum. Since ${\rm Det}(\mathcal{Q}U)=0$ due to $U$ being antisymmetric, no term proportional to ${\eta}/{J_{\eta}}$ arises. 
Only terms linear and quadratic in $J_{\eta}/\eta$, and terms not featuring $J_{\eta}/\eta$ at all show up.
Moreover, we have $U^{\sharp}=uu^T$ where $u$ is the three-vector associated to the antisymmeric matrix to $U$ (see the third formula of Eq.(\ref{fontidefi})).
After introducing the antisymmetric part of the matrix $\mathcal{Q}V^T$, i.e.
\beq
H_{\alpha\beta}=
{\mathcal{Q}_{[\alpha}}^{\overline{\gamma}}V_{\overline{\gamma}\vert\beta]},\qquad h^{\alpha}=
\frac{1}{2}\varepsilon^{\alpha\beta\gamma}H_{\beta\gamma},
\label{antimetric}
\eeq
one obtains the alternative formula
\beq
\mathcal{J}(\mathcal{P})=-\frac{1}{\eta^2}(J_{\eta}^2+{\rm Det}\mathcal{Q}){\rm Det}V+\frac{1}{\eta^2}h^TGh-\frac{J_{\eta}}{2\eta}{\rm Tr}(UV^{\sharp}\mathcal{Q})-\frac{1}{4}u^T\mathcal{Q}^{\sharp}Vu.
\label{cute2}
\eeq
\noindent
Now it is well-known from studies concerning the correspondence between 4D and 5D black hole solutions that\cite{Pioline,Freudual} 
\beq
\mathcal{D}(P)=\frac{4}{\eta^2}(J_{\eta}^2+\rm{Det}\mathcal{Q})
\eeq
\noindent
as can be checked using the definitions (\ref{Cayleygen}), (\ref{angularmom}) and the identity (\ref{detes}).
Notice that this interpretation also identifies the physical meaning of the quantities $\eta$, $\mathcal{Q}$ and $J_{\eta}$ 
as the NUT charge, the charge and angular momentum of the 5D spinning black hole\cite{Gaiotto,Pioline,Freudual}.
The first term in our final formula
\beq
\mathcal{J}(\mathcal{P})=-\frac{1}{4}\mathcal{D}(P)\cdot{\rm Det}V
-\frac{1}{4}u^T\mathcal{Q}^{\sharp}Vu
+\frac{1}{\eta^2}h^TGh-\frac{J_{\eta}}{2\eta}{\rm Tr}(UV^{\sharp}\mathcal{Q})
\label{cute3}
\eeq
\noindent
shows the desired factorization of the seventh-order invariant to a quartic and cubic ones. 
The remaining terms are to be considered as "interference terms". 

We would like to interpret the vanishing conditions of these "interference terms".
Our decomposition of $\mathcal{P}$ can be interpreted as a means to regarding the seventh dimension of our $\mathcal{M}$ as special. If we consider compactifications of M-theory of the form $\mathcal{M}=S^1\times M$, where $M$ is a six-manifold, one can imagine $M$ as manifold also equipped with a symplectic form $\omega$ for which only the off-diagonal blocks of its $\omega_{ab}$ matrix are nonvanishing, i.e. only the $\omega_{\alpha\overline{\beta}}$ terms are nonzero.
These are the terms corresponding to the matrix on the left-hand side of Eq.(\ref{szamok}) and used in Eq.(\ref{MerminGrid}).
In this case the physical meaning of the matrix $V\equiv \omega_M$ is clear: its components define the symplectic form on $M$.
On the other hand, $U$ is just $2/\eta$ times the antisymmetric part of the matrix $Y\omega_M$, and $H$ and $G$ are the antisymmetric and symmetric parts of $\mathcal{Q}\omega_M^T$.
Hence, in this special case our invariant is
\beq
\mathcal{J}(\mathcal{P})=\frac{1}{4}\mathcal{D}(P)\cdot{\rm Pf}(\omega)
-\frac{1}{4}u^T\mathcal{Q}^{\sharp}\omega_Mu
+\frac{1}{\eta^2}h^TGh-\frac{J_{\eta}}{2\eta}{\rm Tr}(U\omega_M^{\sharp}\mathcal{Q}).
\label{cute3}
\eeq
\noindent
Clearly, if the extra 
conditions that the matrices $Y\omega_M$ and $\mathcal{Q}\omega_M^T$ are symmetric hold then the interference terms are vanishing.
It can be shown that these conditions are equivalent to the one
\beq
\omega\wedge P=0
\eeq
meaning that the $P$ part of $\mathcal{P}$ is primitive with respect to $\omega$. This condition disentangles the seventh-order
invariant\cite{FormTH} as
\beq
\mathcal{J}(\mathcal{P})=\frac{1}{4}{\rm Pf}(\omega)\mathcal{D}(P).
\eeq
\noindent
It means that the invariants of the $GQ(2,2)$ and $EGQ(2,1)$ parts factorize the seventh-order invariant.
For a single residue and its corresponding grid only the triple $(\omega_{\alpha\overline{\beta}},\eta,{X_{\alpha}}^{\overline{\beta}})$ contributes. In this case
\beq
\mathcal{J}(\mathcal{P})=-\eta{\rm Det}{\bf X}{\rm Det}\omega_M+\frac{1}{\eta^2}h^TGh
\eeq
so factorization is achieved when ${\bf X}\omega_M^T$ is symmetric.

It would be interesting to see whether there is an analogue of formula (\ref{abstract}) in this $G_2$-Hitchin invariant case.
In order to find this formula, as suggested by the pattern of (\ref{szamok}), a seventh-order polynomial defined for antipodal residues and their corresponding grid is needed.
However, in order to find this interpretation of the seventh-order invariant a deeper understanding of the "entanglement" between the $GQ(2,2)$ and $EGQ(2,1)$ geometries is needed.

\subsection{A note on extended generalized quadrangles of type {\bf EGQ(2,2)}}

We have demonstrated that the structure of our magic Veldkamp line is based on different combinations of generalized quadrangles $GQ(2,1)$, $GQ(2,2)$, $GQ(2,4)$ and their extensions $EGQ(2,1)$ and $EGQ(2,2)$. To these geometric structures one can associate in a natural manner invariants of physical significance. These invariants are of cubic type for the generalized quadrangles, i.e. the determinant (\ref{szukul}), the Pfaffian (\ref{Pfaff2}), and Cartans cubic invariant (\ref{Cartaninv}). For the extended generalized quadrangles they are of quartic type: Hitchin's invariant (\ref{Cayleygen}) and the generalized Hitchin invariant (\ref{GenHInv}).

Here we would like to point out that one can extend $GQ(2,2)$, comprising the core configuration of our Veldkamp line,
in different ways, hence the $EQ(2,2)$ structure underlying the generalized Hitchin invariant is just one of other possible extensions. 

\begin{figure}[pth!]
\centerline{\includegraphics[width=6truecm,clip=]{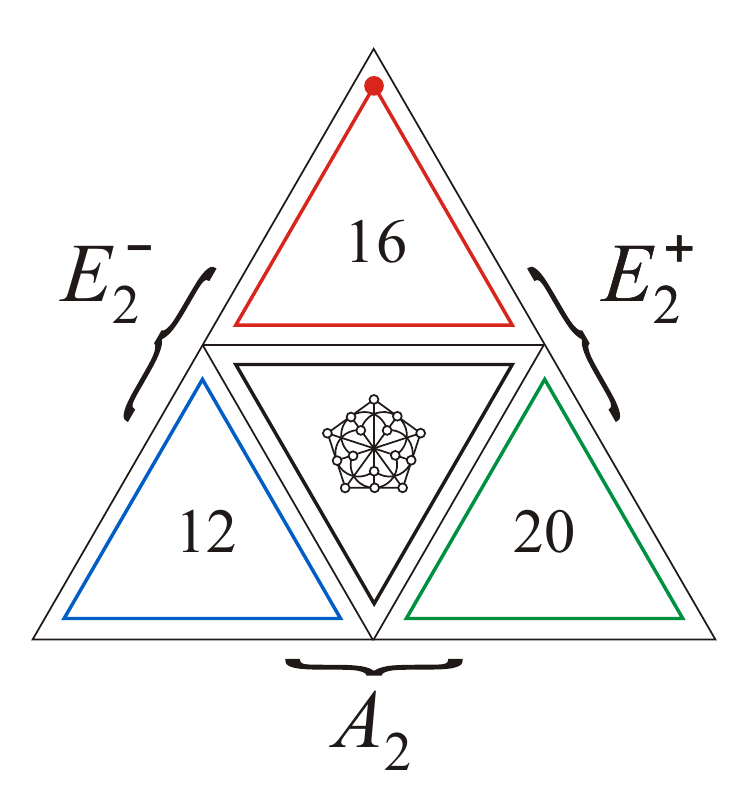}}
\caption{Different extensions $EGQ(2,2)$ of the doily $GQ(2,2)$ living inside our magic Veldkamp line.}
\label{egq22s}
\end{figure}

The different extensions are coming from a construction based on affine polar spaces\cite{Buek1}.
According to this result there are $EGQ(2,t)$s (of necessity $t=1,2,4$) of ten different types.
For $t=2$ we have three different extensions conventionally denoted by the symbols $A_2,E_2^+,E_2^-$. 
They are having points: $32,36,28$ respectively.
Surprisingly, \emph{all} of these extensions of the doily can be accomodated in our Veldkamp line in a natural manner.
Indeed, as shown in Figure \ref{egq22s}, the three different pairs of coloured triangles produce the right count for the number of points of these extensions. Namely, the blue and green triangles produce $A_2$, and the red and green and red and blue ones give rise to $E_2^+$ and $E_2^-$ respectively.
For the $A_2$-part we have already verified the $EGQ(2,2)$ structure in connection with the generalized Hitchin functional. Here we check the $EGQ(2,2)$ structure of type $E_2^-$.

In order to verify the $E_2^-$-structure, we notice that according to Figure \ref{gods-eye-clifford} the relevant red and blue triangles are labelled as $\{7,ab,a,a7\}$. These labels can be mapped to the $28$ weights of the $28$-dimensional irrep of $SU(8)$.
Indeed, one can label the corresponding $A_7$ Dynkin diagram by formally adjusting an extra label as follows: $\{78,ab,a8,a7\}$. Then the seven nodes of the Dynkin diagram are labelled by $12,23,34,45,56,67,78$.
The Dynkin labels\cite{Slansky} of the $28$ of $A_7$ are $(0100000)$ and the weights can be constructed in the usual manner. The result is
\beq
\Lambda^{(\hat{I}\hat{J})}=e_{\hat{I}}+e_{\hat{J}}-\frac{1}{4}n, \qquad n=(1,1,1,1,1,1,1,1)^T,\qquad 1\leq \hat{I}<\hat{J}\leq 8.
\eeq
\noindent 
Now, the $28$ weights correpond to the points of our $E_2^-$ and the blocks are coming from quadruplets of different weights with their sum giving the zero vector. Clearly, these quadruplets are the ones whose labels partition the set $S=\{1,2,3,4,5,6,7,8\}$ into $4$ two-element sets. By Example 9.8 of Ref.\cite{Cameron}, the structure whose points are unordered $2$-sets of $S$ and whose block are partitions of $S$ into four $2$-sets is precisely $E_2^-$. This gives a representation theoretic realization of the $E_2^-$-structure.
On the other hand, after reinterpreting the labels as observables $\{\Gamma_7,\Gamma_a,i\Gamma_a\Gamma_b,i\Gamma_a\Gamma_7\}$ the blocks will correspond to quadruplets of pairwise commuting observables with their products being the $\pm$ identity.
This gives a physically interesting realization of the $E_2^-$-structure in terms of three-qubit Pauli observables.
Note that an alternative interpretation can be given in terms of the $28$ of $SO(8)$. In this case the the $8\times 8$ matrices $\{\Gamma_7,\Gamma_a,\Gamma_a\Gamma_b,\Gamma_a\Gamma_7\}$ are directly related to the generators of $SO(8)$.

According to our basic philosophy now we can look after an invariant whose stucture is encapsulated in the $E_2^-$-structure. Clearly, this invariant is just the (\ref{buvos}) Pfaffian of an $8\times 8$ antisymmetric matrix.
In the $SO(8)$ basis we have already found the physical meaning  of this invariant. Indeed, in the black hole context, according to Eq.(\ref{quarticCartan}) this invariant is appearing as a substructure of the $E_{7(7)}$-symmetric entropy formula\cite{Julia,KK96}.
We have already seen this phenomenon in connection with the generalized Hitchin invariant, i.e. the $A_2$-part.
According to Eq.(\ref{kapcsolatHitch}) this part is living naturally inside the $E_{7(7)}$-invariant via truncation to the RR-sector.
Now the invariant of the $E_2^-$-part produces another truncation of this $E_{7(7)}$-invariant. 
In order to see this we just have to recall that in the case of toroidal ($T^6$) compactifications one can use either the full set of $28$ components of the $x^{\hat{I}\hat{J}}$, or the
$y_{\hat{I}\hat{J}}$ matrix corresponding to Eq.(\ref{Dbranecharges}). In the first case we are content with $15$ $D2$ branes, and a $D6$ brane (red triangle of Figure \ref{gods-eye-clifford}), $6$ fundamental string windings and $6$ wrapped KK5 monopoles\cite{Pioline} (blue triangle of Figure \ref{gods-eye-clifford}). 
In the second case we restrict our attention to subconfigurations of $15$ $D4$ branes and a $D0$ brane, $6$ wrapped $NS5$
branes and $6$ Kaluza-Klein momenta\cite{Pioline}.
The fact that only truncations of the $E_{7(7)}$-invariant can be accomodated into our Veldkamp line
and not the full formula indicates that for an implementation of this invariant we should embed our Veldkamp line within another one on four-qubits. We postpone the discussion on this interesting issue to our last and concluding section.

Let us also verify that the red and green triangle parts of Figure \ref{egq22s} indeed represent an $E_2^+$-structure.
For this we merely have to verify that the structure of subsets of the form $\{7,ab,abc\}$,  equipped with the adjacency relation given by the vanishing of the symplectic form of Eq.(\ref{alldjacentsympl}), is just  the point graph of $E_2^+$. This graph is a strongly regular graph\cite{Cameron} with parameters $(36,15,6,6)$. These parameters are in order: the number of vertices, the valency, the number of common neighbors of adjacent and non-adjacent pairs of vertices.
First, using (\ref{gamszorz}), instead of $\{7\}$ we write $\{123456\}$.
We have then $36$ residues belonging to $3$ types of cardinality $1,15,20$. It is enough to examine one from each type.
The first type is trivial: it is just the residue of $\{123456\}$, which consists of the $15$ vertices of the form $\{ab\}$, $a<b$.
For the representatives of the remaining types we consider the residues of $56$ and $456$. We order them in a way compatible with the duad labelling of the doily in lexicographic order: 
\beq
\{12,13,14,234,156,23,24,134,256,34,124,356,123,456,123456\},
\eeq
\noindent
\beq
\{12,13,234,235,236,23,134,135,136,124,125,126,45,46,56\}.
\eeq
\noindent
One can check that the adjacent vertices labelled by $45$ and $456$ have $6$ common neighbors, and the nonadjacent ones $123456$ and $456$ have $6$ common neighbors as well.
There is an action of $S_6$ on these subset labels. The number of residues of the first type is $15$ and of the second one $20$. These copies can be obtained by the action of elements taken from $S_6$ not stabilizing the labels $56$, $456$ and their complements. Due to the permutational symmetry, the point graph structure can be checked easily via looking merely at representative cases. The blocks are containing four points, and the total number of blocks is $36\times 15/4=135$.
Hence the commutation properties of the observables
 $\{\Gamma_7, i\Gamma_a\Gamma_b,i\Gamma_a\Gamma_b\Gamma_c\}$ give rise to a realization of the geometry
 $E_2^+$.
Again its blocks will correspond to quadruplets of pairwise commuting observables with  products being $\pm$ the identity.

What about an invariant associated with this part?  
In order to attempt finding an answer to this question we turn to our last section.

\subsection{Connecting the Magic Veldkamp Line to $Spin(14)$} 

In our finite geometric investigations of the Magic Veldkamp Line (MVL) we managed to associate to its different \emph{parts} incidence structures, representations and invariants with physical meaning.
Now a natural question to be asked is the following: what is the finite geometric, representation theoretic and physical meaning of our MVL  \emph{as a whole}?

As far as \emph{group representations} are concerned, the answer to this question is easy to find: our MVL encapsulates information on the $64$-dimensional spinor representation of the group $Spin(14)$  (type $D_7$) of odd chirality.
Indeed, including also the identity we have $64$ three-qubit Pauli operators, hence after associating to the odd chirality spinor of $Spin(14)$ a polyform
\beq
\sigma=u_Idx^I+\frac{1}{3!}\mathcal{P}_{IJK}dx^{IJK}+\frac{1}{2!5!}v^{IJ}\varepsilon_{IJKLMNR}dx^{KLMNR}+\zeta dx^{1234567}
\label{mvlampl}
\eeq
\noindent
there is a corresponding three-qubit observable of the form
\beq
\Sigma=u_I\Gamma_I+\frac{i}{3!}\mathcal{P}_{IJK}\Gamma_I\Gamma_J\Gamma_K+\frac{1}{2!5!}v^{IJ}\varepsilon_{IJKLMNR}\Gamma_K\Gamma_L\Gamma_M\Gamma_N\Gamma_R+\zeta \mathbf{1},
\eeq
\noindent
where $1\leq I<J<\dots <R\leq 7$ and $\mathbf{1}$ is the identity realized by the $8\times 8$ identity matrix.

In order to show that we are on the right track let us  first give a special status to the Klein quadric part of our MVL. This part is related to the $35$ of $A_6$, i.e. the green parallelogram of Figure \ref{gods-eye-composition}. Under the decomposition
$so(14)\supset su(7)\oplus u(1)$ we have
\beq
64=\overline{7}(3)\oplus \overline{35}(1)\oplus 21(-3)\oplus 1(-7)
\eeq
\noindent
corresponding to the structures above.
Using that under $su(7)\supset su(6)\oplus u(1)$ we have 
\beq
\overline{7}=\overline{6}(-5)\oplus 1(6),\qquad \overline{35}=20(-3)\oplus\overline{15}(4)
\eeq
\noindent
and neglecting one of the $u(1)$s, we obtain
\beq
64=\overline{6}(-5)\oplus 6(-1)\oplus \overline{15}(4)\oplus 20(-3)\oplus 15(2)\oplus 1(6)\oplus 1,
\label{MVLdec}
\eeq
\noindent
which after taking into account (\ref{gamszorz}) reproduces our split of Figure \ref{gods-eye-clifford}.

Let us now give a special status to the $32$ of $D_6$-part, a combination of the blue and green triangles of Figure \ref{gods-eye-composition}.
In this case the relevant decomposition is: $so(14)\supset so(12)\oplus u(1)$
\beq
64=32(1)\oplus 32^{\prime}(-1).
\eeq
\noindent
Under the decomposition $so(12)\supset su(6)\oplus u(1)$ one has
\beq
32=6(-2)\oplus 20(0)\oplus \overline{6}(2),\qquad 32^{\prime}=15(-1)\oplus\overline{15}(1)\oplus 1(3)\oplus 1(-3).
\eeq
\noindent
Combining these we obtain again the usual decomposition of Eq.(\ref{MVLdec}).

As far as \emph{finite geometry} is concerned, our analysis clearly demonstrated that the structure of the MVL nicely encodes an entangled structure of generalized and extended generalized quadrangles. The MVL naturally connects information concerning incidence structures to the one hidden in weight systems of certain subgroups appearing in branching rules of $Spin(14)$. Alternatively, this information on incidence geometry is revealed by the commutation properties of special arrangements of three-qubit observables like Mermin squares and pentagrams.

Finally, the unified picture of the MVL combined with the appearance of $Spin(14)$ naturally hints at an invariant
\emph{unifying all the invariants} we came accross in our investigations.
As emphasized in \cite{LevSar}, the invariants connected to Hitchin functionals are associated to prehomogeneous vector spaces\cite{SatoKimura}.
These objects are triples $(V,G,\varrho)$ of a vector space $V$, a group $G$ and an irreducible representation $\varrho$ acting on the vector space, such that we have a dense orbit of the group action in the Zariski topology. This property is crucial for the stability property needed for the variational problem of the corresponding Hitchin functional to make sense\cite{Hitchin,Hitchin2,Hitchin3,FormTH}.
In the black hole/qubit correspondence\cite{BHQC,LevSar} these invariants give rise to measures of entanglement and the prehomogenity property ensures the existence of a special GHZ-like entanglement class playing a crucial role in the subject.
From the table of regular prehomogeneous vector spaces\cite{SatoKimura} one can see that the highest possible value of $n$ such that the spinor irrep for the group $Spin(n)$ is a regular prehomogoneous vector space is $n=14$.
In this case, 
under the group $\mathbb{C}^{\times}\otimes Spin(14)$, 
one has a unique relative invariant $J_8$ of order \emph{eight}.
According to the general formula obtained in \cite{Gyoja}
it is of the form
\beq
J_8(\sigma)=J_6(z)\zeta^2+4J_7(z)\zeta,
\label{je8}
\eeq
\noindent
where
\beq
z=e^{v/\zeta}\llcorner\sigma=z_Idx^I+z_{IJK}dx^{IJK}+\zeta dx^{1234567}\qquad v=\frac{1}{2}v^{IJ}e_{IJ}, \qquad dx^I(e_J)= \delta^I_J,
\eeq
\noindent
and for the explicit expressions of $J_{6,7}$ we orient the reader to \cite{Gyoja}.

Let us first consider the truncation of $\sigma$
 when only the $36$ components corresponding to the pair      
$(\zeta,\mathcal{P}_{IJK})$
are nonzero. 
The $(\mathcal{P}_{IJK})$-part corresponds to the Klein quadric part of the MVL. The $\zeta$-part can be interpreted as an extra point. At the level of observables this is just the term proportional to the identity observable commuting with \emph{all} observables.
Geometrically, the $63+1$ structure of the MVL plus an extra point arising in this way can be regarded as a one point extension\cite{Fisher} of the symplectic polar space $\mathcal{W}(5,2)$ defined in the paragraph following Eq.(\ref{szimpltulajd}).
We write
\beq
\sigma_1=
\left(\frac{1}{2!}\omega_{ab}dx^{ab}+\zeta dx^{123456}\right)\wedge dx^7+\frac{1}{3!}P_{abc}dx^{abc}.
\label{sigma1}
\eeq
\noindent
where $\omega_{ab}=P_{ab7}$.
The six-dimensional interpretation of this configuration is that of a $D0D4$-system combined with a $D3$-one.
In this case\cite{SaroPhD}
\beq
J_8(\sigma_1)=16\zeta\mathcal{J}(\mathcal{P}),
\label{beszurkal}
\eeq
\noindent
where $\mathcal{J}(\mathcal{P})$ is the seventh-order invariant of Eqs (\ref{jeinv}).
Notice that if, in addition, the constraint $\omega\wedge P=0$ holds, then according to (\ref{cute3}) our formula simplifies to
\beq
J_8(\sigma_1)=4\zeta\rm{Pf}(\omega)\mathcal{D}(P),\qquad \omega\wedge P=0.
\label{Tdual1}
\eeq
\noindent

Next we consider a truncation corresponding to the unclarified case of $E_2^+$ of the previous section.  Now we keep the $36+1$ quantities: $(u_7,P_{abc},v^{ab},\zeta)$ with the $(u_7,P_{abc},v^{ab})$-part labelling the $E_2^+$-part and the $\zeta$-part indicates the extra point.
In this case one can write
\beq
\sigma_2=\left(u_7\mathbf{1}+\frac{1}{2!4!}v^{ab}\varepsilon_{abcdef}dx^{cdef}+\zeta dx^{123456}\right)\wedge dx^7+\frac{1}{3!}P_{abc}dx^{abc}\equiv\varphi\wedge dx^7+\psi.
\eeq
\noindent
Clearly, this arrangement can be related to a $D0D4D6$-brane system ($\varphi$) combined with a $D3$-brane one ($\psi$).
Suppose now that $\varphi$ and $\psi$ are \emph{nondegenerate}, i.e. $\mathcal{D}(\varphi)\mathcal{D}(\psi)\neq 0$.
Then, using the results of \cite{SaroPhD} a straightforward calculation shows that the special condition $v\llcorner P=0$ is a sufficient and necessary one for obtaining 
\beq
J_8(\sigma_2)=(\zeta^2 {u_7}^2+4u_7{\rm Pf}(v))\mathcal{D}(P)=\mathcal{D}(\varphi)\mathcal{D}(\psi)
\label{fakt}
\eeq
\noindent
provided that $\rm{Pf}(v)\neq 0$.
This result means that $J_8$ in this case is factorized to the quartic invariants of the even and odd chirality spinor representations of $Spin(12)$. These can be regarded as two truncations of the generalized Hitchin invariant of the Eq.(\ref{GenHInv}) form, where the $D3$-brane part is T-dualized to a $D0D2D4D6$ system using Eqs.(\ref{ixtilde})-(\ref{iytilde}). For $\zeta=0$ 
\beq
J_8(\sigma_2)=4u_7{\rm Pf}(v)\mathcal{D}(P), \qquad v\llcorner P=0.
\label{fakt2}
\eeq
\noindent

Let us now compare the case of $\sigma_1$ with the one of $\sigma_2$ constrained by $\zeta=0$.
Both cases have $36$ components related to incidence structures on $36$ points.
However, their underlying finite geometries are different: 
the $\sigma_1$ case has the one-point extended Klein quadric, and the $\sigma_2$ one the extended generalized quadrangle $E_2^+$.
In spite of their different underlying geometries, their (\ref{Tdual1}) and (\ref{fakt2}) $J_8$ invariants are the same provided we make the substitutions:
\beq 
u_7\leftrightarrow \zeta,\qquad v \leftrightarrow \omega,\qquad v\llcorner P=0\leftrightarrow \omega\wedge P=0.
\label{Tdualka}
\eeq
\noindent
This shows the duality of the $D0D4$ and $D2D6$ system well-known from string theory.
 
Our example of a duality shows that one can find the quartic and cubic invariants of Eqs. (\ref{Cayleygen}) and (\ref{Pfaff}) inside the eight-order one in many different ways. 
Obtaining the same structure of $J_8$ up to field redefinitions indicates that as form theories of gravity\cite{FormTH} these truncations are the same, though their geometric underpinnings are wildly different.

Finally, from Eq.(\ref{fakt}) one can also see that $Pf(v)$, as an invariant, is associated to the residue of $u_7$
familiar from the $E_2^+$ setup. Of course, relaxing the condition $v\llcorner P=0$ we discover the $35$ other residues embedded in the complicated structure of $J_8$.
That a sum of terms with Pfaffians corresponding to residues is showing up in this way is also obvious form Eq.(5.2.) of Ref.\cite{Gyoja}.  
Although the detailed finite geometric understanding of $J_8$ is yet to be achieved, these observations at least indicate that even the  $E_2^+$-part of our MVL is also featuring a natural invariant in the form of a truncation of $J_8$.

\section{Conclusions}

In this paper we have investigated the structure of the three-qubit magic Veldkamp line (MVL).
We have shown that apart from being a fascinating mathematical structure in its own right, this object provides a unifying
finite geometric underpinning for understanding the structure  of functionals used in form theories of gravity and black hole entropy.
We managed to clarify the representation theoretic, finite geometric, and invariant theoretic meaning of the different parts of our MVL.
The upshot of our considerations was that the basic finite geometric objects underlying the MVL are  the \emph{unique} generalized quadrangles $GQ(2,1),GQ(2,2)$ and $GQ(2,4)$, and their \emph{non-unique} extensions: of type $EGQ(2,1),EQG(2,2)$ and $EGQ(2,4)$.

In \cite{Levfin2} we have connected generalized quadrangles to structures already familiar from magic supergravities. They are the cubic Jordan algebras defined over the complex numbers the quaternions and the octonions. Their associated cubic invariants are related to entropy formulas of black holes and strings in \emph{five} dimensions.
In this paper we extended this analysis to also providing a finite geometric understanding of \emph{four}-dimensional black hole entropy formulas and their underlying Hitchin functionals of form theories of gravity. From the algebraic point of view, this extension is the one of moving from cubic Jordan algebras to the Freudenthal systems based on such algebras\cite{Krutelevich}.

Indeed, in this picture $GQ(2,1)$ is associated with the complex cubic Jordan algebra. The cubic invariant is the determinant of a $3\times 3$ matrix. The extension of $GQ(2,1)$ is an $EGQ(2,1)$
which is denoted by $D_2^+$ in \cite{Cameron} and associated with the corresponding complex Freudenthal system.
The quartic invariant in this case is the one underlying the Hitchin functional.
Similarly, $GQ(2,2)$ is associated with the quaternionic cubic Jordan algebra. The cubic invariant is the Pfaffian of a
$6\times 6$ antisymmetric matrix. The extension of $GQ(2,2)$ is an $EGQ(2,2)$
which is denoted by $A_2$ in \cite{Cameron}, corresponding to the quaternionic Freudenthal system.
The quartic invariant is the one underlying the generalized Hitchin functional.
The next item in the line is $GQ(2,4)$ which is associated with the split octonionic cubic Jordan algebra.
The cubic invariant in this case is Cartan's cubic one. However, in this case the extension of $GQ(2,4)$, which is an $EGQ(2,4)$
and which is denoted by $D_2^-$ in \cite{Cameron}, is \emph{not} showing up in our MVL!
Although we have already made use of \emph{truncations} of the corresponding quartic invariant of Eq.(\ref{quarticCartan}), in our considerations no part displaying the full structure of this invariant has shown up yet.
This quartic invariant is the one underlying the $E_{7(7)}$ symmetric black hole entropy formula\cite{CJ79,KK96}, the corresponding functional for form theories is the one used in connection with generalized exceptional geometry\cite{Hull,Grana} and it has an interesting interpretation as the tripartite entanglement of seven qubits\cite{Duff,LevFano,Geemen}.
Hence, it would be desirable to find a place for this important invariant in our finite geometric picture.

Clearly our MVL on \emph{three qubits} is not capable of accomodating this structure.
However, in closing this paper we show that our MVL with its associated stuctures, \emph{taken together} with the missing $D_2^-$ part,
is naturally embedded in a Veldkamp line for \emph{four qubits}. 
In order to achieve a similar level of understanding as for the MVL, we have to employ an eight-dimensional Clifford algebra with
generators $\gamma_{\hat{I}}$, where $\hat{I}=1,2,\dots 8$.
This algebra can be given a realization in terms of \emph{antisymmetric} four-qubit Pauli operators, see e.g. \cite{Richter}.
An alternative realization, more convenient for our purposes, is obtained by modifying our original seven-dimensional Clifford algebra of Eq.(\ref{choice}) as
\beq
\gamma_I=\Gamma_I\otimes X,\qquad \gamma_8={\bf 1}\otimes Y,\qquad I=1,2,\dots 7.
\label{8Cliff}
\eeq
\noindent

\begin{figure}[pth!]
\centerline{\includegraphics[width=7truecm,clip=]{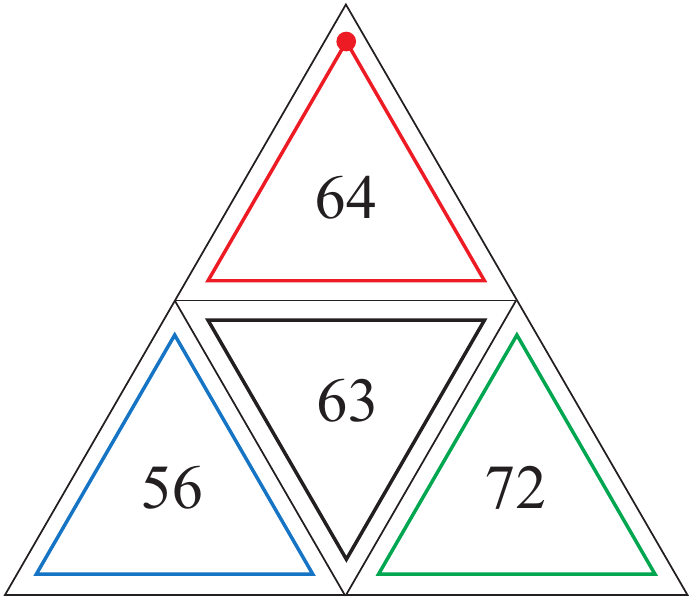}}
\vspace*{0.5cm}
\centerline{\includegraphics[width=10truecm,clip=]{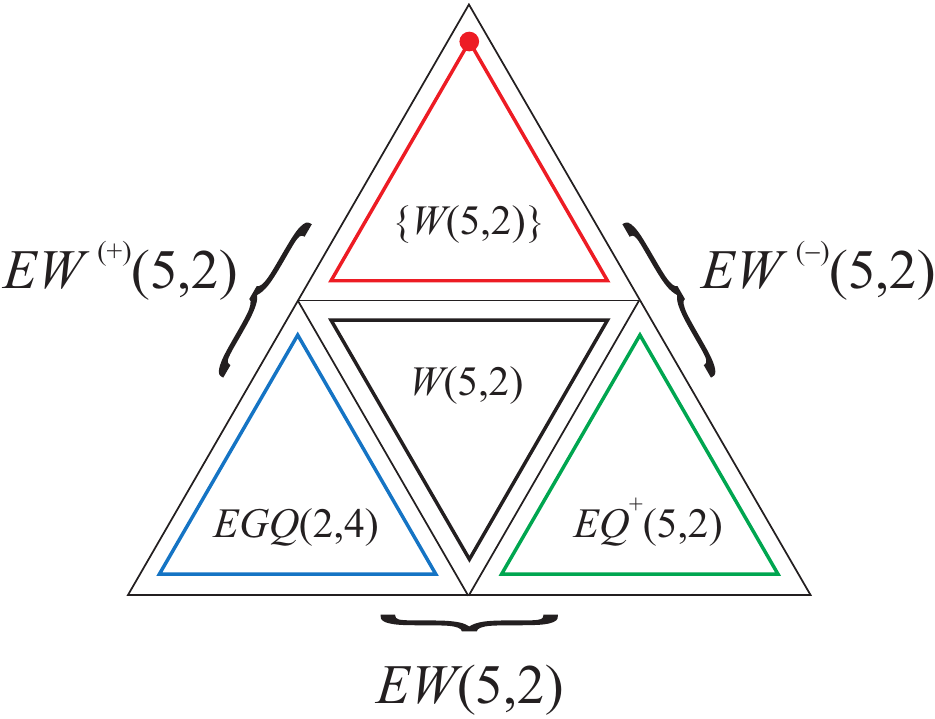}}
\caption{The characteristic numbers ({\it top}) and finite-geometric structures ({\it bottom}) for the decomposition of the four-qubit MVL (compare with Figures 5 and 14). Both the extended generalized quadrangle $EGQ(2,4)$ (of type $D_2^{-}$, see Example 9.6(i) in \cite{Cameron}) and the extended Klein quadric $EQ^{+}(5,2)$ ({\it alias} the extended dual 2-design with residues isomorphic to the duals of $PG(3,2)$, see Example 7.3(b) in
\cite{hughes}) have the property that for every point there exists a unique antipodal point; hence, they both have unique quotients isomorphic, respectively, to a one-point extension of $GQ(2,4)$ (see, e.\,g., Example 9.7 in \cite{Cameron}) and a one-point extension of the Klein quadric $Q^{+}(5,2)$ (see, e.\,g., \cite{Fisher}). For the three extensions of symplectic polar space $\mathcal{W}(5,2)$ we use our own symbols, reflecting the fact whether the point-set of the extension lies in the complement of the hyperbolic quadric ($EW^{(+)}(5,2)$), of the elliptic quadric ($EW^{(-)}(5,2)$) or of the quadratic cone ($EW(5,2)$) of $\mathcal{W}(7,2)$. The symbol $\{\mathcal{W}(5,2)\}$ stands for the projection of the core $W(5,2)$ from the vertex of the cone.}
\label{Metodfig}
\end{figure}

Let us now repeat the construction of a Veldkamp line $(C_p,H_q,H_{p+q})$ featuring a perp set, a hyperbolic and an elliptic quadric! We choose: $p\leftrightarrow\gamma_8=IIIY$ and $q\leftrightarrow IIII$. Hence, $C_p$ is comprising the operators commuting with $\gamma_8$, $H_q=H_0$ 
is consisting of the $135$ symmetric four-qubit observables not counting the identity, and finally,
$H_{p+q}$ is consisting of 
the ones that are either symmetric and commuting, or skew-symmetric and anti-commuting with $\gamma_8$.
The cardinalities of the characteristic sets of this Veldkamp line are shown in Figure \ref{Metodfig}. We have $64+63$
elements of $C_p$ and the elements of $H_{q}$ and $H_{p+q}$ split as $72+63$ and $56+63$, respectively, with the core set
having the geometry of a $\mathcal{W}(5,2)$ with $63$ points.

A subset of particular interest for us is the blue triangle of Figure \ref{Metodfig}.
This subset of cardinality $56$ has a $28+28$ split, which can be described by the following set
 of \emph{skew-symmetric} operators
\beq
\{\gamma_I,\gamma_I\gamma_J\gamma_K\gamma_L\gamma_M\}\oplus\{\gamma_I\gamma_8,\gamma_I\gamma_J\gamma_K\gamma_L\gamma_M\gamma_8\},\qquad 1\leq I<J<K<L<M\leq 7.
\label{56ofe7gamma}
\eeq
\noindent
Notice that according to Eq.(\ref{choice}) the first of these two sets comprises $28$  Hermitian observables of the form
$A\otimes X$ and the second $28$ skew-Hermitian ones of the form $iA\otimes Z$, where the three-qubit ones are skew-symmetric i.e. $A^T=-A$. 
These can be used to label the weights of the $56$-dimensional irrep of $E_7$. 
In order to see this one has to label the  $E_7$ Dynkin diagram as follows.
Add an extra node to the right of Figure \ref{E6Dynk} labelled by the pair $67$ and replace the label $456$ by $45678$.
Then, starting from the highest weight $78$, and applying transvections corresponding to the simple roots the weight diagram of the $56$ of $E_7$ is reproduced.
Note that $\gamma_8$ is anticommuting with all elements in these sets, hence, the lift of the corresponding transvection
(see $\Gamma_7$ in a similar role in the first expression of (\ref{unitaries})) acts as an involution exchanging the two $28$-element sets. 
It is easy to see that this transformation implements the involution of electric-magnetic duality we are already familiar with.

Let us now consider $\gamma_7$ as a special operator. 
It is commuting with the following set of $27=6+6+15$ operators
\beq
\{\gamma_a\gamma_8,\gamma_a\gamma_b\gamma_c\gamma_d\gamma_e\gamma_8,\gamma_a\gamma_b\gamma_c
\gamma_d\gamma_7\}\qquad 1\leq a<b<c<d<e\leq 6.
\label{kanrse7}
\eeq
\noindent  
Regarding this set of cardinality $27$ as a residue of a point represented by $\gamma_7$ it can be shown that
this set can be given the structure of a $GQ(2,4)$. Moreover, this property remains true for choosing an \emph{arbitrary point} from our $56$-element set. Continuing in this manner one can convince oneself that the blue part of Figure \ref{Metodfig} is a copy of $D_2^-$, i.e. an $EGQ(2,4)$, our missing extended generalized quadrangle.
Now, it is natural to conjecture that the quartic $E_{7(7)}$-invariant, i.e. the one associated with the Freudenthal system of the split octonionic case, can be given a form similar to the quartic ones of Eqs.(\ref{abstract})
and (\ref{errorcorr}) related to the complex and quaternionic Freudenthal ones.
In this case the relevant coset should be  $G/H$ where $G=W(E_7)/{\mathbb Z}_2\simeq Sp(6,2)$ and $H=W(E_6)$.
In order to prove this conjecture one only has to find an appropriate labelling of this coset whose identity element
is leaving invariant the canonical residue defined by Eq.(\ref{kanrse7}).
Notice also that the generalization of the polynomials (\ref{atom}) and (\ref{atomgenhitch}) in this case is trivially dictated by the Freudenthal structure. 
It is also clear that this invariant is the one encapsulating the stucture of the $D_2^-$-part of a four-qubit Veldkamp line.

One can also show that the green triangle part of Figure \ref{Metodfig}, of cardinality $72$, comprises the extension of the Klein quadric familiar from the MVL. 
Indeed, a parametrization of this part in terms of Clifford algebra elements is provided by the sets
\beq
\{\gamma_I\gamma_J\gamma_K,\gamma_1\gamma_2\gamma_3\gamma_4\gamma_5\gamma_6\gamma_7\}\oplus
\{\gamma_I\gamma_J\gamma_K\gamma_8,\gamma_1\gamma_2\gamma_3\gamma_4\gamma_5\gamma_6\gamma_7\gamma_8\}.
\eeq
\noindent
This part is decomposed into two subsets of cardinality $36=1+35$ exchanged by the lift of the transvections
generated by $\gamma_8$. Any of these subsets can be regarded as a one-point extension of the Klein quadric $Q^+(5,2)$. Since to the Klein quadric part one can naturally associate the seventh-order invariant giving rise to Hitchin's $G_2$-functional, it is an interesting question whether one can associate to this part a natural invariant of order eight based on the extension $EQ^+(5,2)$. And, if the answer is yes, what could be its physical meaning?
Based on the decomposition of Eq.(\ref{sigma1}) featuring $1+35$ quantities and giving rise to the eight-order invariant of Eq.(\ref{beszurkal}), it is natural to conjecture that the underlying physics is somehow connected to \emph{two} copies of such decompositions and two copies of the seventh-order $G_2$-invariants. 

Motivated by our success with the group $Spin(14)$ in the MVL case, one can try to arrive at a group theoretical understanding of the structure of Figure \ref{Metodfig} based on the group $Spin(16)$. Indeed it is known\cite{LevHolw} that $n$-qubits can be naturally embedded into spinors of $Spin(2n)$.  For four qubits $Spin(16)$ has two spinor representations of even or odd chirality of dimension $128$ and $128^{\prime}$ and also irreps of dimension $135$ and $120$. 
Under the decomposition of $spin(16)\supset su(8)\oplus u(1)$ we have for the even and odd chirality spinor irreps
\beq
128=1(-4)\oplus 28(-2)\oplus 70(0)\oplus\overline{28}(2)\oplus 1(4),\quad
128^{\prime}=8(-3)\oplus 56(-1)\oplus \overline{56}(1)\oplus \overline{8}(3),
\label{128lab}
\eeq
and for the last two ones
\beq
135=36(2)\oplus 63(0)\oplus \overline{36}(-2),\qquad
120= 28(2)\oplus 63(0)\oplus\overline{28}(-2)\oplus 1(0).
\eeq
\noindent
The $135$-dimensional representation can be related to the hyperbolic quadric part (symmetric $16\times 16$ matrices), and the $120$-dimensional one to the blue and red triangle parts (skew-symmetric $16\times 16$ matrices) of Figure \ref{Metodfig}. However, the naive identification of the blue and green triangle parts with a spinor representation fails, since according to Eq.(\ref{56ofe7gamma}) these parts are containing both an even and an odd number of gamma matrices, hence they are not having a definite chirality. 
This is to be contrasted with the situation of the blue and green triangles of the MVL of Figure \ref{gods-eye-clifford}. Indeed, we could identify these parts as the $32$ dimensional spinor irrep of $spin(12)$ of negative chirality.
The reason for our success in that case was that, by virtue of Eq.(\ref{gamszorz}), it was possible to convert the $\Gamma_a\Gamma_7$ part of Figure \ref{gods-eye-clifford} to the one containing an odd number of gamma matrices. 
Notice also that the one-point extended MVL corresponds to an irrep, the spinor one of negative chirality, of $Spin(14)$. A similar identification of the four-qubit Veldkamp line of Figure \ref{Metodfig} with a \emph{single} irrrep is not possible.
Hence our Veldkamp line of Figure \ref{gods-eye-clifford} is a magical one also in this respect, since it incorporates very special
representation theoretic structures. Of course this is as it should be, since our MVL is a special collection of representation theoretic data related to prehomogeneous vector spaces.

Finally let us comment on the possible physical role of Mermin squares playing in our considerations.
Throughout this paper we emphasized that grids, labelled by
Pauli observables, alias Mermin squares, are the basic building blocks of our MVL.
From the finite geometric point of view such grids are underlying the extension procedure based on residues. For example, when producing our simplest extended generalized quadrangle $EGQ(2,1)$
we used the residues of Eqs.(\ref{kanvan})-(\ref{kanvananti}).
Grids also define invariants with physical meaning (Hitchin's invariant) via the "averaging" trick of Eq.(\ref{baromiklassz}).
Moreover, from the discussion following Eq.(\ref{szamok}) we see that the grids underlying the structure of $EGQ(2,1)$
are related to the grids of the doily residing in the core of the MVL. In particular, the two antipodal residues of Eqs.(\ref{kanvan})-(\ref{kanvananti}) give rise to the \emph{same} grid of the core doily. 
Via Fano heptads, like the one of Eq.(\ref{specifanok}),
this relationship also connects the stucture of the seventh-order invariant underlying Hitchin's $G_2$-functional, to the one of the Klein-quadric. Continuing in this manner we have seen that, using the idea of extended geometries, one can build up the whole MVL.
These considerations show the fundamental nature of the $10$ grids, similar to the ones of Figure \ref{mermins}, residing in the core doily.
Recall also that apart from incidence, the Mermin squares also incode information on \emph{signs}.
These signs are implemented into the structure of invariants via the (\ref{unitaries}) lifts of the transvections, which represent the generators of the automorphism groups of the finite geometric structures (see e.g. Eq.(\ref{antipodalmapampl})).
According to Figure \ref{mermins} there are $10$ Mermin squares inside the doily. For a particular three-qubit labelling they represent different embeddings of these objects as geometric hyperplanes inside the embedding geometry. Geometric hyperplanes (Mermin squares) in some sense act like codewords embedded into the larger environment of the mother geometry.
What kind of information might grids, when regarded as Mermin squares, encode?

Within the context of black hole solutions arising from wrapped brane configurations one possible answer to this question is as follows. In the type IIA duality frame the $15$ lines of the doily correspond to the 15 possible two cycles of a $T^6$ 2-branes can wrap. Hence, there are normally $15$ different brane charges. However, when we are considering merely supersymmetric configurations only $9$ from the charges are non-vanishing\cite{Vijay}. These charges can be assembled into a charge matrix which has the index structure $q_{\alpha\overline{\beta}}$ with respect to a fixed complex structure of the $T^6$. This structure is similar to the index structure of the expansion coefficient of the observable of Eq.(\ref{MerminGrid}). In particular, this structure refers to an $U(3)\times U(3)$ subgroup
of $SO(6)$ which encapsulates the possible rotations in the fixed complex structure of $T^6$. In the quantum theory the automorphism group of the grid should then correspond to the relevant discrete subgroup of this group.
The classical supersymmetric black hole solution with the above features has been constructed in \cite{Vijay}. It turns out that this solution is characterized by an extra charge: the $D6$-brane charge, hence we altogether have a $10$ parameter seed solution.
Some discrete data of this solution possibly can be regarded as a codeword, or in finite geometric terms a residue.  
The remaining $6$ parameters describe how the seed solution is embedded into the full $15$ parameter one.
These remaining $6$ parameters arise from global $SO(6)/U(3)$
rotations that deform the complex structure of the solution. At the quantum level these transformations should boil down to the discrete set of transformations generated by the transvections showing up in Eq.(\ref{baromiklassz}).
According to \cite{Vijay} it is not possible to add additional $2$-branes that lie along these additional $6$ cycles
consistently with supersymmetry. So if we regard the discrete information (e.g. the distribution of signs of the $9$ brane charges) embedded into the full 15 parameter solution, as some message, then this information is in some sense protected from errors of a very particular kind, namely global rotations of the complex structure.
These observations might lead to a further elaboration of the analogy already noticed between error correcting codes and the structure of BPS and non-BPS STU black hole solutions\cite{LevayBPSnonBPS}.
We would like to explore these interesting ideas in a subsequent publication.

\section{Acknowledgements}

This work is supported  by the French ``Investissements d'Avenir'' program, project ISITE-BFC (contract ANR-15-IDEX-03).
This work was also supported by the Slovak VEGA Grant Agency, Project No. 2/0003/16, as well as the Franche-Comt\'e Conseil R\'egional Research Project RECH-MOB15-000007. We are extremely grateful to our friend Petr Pracna for his help with several figures.

\end{document}